\documentclass[11pt]{article}
\pdfoutput=1

\usepackage[usenames,dvipsnames,svgnames,table]{xcolor} % use color
\usepackage[obeyspaces,hyphens,spaces]{url}
\usepackage{jcapmod}

\usepackage{shorthand}
\usepackage{mathtools}
\usepackage{booktabs}
\usepackage[english]{babel}
\usepackage{amsmath,amssymb,amsbsy,amstext, amsthm, simplewick, amsfonts,braket}
\usepackage{hyperref}
\usepackage{graphicx}
\usepackage[small]{caption}
\usepackage{slashed}
\usepackage[separate-uncertainty=true,multi-part-units=single]{siunitx}
\usepackage{upgreek}
\usepackage{framed}
\usepackage{wrapfig}
\usepackage{multirow}
\usepackage{bbm}

\usepackage[utf8x]{inputenc}
\usepackage{selinput}
\usepackage{bm}
\usepackage{float}
\usepackage{geometry}
\usepackage{yfonts}
\usepackage{caption}
\usepackage{subcaption}
\usepackage{sidecap}
\usepackage{longtable}
\usepackage{anyfontsize}
\usepackage{dsfont}
\usepackage{tikz}
\usepackage{relsize}
\usepackage{tcolorbox}

\newcommand{\ud}{\mathrm{d}}
\newcommand{\lab}[1]{{\mathrm{#1}}}
\newcommand{\slab}[1]{{\textsc{#1}}}
\newcommand{\mb}[1]{{\mathbf{#1}}}
\newcommand{\minus}{{\scalebox {0.75}[1.0]{$-$}}}
\newcommand{\sminus}{{\scalebox {0.6}[0.85]{$-$}}}
\newcommand{\sperp}{{\scalebox{0.7}{$\perp$}}}

\def\r{{\bf{r}}}

\newcommand{\floq}[1]{{\scalebox{0.65}{$(#1)$}}}

\newcommand{\subp}{{\scriptscriptstyle +}}
\newcommand{\subm}{{\scriptscriptstyle -}}
\newcommand{\subpm}{{\scriptscriptstyle \pm}}

\usepackage{colortbl}
\definecolor{lightgreen}{cmyk}{0.2, 0, 0.2, 0.2}
\definecolor{lightgray2}{cmyk}{0.1,0.1,0,0.1}
\definecolor{Red2}{RGB}{214, 39, 40}
\definecolor{Blue2}{RGB} {31, 119, 180}
\definecolor{Orange2}{RGB}{255, 127, 14}
\definecolor{Green2}{RGB}{44, 160, 44}
\definecolor{greyish2}{rgb}{.96,.96,.96}

\makeatletter
\newlength{\apb@width}
\newcommand{\autoparbox}[2][c]{\settowidth{\apb@width}{#2}\parbox[#1]{\apb@width}{#2}}

\makeatother


\allowdisplaybreaks[1]
\newcommand{\ped}[1]{\textormath{\textsubscript{#1}}{_{\mathrm{#1}}}}

\DeclarePairedDelimiter{\abs}{\lvert}{\rvert}

\renewcommand{\vec}[1]{\boldsymbol{\mathbf{#1}}}

\newcommand{\dd}{\mathop{\mathrm{d}\!}{}}

\DeclareMathOperator\sgn{sgn}

\DeclareMathOperator\erf{erf}
\DeclareMathOperator\Li{Li}
\DeclareMathOperator\dilog{dilog}
\DeclareMathOperator\HeunC{HeunC}
\DeclareMathOperator\erfc{erfc}

\newcommand{\rBC}{r\ped{c}}

\newcommand{\es}{\hspace{0.5pt}}

\renewcommand{\Im}{\operatorname{Im}}
\renewcommand{\Re}{\operatorname{Re}}

\newcommand{\compell}{\ell_*}
\newcommand{\compm}{m_*}

\definecolor{blue2}{cmyk}{1, 0.1, 0.1, 0.1}

\definecolor{pyBlue}{RGB}{31, 119, 180}
\definecolor{pyRed}{RGB}{214, 39, 40}
\definecolor{pyGreen}{RGB}{44, 160, 44}
\definecolor{pyBlue2}{RGB}{0, 111, 237}
\definecolor{pyRed2}{RGB}{224, 52, 36}
\definecolor{Mathematica1}{rgb}{0.368417, 0.506779, 0.709798}
\definecolor{Mathematica2}{rgb}{0.880722, 0.611041, 0.142051}

\def\beq{\begin{equation}}
\def\eeq{\end{equation}}

\title{Inspiral Inside a Gravitational Atom}
\date{\today}

\begin{document}

\pagenumbering{roman}
\begin{titlepage}
\baselineskip=15.5pt \thispagestyle{empty}

\phantom{h}
\vspace{1cm}
\begin{center}
{\fontsize{22}{24}\selectfont  \bfseries  Ionization of Gravitational Atoms}
\end{center}

\vspace{0.5cm}
\begin{center}
{\fontsize{12}{18}\selectfont Daniel Baumann,$^{1,2,3}$  Gianfranco Bertone,$^{1}$ John Stout$^{4}$ and  Giovanni Maria Tomaselli$^{1}$} 
\end{center}

\vspace{-0.1cm}
\begin{center}
     \vskip 6pt
\textit{$^1$ Gravitation Astroparticle Physics Amsterdam (GRAPPA),\\
University of Amsterdam,  Amsterdam, 1098 XH, The Netherlands}

  \vskip 6pt
\textit{$^2$  Center for Theoretical Physics, National Taiwan University, Taipei 10617, Taiwan}

  \vskip 6pt
\textit{ $^{3}$ Physics Division, National Center for Theoretical Sciences, Taipei 10617, Taiwan} \\

  \vskip 6pt
\textit{$^4$  Department of Physics, Harvard University, Cambridge, MA 02138, USA}
\end{center}

\vspace{1.2cm}
\hrule \vspace{0.3cm}
\noindent {\bf Abstract}\\[0.1cm]
Superradiant instabilities may create clouds of ultralight bosons around rotating black holes, forming so-called “gravitational atoms.”  It was recently shown that the presence of a binary companion can induce resonant transitions between bound states of these clouds, whose backreaction on the binary's orbit leads to characteristic signatures in the emitted gravitational waves. In this work, we show that the interaction with the companion can also trigger transitions from bound to unbound states of the cloud---a process that we refer to as ``ionization'' in analogy with the photoelectric effect in atomic physics. The orbital energy lost in the process overwhelms the losses due to gravitational wave emission and contains sharp features carrying information about 
the energy spectrum of the cloud.
Moreover, we also show that if the companion is a black hole, then the part of the cloud impinging on the event horizon will be absorbed. This ``accretion'' leads to a significant increase of the companion's mass, which alters the dynamical evolution and ensuing waveform of the binary. We argue that a combined treatment of resonances, ionization, and accretion is crucial to discover and characterize gravitational atoms with upcoming gravitational wave detectors.

\vskip10pt
\hrule
\vskip10pt

\end{titlepage}

\thispagestyle{empty}
\setcounter{page}{2}
\tableofcontents

\newpage
\pagenumbering{arabic}
\setcounter{page}{1}

\clearpage

\section{Introduction}

Black holes are remarkably simple objects. The spacetime around a black hole is uniquely determined by its mass and spin, and the gravitational waves (GWs) released in the merger of two black holes can be predicted very precisely. This makes black holes exceptionally clean environments to probe the fundamental laws of nature~\cite{Barack:2018yly,Barausse:2020rsu,Bertone:2019irm,Chia:2020dye}, with any deviation from the predictions of general relativity being an indication of new physics.  

\vskip 4pt
A particularly well-studied example of new physics, %JS removed "that is"
accessible with future GW observations, are ultralight bosons. Such bosons can be generated by superradiance~\cite{Brito:2015oca}, forming long-lived condensates (``clouds") around rotating black holes~\cite{Arvanitaki:2009fg,Arvanitaki:2010sy,East:2017ovw,East:2018glu}. Due to their strong similarity with the hydrogen atom, such systems have been called ``gravitational atoms."
For isolated gravitational atoms, there are essentially two ways of inferring the presence of these boson clouds. First, rotating clouds will emit gravitational waves~\cite{Arvanitaki:2010sy} that can be looked for in continuous-wave searches~\cite{LIGOScientific:2021jlr}. Second, the clouds extract spin from their parent black holes and this spin-down can be inferred statistically in a population of rotating black holes~\cite{Arvanitaki:2010sy, Brito:2014wla, Brito:2017zvb,Ng:2019jsx,Ng:2020ruv,Fernandez:2019qbj}. 
The existence of rapidly spinning black holes would then rule out ultralight bosons in a certain mass range. Unfortunately, neither of these effects is very distinctive, so it is hard to use them as a way of unambiguously discovering gravitational atoms in the sky. 

\vskip 4pt
Recently, a new avenue for detecting gravitational atoms has been explored which exploits their effects in binary systems.
When a gravitational atom is part of a binary it gets 
perturbed by the companion. As was shown in~\cite{Baumann:2018vus}, the gravitational interaction between the companion and the boson cloud is resonantly enhanced when the orbital frequency matches the energy difference between two eigenstates of the cloud; see \cite{Zhang:2019eid,Berti:2019wnn,Zhang:2018kib,Su:2021dwz,Tong:2021whq,Ding:2020bnl,Takahashi:2021eso,Choudhary:2020pxy, Asali:2020wup, Ikeda:2020xvt, Wong:2020qom, Cardoso:2020hca} for related work. This leads to  an analog of the 
Landau--Zener transition in quantum mechanics~\cite{landau1932theorie,zener1932non, Baumann:2019ztm}, where the companion forces the cloud to smoothly transition from one state to another.
These transitions are  
a distinctive fingerprint of a boson cloud. 

\begin{figure}[t!]
            \centering
            \includegraphics[trim = {0 14pt 0 0}]{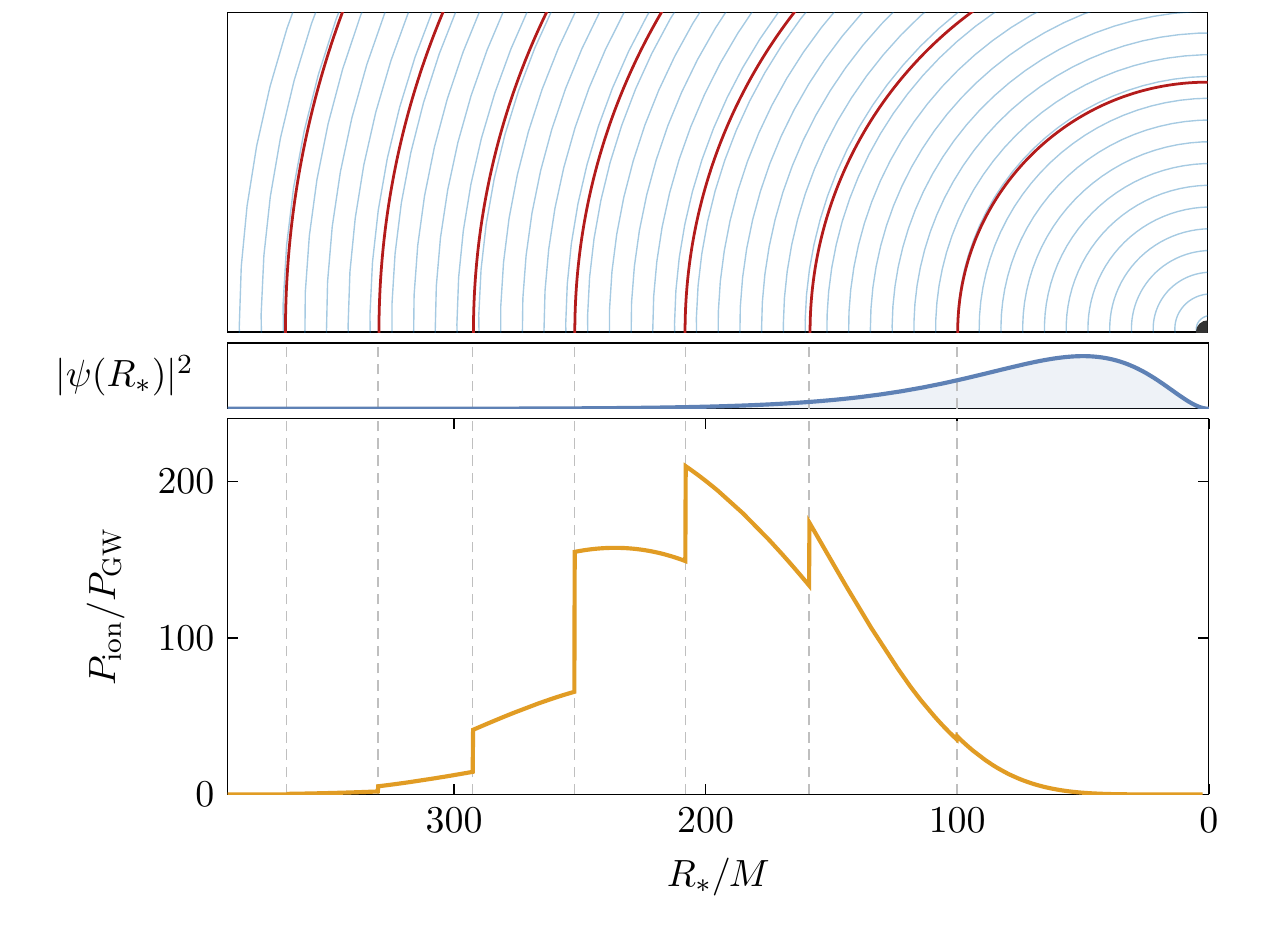}
            \caption{Schematic illustration of the ionization of the gravitational atom. In the bottom panel, we plot the ratio of the ionization power $P_\lab{ion}$ and the power lost due to GW emission $P_\slab{gw}$. We see that the energy loss due to ionization can overwhelm that due to GW emission and hence dominate the binary's dynamics.
           The signal has sharp features when the bound state begins to resonate with the continuum, which occurs at specific separations $R_*$. Shown also is the density profile of the cloud, $|\psi(R_*)|^2$, for a $|211\rangle$ bound state.  }
            \label{fig:IonizationSchematic}
        \end{figure}

\vskip 4pt
In this paper, we extend the treatment of~\cite{Baumann:2018vus,Baumann:2019ztm} to allow for transitions to unbound states of the cloud.
When these transitions are effective, the cloud escapes from the parent black hole and the gravitational atom gets ``ionized," like in the photoelectric effect for ordinary atoms. 
Figure~\ref{fig:IonizationSchematic} illustrates the main result of this analysis. Shown is the ionization power $P_\lab{ion}$, the rate of energy lost by the binary due to ionization, as a function of the separation $R_*$ between the parent black hole and companion for typical parameters of the system.
We see that the effect of ionization can be orders of magnitude larger than the rate of energy lost due to GW emission, $P_\slab{gw}$, and therefore dominate the binary's dynamics.
Moreover, this ionization power is not a smooth function of $R_*$, but contains interesting ``discontinuities''
at specific separations.  These sharp features arise when the bound state begins to resonate with the continuum. Like the resonant transitions between bound states~\cite{Baumann:2018vus,Baumann:2019ztm}, the ionization signal therefore contains very distinctive information about the microscopic structure of the cloud. We will show how these features are imprinted in the binary's dynamics, and thus in the emitted gravitational waves.

\vskip 4pt
For the first time, we also include the accretion of the cloud onto the companion. Although accretion is suppressed for the wave-like boson cloud (compared to an equal density of particles), it is nevertheless a large effect, since the typical densities of the boson clouds are large. In many cases, the mass of the companion can change by up to an order-one fraction during the inspiral, leading to a significant speed-up of the merger compared to the vacuum evolution. Unlike the ionization signal, the effect of accretion is a smooth function of the separation $R_*$ and hence more degenerate with changes in the source parameters.

\vskip 4pt
We conclude that ionization and accretion play a critical role in the phenomenology of gravitational atoms in binaries. Rather remarkably, the effects can be so large that they overwhelm the energy lost due to GW emission and therefore drive the inspiral (rather than just being a small perturbation). A consistent treatment of these systems must therefore take these effects into account, as well as their interplay with the resonances between bound states studied in~\cite{Baumann:2018vus,Baumann:2019ztm}.

\paragraph{Outline} The outline of the paper is as follows: In Section~\ref{sec:review}, we review the energy eigenstates of the gravitational atom and describe the resonant transitions between bound states. In Section~\ref{sec:ionization}, we study transitions to unbound states and describe the ionization of the boson cloud. 
Special attention is paid to the sharp resonance features in the ionization power.
In Section~\ref{sec:accretion}, we compute the accretion of the wave-like boson cloud onto the companion black hole.
In Section~\ref{sec:binary-evolution}, we show how both ionization and accretion change the dynamics of the binary. We present numerical results for a few representative examples. Finally, we state our conclusions and discuss open problems in Section~\ref{sec:conclusions}.  
 
 \newpage
 A number of appendices contain technical details. In Appendix~\ref{app:approx}, we describe the various approximations that are used in Section~\ref{sec:ionization} to integrate out the continuum states and derive  the effective dynamics of the bound states.  Such a description only holds in the Markov approximation, whose validity we discuss in Appendix~\ref{app:Markov}, and we argue there that this approximation applies to the systems we consider in the main text. In Appendix~\ref{app:ionizedEnergy}, we derive an approximation for the ionization power, which measures how quickly the companion transfers energy from bound to unbound states. In Appendix~\ref{app:zeroMode}, we discuss the low-energy limit of the unbound states and describe under which conditions the discontinuities seen in Figure~\ref{fig:IonizationSchematic} appear. Finally, in Appendix~\ref{app:heunc}, we describe the exact solutions of the Klein--Gordon equation in the Kerr geometry, and discuss an approximation relevant for our derivation of the accretion rate in Section~\ref{sec:accretion}.

 \paragraph{Notation and conventions} Our metric signature will be $(-,+, +, +)$ and, unless stated otherwise, we will work in natural units with $G = \hbar = c = 1$. Greek letters will stand for spacetime indices.   
  Quantities associated to the boson clouds will be denoted by the subscript~$\lab{c}$.
  For example, the initial mass and angular momentum of the cloud are $M_\lab{c}$ and $S_\lab{c}$, respectively.
  The gravitational fine-structure constant, $\alpha = \mu M$, is the ratio of the gravitational radius of the black hole (which in natural units is simply $r_\lab{g} = M$) and the (reduced) Compton wavelength of a boson field, $\lambda_\slab{c}=\mu^{\sminus 1}$, where $\mu$ is the mass of the field.
 
  \vskip 4pt
  The Kerr metric for a black hole of spin $J$ is 
  \begin{equation}
  \ud s^2 = - \frac{\Delta}{\rho^2}\left(\ud t - a \sin^2 \theta\, \ud \phi \right)^2 + \frac{\rho^2}{\Delta} \es \ud r^2 + \rho^2\, \ud \theta^2 + \frac{\sin^2 \theta}{\rho^2}  \left(a\, \ud t - (r^2 + a^2) \, \ud \phi \right)^2\, , \label{equ:Kerr}
  \end{equation}
where $a\equiv J/M$, $\Delta \equiv r^2 -  2 Mr +a^2$ and $ \rho^2 \equiv r^2 + a^2 \cos^2 \theta$.  
  The roots of $\Delta$ determine the inner and outer horizons, located at $r_\subpm = M \pm \sqrt{M^2 -a^2}$, and the angular velocity  at the outer horizon is $\Omega_+ \equiv a/ 2 M r_\subp$. Dimensionless quantities, defined with respect to the black hole mass~$M$, are labeled by tildes. For example, the dimensionless spin of the black hole is~$\tilde{a} \equiv a/M$. We use an asterisk to denote quantities associated to the black hole companion; for instance, $M_*$ and $a_*$ are the mass and spin of the companion, while $q = M_*/M$ is the ratio of the black hole masses.

\newpage
\section{Gravitational Atoms in Binaries}
\label{sec:review}

We begin with a brief review of the structure of the gravitational atom.
We start, in Section~\ref{sec:spectra}, by describing the bound and unbound spectra of the atom in isolation. 
 In Section~\ref{sec:Perturbed}, we 
  explain how a binary companion perturbs this atom, mediating transitions between different states.  We then describe the case of resonant transitions between bound states in Section~\ref{sec:resonances}.

    \subsection{Scalar Field around Kerr} \label{sec:spectra}

\begin{figure}[t!]
      \centering
      \includegraphics[trim={0 0pt 0pt 0}]{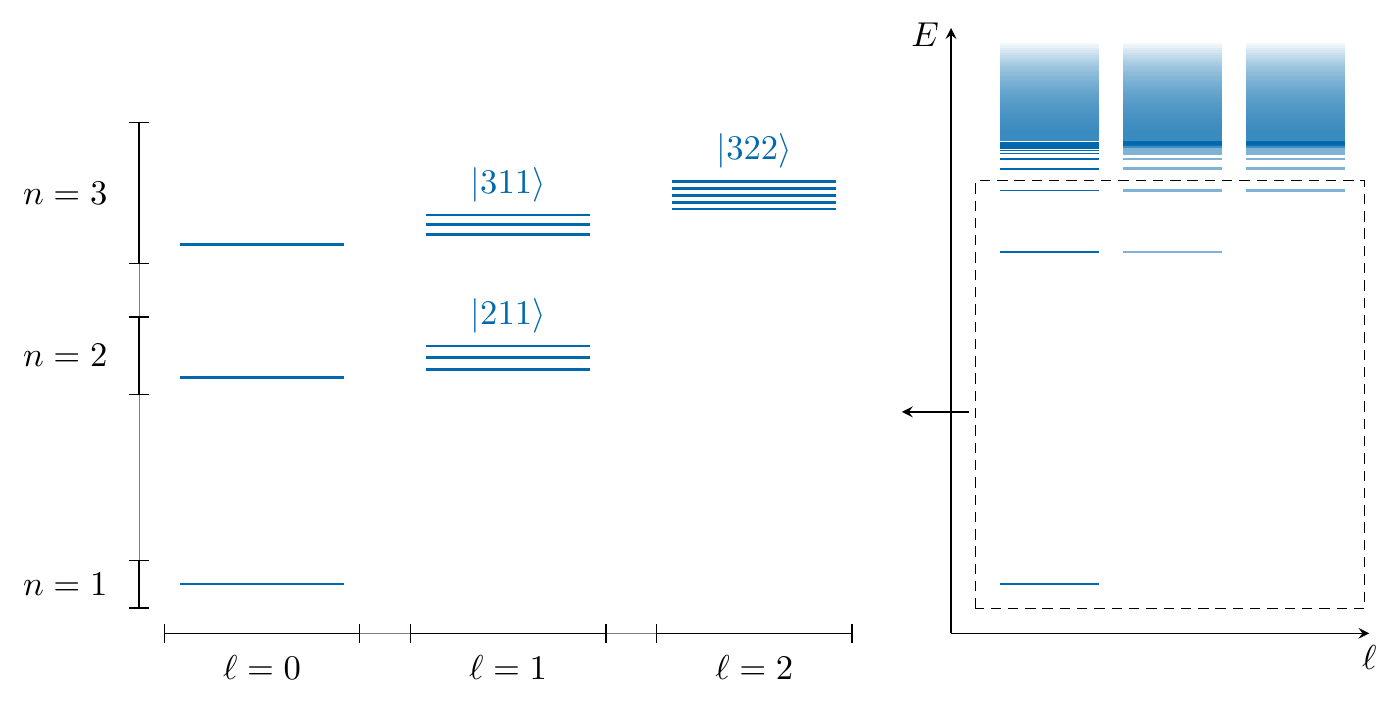}
      \caption{Illustration of the spectrum of bound and unbound states of the gravitational atom. \label{fig:states}}
    \end{figure}

    The Klein--Gordon equation for a scalar field of mass $\mu$ in a curved spacetime is
    \begin{equation}
        \left(g^{\alpha \beta} \nabla_\alpha \nabla_\beta - \mu^2\right) \Phi(t, \mb{r}) = 0\,. \label{eq:kgEom}
    \end{equation}
    As is well known, in the Kerr background (\ref{equ:Kerr}), the Klein--Gordon equation admits bound state solutions that are remarkably similar to those of the hydrogen atom. When the Compton wavelength of the field is much larger than the gravitational radius of the black hole, $\alpha \equiv r_\lab{g}/\lambda_\slab{c} \ll 1$, it is useful to consider the following ansatz
    \begin{equation}
        \Phi(t, \mb{r}) = \frac{1}{\sqrt{2 \mu}} \left[\psi(t, \mb{r}) e^{-i \mu t} + \psi^*(t, \mb{r}) e^{+i \mu t}\right] , \label{eq:nrField}
    \end{equation}
    where $\psi$ is a complex scalar field which varies on timescales much longer than $\mu^{\sminus 1}$, see e.g.~\cite{Baumann:2019eav}. If $\Phi$ is itself a complex scalar field, then we only use the first term in (\ref{eq:nrField}). We will often refer to $\psi$ as the wavefunction of the cloud.
  Far from the black hole and at leading order in~$\alpha$, the Klein--Gordon equation (\ref{eq:kgEom}) is then identical to the Schr\"odinger equation for the hydrogen atom,
    \begin{equation}
        i \frac{\partial}{\partial t} \psi(t, \mb{r}) = \left(-\frac{1}{2 \mu} \nabla^2 - \frac{\alpha}{r}\right) \psi(t, \mb{r})\,. \label{eq:schrodinger}
    \end{equation}
   In this limit, the scalar field can be studied using standard techniques of nonrelativistic quantum mechanics. This Schr\"{o}dinger equation permits two qualitatively different sets of eigenstates (see Fig.\,\ref{fig:states}), whose properties we will now review.

    \subsubsection*{Bound states}
    
 We first consider the familiar bound state solutions, which are labeled by three integers: a principal ``quantum number'' $n$, orbital angular momentum $\ell$, and azimuthal angular momentum~$m$. At leading order in $\alpha$, these bound state solutions have the form
    \begin{equation}
        \psi_{n \ell m}(t, \mb{r}) = R_{n \ell}(r) Y_{\ell m}(\theta, \phi) e^{-i(\omega_{n \ell m} - \mu) t}\,, \label{eq:boundStates}
    \end{equation}
    where $Y_{\ell m}(\theta, \phi)$ are spherical harmonics and $R_{n \ell}(r)$ are the hydrogenic radial functions. 
    The latter are given by
    \begin{equation}
        R_{n \ell}(r) = \sqrt{\left(\frac{2 \mu \alpha}{n}\right)^3 \frac{(n - \ell - 1)!}{2 n (n+ \ell)!}} \left(\frac{2 \alpha \mu r}{n}\right)^\ell \exp\!\left(\minus \frac{\mu \alpha r}{n}\right) L_{n - \ell - 1}^{2 \ell +1}\!\left(\frac{2 \mu \alpha r}{n}\right), \label{eq:boundWavefunctions}
    \end{equation}
    where $L_{n- \ell -1}^{2 \ell+1}(x)$ is the associated Laguerre polynomial. For small values of $\alpha$, the radial profile peaks at a multiple of the ``Bohr radius''  $r_\lab{c} \equiv (\mu \alpha)^{\sminus 1}$ and decays exponentially as $r \to \infty$.
    These bound state solutions are defined for $n \geq \ell+1$, $\ell \geq 0$, and $\ell \geq |m|$. For notational simplicity, it will be convenient to lean on the quantum mechanical analogy and represent (\ref{eq:boundStates}) using the bra-ket notation $|n \es \ell \es m \rangle$. The  normalization of the bound states is chosen so that 
    \begin{equation}
        \langle n \es \ell \es m | n' \es \ell' \es m' \rangle = \int\!\ud^3 r\, \psi^*_{n \ell m}(t, \mb{r}) \psi_{n' \ell' m'}(t, \mb{r}) = \delta_{nn'} \delta_{\ell \ell'} \delta_{m m'}\,.
    \end{equation}
    The amplitude of (\ref{eq:boundStates}) is determined by the total mass of the cloud and will be restored when necessary.  

\vskip 4pt
       There is one important difference between the hydrogen atom and the gravitational atom. While the wavefunctions of the former are regular at $r = 0$, the latter must be purely ingoing at the black hole's outer horizon since no physical mode can escape from the black hole. This ``dissipative'' boundary condition forces the bound state eigenfrequencies of the boson cloud  to be complex,
    \begin{equation}
        \omega_{n \ell m} = E_{n \ell m} + i \Gamma_{n \ell m}\,,
    \end{equation} 
    where $E_{n \ell m}$ and $\Gamma_{n \ell m}$ denote the energies and instability rates, respectively. At leading order in $\alpha$, these are 
    \begin{align}
        E_{n \ell m} &= \mu\left(1 - \frac{\alpha^2}{2 n^2}  + \mathcal{O}\big(\alpha^4\big)\right) ,\label{eq:boundEnergies}\\
        \Gamma_{n \ell m} &= 2 \tilde{r}_\subp C_{n \ell} \hskip 1pt g_{\ell m}(\tilde{a}, \alpha, \omega)(m \Omega_\subp - \omega_{n \ell m}) \es \alpha^{4 \ell + 5} + \mathcal{O}\big(\alpha^{4 \ell + 7}\big)\,,
    \end{align}
    where the numerical coefficients $C_{n \ell}$ and $g_{\ell m}$ can be found in \cite{Baumann:2019eav}. As discussed there, these bound states are still labeled by the ``quantum'' numbers $n$, $\ell$, and $m$, and the latter two reduce to the orbital and azimuthal angular momenta of the cloud in the $\alpha \to 0$ limit. Crucially, the nonzero instability rates allow a rapidly spinning black hole to spontaneously shed a sizable fraction of its mass and angular momentum to form the boson cloud. Even though these rates are highly suppressed for $\alpha \ll 1$, the cloud can still grow very quickly on astrophysical timescales. The gravitational atom is the endpoint
     of this process. Since the state $|2 \es 1 \es 1 \rangle$ grows fastest, we will take this as the initial configuration of the cloud when the binary inspiral begins.

\vskip 4pt
    In the non-relativistic limit, a cloud in a state $|\psi \rangle$, with wavefunction $\psi(t, \mb{r})$, has mass density  
    \begin{equation}
        \rho(t,\mb{r}) = \left\{ \begin{array}{ll} M_\lab{c} |\psi(t,\mb{r})|^2 & \quad  \text{(complex field)\,,}\\[6pt]
         2 M_\lab{c}|\!\es\es \Re \psi(t,\mb{r})\es|^2   & \quad  \text{(real  field)\,,}
         \end{array}  \right. \label{eq:massDensity}
    \end{equation}
    where $M_\lab{c}$ is the initial mass of the cloud. By convention, we require that the cloud's wavefunction is initially unit normalized, $\langle \psi | \psi \rangle = 1$. Superradiant growth can be quite efficient and, depending on the initial spin of the parent black hole, the mass of the cloud $M_\lab{c}$ can be a significant fraction of the total mass of the system (up to $0.1 M$, where $M$ is the mass of the central black hole).    Since the typical size of the cloud $r_\lab{c}$ is between 10 and $10^3$ times the Schwarzschild radius of the parent black hole for typical values of $\alpha$, the cloud can be an exceptionally dense region of matter compared to other astrophysical environments. For example, if the cloud sits around a stellar mass black hole with $M = 10 M_\odot$, then the average mass density is between~$10^8$~and~$10^{12}\,\lab{kg/m^3}$. 
     On the other hand, around an intermediate mass black hole $M = 10^{5} M_{\odot}$, the cloud can be much more spread out so that its average mass density is ``only''~$1$~to~$10^{4} \, \lab{kg/m^3}$. As a point of reference, the density of water is~$\rho_{\lab{H}_2\lab{O}} = 10^{3}\, \lab{kg/m^3}$, so an inspiralling  black hole companion moving through the cloud encounters a medium that can be potentially much denser than water. As we will see, the associated large flux of mass through the companion's horizon can strongly impact the dynamics of the inspiral.

    \subsubsection*{Continuum states}

The Schr\"{o}dinger equation (\ref{eq:schrodinger}) also permits continuum state solutions. In addition to the orbital and azimuthal angular momentum 
$\ell$
 and $m$, these solutions are labeled by a positive, real-valued wavenumber $k$,
    \begin{equation}
        \psi_{k; \ell m}(t, \mb{r}) = R_{k;\ell}(r) Y_{\ell m}(\theta, \phi) e^{\sminus i \epsilon_{\ell m}(k) t}\,.
    \end{equation}
    We distinguish the continuous index by a trailing semicolon and use the bra-ket notation~$|k;\ell \es m \rangle$. In the hydrogen atom, these continuum states represent those states  in which the electron has been unbound from the proton, and can thus be thought of as scattering states. A similar interpretation applies to the gravitational atom: these states represent the situation in which the scalar field is not bound to the black hole. The continuum radial functions are given by
    \begin{equation}
        R_{k; \ell}(r) = \frac{2 k e^{\frac{\pi \mu \alpha}{2 k}} |\Gamma(\ell + 1 + \tfrac{i \mu \alpha}{k})|}{(2 \ell+1)!} (2 k r)^{\ell} e^{\sminus i k r} {}_1 F_{1}(\ell + 1 + \tfrac{i \mu \alpha}{k}; 2 \ell + 2; 2 i k r)\,, \label{eq:contWavefunctions}
    \end{equation}
    where ${}_1 F_1(a; b; z)$ is the Kummer confluent hypergeometric function. In contrast to the bound states, these continuum states do not decay exponentially  as $r \to \infty$ and are not unit-normalizable. The normalization is instead chosen so that
    \begin{equation}
        \langle k; \ell \es m | k'; \ell' \es m' \rangle = \int\!\ud^3 r\, \psi^*_{k; \ell m}(t, \mb{r})\, \psi_{k'; \ell' m'}(t, \mb{r}) = 2 \pi \delta(k - k') \delta_{\ell \ell'} \delta_{m m'}\,, \label{eq:contNormalization}
    \end{equation}
    i.e. these continuum states are $\delta$-function normalized.

\vskip 4pt  
    Since the boundary conditions for these continuum states are much less restrictive than those for the bound states, the exact eigenfrequencies are known and are purely real $\omega(k) = \sqrt{\mu^2 + k^2}$, with~$k \in [0, \infty)$. We will work  in the non-relativistic limit, $k \ll \mu$, where the dispersion relation for the continuum states is
    \begin{equation}
        \epsilon(k) \equiv \sqrt{\mu^2 + k^2} - \mu \approx \frac{k^2}{2 \mu}\,. \label{eq:contEnergies}
    \end{equation}
    In Section~\ref{sec:realisticIonization}, we will find that only the continuum states with $k \sim \mathcal{O}\big(\mu \alpha^2\big)$ play an important dynamical role and, since we will always work in the limit $\alpha \ll 1$, we will not need to consider corrections to the non-relativistic approximation. 

\vskip 4pt
According to the normalization condition (\ref{eq:contNormalization}), the continuum states are linearly distributed in $k$; that is the density of states behaves as $\ud n \propto \ud k$. However, in terms of the energy $\epsilon$, this density of states \emph{diverges} as $\epsilon \propto k^2 \to 0$:
  \begin{equation}
    \ud k = \frac{\mu \, \ud \epsilon}{k(\epsilon)}\,. \label{eq:densityOfStates}
  \end{equation} 
  This diverging density of states at low energies will play a crucial role in the ionization effects we describe in the next section.

\vskip 4pt
    An important related property of the continuum wavefunctions is that they vanish as $\sqrt{k}$ in the soft limit~$k \to 0$. As we discuss in Appendix~\ref{app:zeroMode}, this behavior is ultimately due to the long-range nature of the gravitational potential, and we show there that 
    \begin{equation}
        R_{k; \ell m}(r) \sim \sqrt{\frac{4 \pi k}{r}} J_{2 \ell +1}\big(2 \sqrt{2 \alpha \mu r}\hskip 1pt \big) \,, \mathrlap{\qquad k \to 0\,,} \label{eq:zeroModeAsympt}
    \end{equation}
    where $J_{\nu}(z)$ is the Bessel function of the first kind. In contrast to the free particle, the long-range Coulombic potential localizes the zero mode to a Bohr radius-sized region around $r = 0$, instead of spreading out over all of space. As we will discuss in Section~\ref{sec:ionization}, this seemingly innocuous behavior, combined with the divergent density of states (\ref{eq:densityOfStates}), is responsible for dramatic discontinuities in the ionization power during the inspiral.

\subsection{Perturbation from the Companion} \label{sec:Perturbed}
    
    Our main goal is to understand the dynamics of the cloud during a binary inspiral. To this end, we must describe the effect that the binary companion has on the cloud through its gravitational field. This can be encoded in an additional potential term in the Schr\"{o}dinger equation~(\ref{eq:schrodinger}) and in this section we will describe both the structure of this potential and the transitions it mediates.

    \begin{figure}
      \centering
      \includegraphics[trim={0 32pt 0pt 0}]{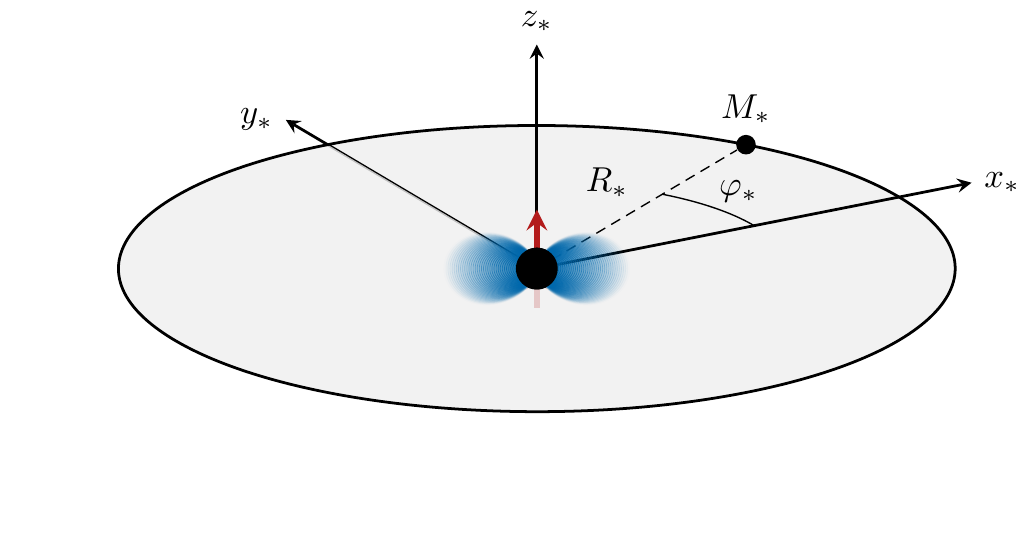}
      \caption{Schematic diagram of an equatorial binary inspiral. The position of the companion with mass~$M_*$ can be described by the distance between the two black holes, $R_*$, and the true anomaly $\varphi_*$, which is the polar angle of the companion in the equatorial plane. \label{fig:bplane}}
    \end{figure}

    \vskip 4pt
    For simplicity, we restrict our attention to inspirals that occur in the equatorial plane of the cloud. As illustrated in Figure~\ref{fig:bplane}, the relative motion of the companion is most conveniently described using the distance between the parent black hole and companion, $R_*$, and the so-called \emph{true anomaly}, $\varphi_*$, which is the companion's polar angle in the equatorial plane. 

    \vskip 4pt
    Denoting the spatial coordinates of the cloud in its Fermi frame with $\mb{r} = \{r,\theta, \phi\}$\footnote{These coordinates coincide with the familiar Boyer--Lindquist coordinates at leading order in the post-Newtonian expansion. See~\cite{Baumann:2018vus,Baumann:2019ztm} for more details.} and working at leading order in $\alpha$, the Schr\"{o}dinger equation (\ref{eq:schrodinger}) is modified by the addition of the companion's gravitational potential
    \begin{equation} 
        \begin{aligned}
            V_*(t) &=  - q \alpha  \!\!\sum_{\substack{\ell_* \geq 2 \\ |m_*| \leq \ell_*}}  \!\!\varepsilon_{\ell_* m_*} e^{\sminus i m_* \varphi_*} \es  Y_{\ell_* m_*}  (\theta, \phi) \Bigg( \frac{r^{\ell_*}}{R^{\ell_* + 1}_*} \es \Theta (R_* - r) + \frac{R_{*}^{\ell_*}}{r^{\ell_* + 1}_*} \es \Theta (r - R_*)  \Bigg)   \,, \label{eq:companionPerturbation}
        \end{aligned}
    \end{equation}
    where $q \equiv M_*/M$ is the mass ratio between the companion and the parent black hole, $\Theta$ is the Heaviside step function, and
    $\varepsilon_{\ell_* m_*} \equiv \frac{4 \pi}{2 \ell_* + 1} Y_{\ell_* m_*}^{*}(\frac{\pi}{2}, 0)$. Importantly, we explicitly exclude the fictitious $\ell_* = 1$ dipole contribution, as it vanishes in the freely falling frame and always eventually cancels in others~\cite{Baumann:2018vus}.

    \vskip 4pt
    This perturbation acts like a periodic driving force whose frequency slowly increases with time. In terms of the instantaneous frequency $\Omega(t) \equiv |\dot{\varphi}_*(t)|$, the true anomaly evolves according to
    \begin{equation}
      \varphi_*(t) = \pm \int_{0}^{t}\!\ud t'\, \Omega(t')\,,
    \end{equation}  
    where $t = 0$ is an initial reference time, and the upper (lower) sign denotes an orbit in which the companion co-rotates (counter-rotates) with the cloud. For the 
     quasi-circular equatorial orbits we consider in this paper, the power emitted by gravitational waves is
    \begin{equation}
 P_\slab{gw} \equiv\frac{\dd E_\slab{gw}}{\dd t}=-\frac{32}5\frac{q^2}{(1+q)^2}M^2R_*^4\Omega ^6\,,
	\label{eq:P_GW}
	\end{equation}
	and the orbital frequency evolves according to~\cite{PhysRev.131.435}
    \begin{equation}
      \frac{\ud \Omega}{\ud t} = \gamma \left(\frac{\Omega}{\Omega_0}\right)^{11/3}, \quad {\rm with} \quad
       \gamma \equiv \frac{96}{5} \frac{q}{(1 + q)^{1/3}} M^{5/3} \Omega_0^{11/3}\,, 
      \label{eq:freqEom}
    \end{equation}
    where $\Omega_0$ is a reference orbital frequency and $\gamma$ is the ``chirp rate''. 
    
\begin{figure}[t!]
      \centering
      \includegraphics[trim={0 0pt 0pt 0}]{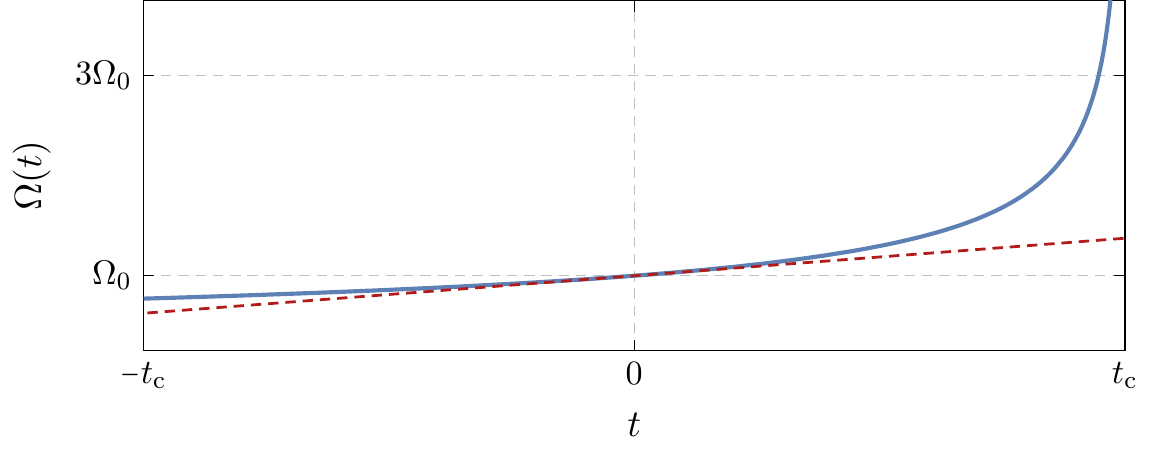}
      \caption{Time dependence of the orbital frequency and its linear approximation. \label{fig:chirp}}
    \end{figure}

\vskip 4pt
    While the equation of motion (\ref{eq:freqEom}) can be solved exactly, $\Omega(t) =  \Omega_0 \big(1 - 8 \gamma t/(3 \Omega_0)\big)^{\sminus 3/8}$, it will be convenient to work on timescales shorter than $\Omega_0/\gamma$ and linearize this solution to
    \begin{samepage}
    \begin{equation}
      \Omega(t) \approx \Omega_0 + \gamma t\,,
    \end{equation}
    so that $\varphi_*(t) \approx \pm (\Omega_0 + \frac{1}{2}  \gamma t) t$. Note that the frequency ``chirps,'' and thus the two black holes\end{samepage} merge at $t = \frac{3}{8}\Omega_0/ \gamma$, so that this linear approximation is useful as long as the inspiral has not reached the merger phase (see Figure~\ref{fig:chirp}).

The chirp rate $\gamma$ is defined in (\ref{eq:freqEom}) with respect to a reference frequency $\Omega_0$. Our primary interest in Section~\ref{sec:ionization} and beyond, is in understanding how the cloud responds to the companion's gravitational perturbation when $\Omega_0$, or an integer multiple of it, matches the energy difference $\Delta E$ between an occupied bound state and one of the continuum bands. Throughout this paper, we will use $\gamma$ to denote the chirp rate for the specific transition under consideration, with reference frequency $\Omega_0 = \Delta E$. This should be contrasted with the instantaneous chirp rate~$\ddot{\varphi}_*(t)$, which is equal to $\gamma$ up to small corrections since the inspiral evolves very slowly. We justify this definition of the chirp rate $\gamma$ in Appendix~\ref{app:nonlinearChirp}.

    \subsection{Resonant Transitions}
\label{sec:resonances}
    
In~\cite{Baumann:2018vus,Baumann:2019ztm}, it was shown that the companion's gravitational perturbation can force the cloud to transition from one bound state to another. We will briefly review these resonant transitions and establish a convenient notation.

\vskip 4pt
Throughout this work, we will denote a generic bound state with a lower-case multi-index, e.g.~$|a \rangle \equiv |n \es \ell \es m \rangle$. The matrix elements $\eta_{ab}(t) = \langle a | V_*(t) | b \rangle$ enable resonant transitions between different bound states when the orbital frequency satisfies a resonance condition.     
Because of the quasi-periodic nature of $\varphi_*$, we can decompose each of the matrix elements into their Fourier coefficients:    
 \begin{equation}
      \eta_{ab}(t) = \sum_{m_\varphi \in \mathbb{Z}} \eta_{ab}^{\floq{m_\varphi}}(t) e^{\sminus i m_\varphi \varphi_*(t)}\,,
    \end{equation}
    where the functions $\eta_{ab}^{\floq{m_\varphi}}(t)$ are slowly varying in time. Since both $|a \rangle$ and $|b \rangle$ have definite angular momentum---say $m_a$ and $m_b$, respectively---the coupling oscillates with a definite frequency
    \begin{equation}
      \eta_{ab}(t) = e^{\sminus i(m_a - \es m_b) \varphi_*(t)} \eta_{ab}^\floq{m_a \es\es \scalebox{0.9}{$-$} \es\es m_b}(t)\,. \label{eq:couplingSelection}
    \end{equation}
When the oscillation frequency matches the energy difference between the two states, 
\begin{equation}
(m_a-m_b)\hskip 2pt \Omega(t)= E_a - E_b\,,
\label{eqn:resonance-condition}
\end{equation}
the binary can resonate with the cloud and we expect that transitions between the two states will be enhanced~\cite{Baumann:2018vus}. Indeed, as the companion slowly moves through the resonance, the cloud is forced to transfer its population from one state to the other~\cite{Baumann:2019ztm}.

\vskip 4pt
This process is the analog of the Landau--Zener transition in quantum mechanics~\cite{landau1932theorie,zener1932non}.  The fraction of the cloud that is transferred from the initial state $|b\rangle$ to the final state $|a \rangle$ is controlled by the dimensionless Landau--Zener parameter $z_{ab} \equiv \eta_{ab}^2/\gamma$. Long after the transition, the total fraction of the cloud populating the state $|a \rangle$ is
\begin{equation}
    \left|\langle a | \psi(\infty) \rangle\right|^2 = 1 - e^{\sminus 2 \pi z_{ab}} \,.
\end{equation}
There are two limiting behaviors of these transitions. 
For $z_{ab} \gg 1$, the transition is \emph{adiabatic}  and the cloud is transferred almost entirely from $|b \rangle$ to $|a \rangle$. On the other hand, for $z_{ab} \ll 1$, the transition is \emph{non-adiabatic}, in which case the system is driven too quickly for it to respond and almost none of the cloud is transferred from $|b\rangle$ to $|a\rangle$.

\vskip 4pt
During these resonant transitions, the cloud's angular momentum changes macroscopically, which must be compensated for by the binary's orbital angular momentum, i.e.~the cloud \emph{backreacts} significantly on the orbital dynamics. If the orbit \emph{gains} angular momentum during this process, it can almost completely balance the angular momentum lost due to GW emission and cause the companion to \emph{float}, temporarily slowing down the inspiral until the transition is completed. On the other hand, if the orbit \emph{loses} angular momentum, then the orbit will \emph{sink}, speeding up the inspiral temporarily. Both types of transitions impart a characteristic signature on the GW signal coming from the inspiral that can be used to detect the presence of a cloud.

\vskip 4pt
In \cite{Baumann:2019ztm}, it was shown that multiple of these transitions occur during the inspiral, leading to a characteristic fingerprint for the cloud that can be used to unambiguously determine the mass and spin of the ultralight boson.  This tree of  transitions  ends when the orbital frequency~$\Omega$ becomes too large and the resonance condition (\ref{eqn:resonance-condition}) between bound states 
can no longer be satisfied. However,  the orbital frequency can then be high enough to ionize the cloud, unbinding it from its parent black hole. Indeed, this process occurs throughout the inspiral and so we will need to understand it, and its backreaction on the orbit, in order to fully characterize the phenomenology of these cloud-binary systems. This ionization process is the subject of the next~section.    

\newpage
\section{Ionization: Exciting Unbound States} \label{sec:ionization}

\begin{figure}[h!]
      \centering
      \includegraphics[trim={0 0pt 0pt 0}]{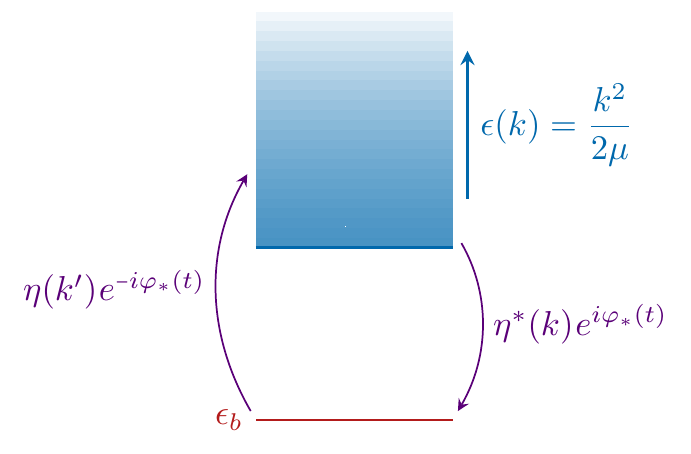}
      \caption{Schematic illustration of the transitions between a bound state and the continuum. \label{fig:transitions}}
    \end{figure}

We will now study transitions between bound and unbound states of the gravitational atom, induced by the gravitational perturbation of the companion (see Figure~\ref{fig:transitions}).\footnote{In principle, the companion can also mediate transitions from one continuum state to another. In this paper, we will ignore these, as we will only be concerned with the leading-order evolution of the cloud that remains bound to the parent black hole. We will justify this approximation in Section~\ref{sec:Unbound}.} 
Since the analysis is somewhat technical, we will start with a simple toy model involving a single bound state interacting with the continuum.  We ignore the interactions between the semi-infinite number of continuum states and also neglect the angular momentum of the continuum states.
This simplified model will capture the main dynamical features of the system without too many technical distractions. After we have gained intuition from this toy model, we will extend it to the real system of interest.

\subsection{A Toy Model} \label{sec:warmup}
          
  Consider a single bound state $|b \rangle$, with energy $\epsilon_b < 0$, interacting
  with a semi-infinite continuum of states $|k \rangle$.  For simplicity, we will assume that the continuum states depend only on the wavenumber~$k$, with dispersion relation $\epsilon(k) = k^2/2 \mu$, and that they do not interact with one another. We will also assume that the interaction between the bound state and the continuum oscillates at a frequency $\dot{\varphi}_*(t)$ that grows slowly and linearly in time, $\ddot{\varphi}_*(t) = \gamma$. This is the simplest generalization of the familiar two-state Landau--Zener system to include the coupling to the continuum. Despite its simplicity, this toy model will illustrate many of the phenomena we will encounter in the more realistic scenario.

\vskip 4pt
       The Hamiltonian of our toy model is\footnote{This is an extension of the Demkov--Osherov model~\cite{Demkov:1968san} to a single bound state interacting with a semi-infinite continuum. A similar model was studied in~\cite{Basko:2017lzs}, but with a different focus and using different techniques.}
        \begin{equation}
            \mathcal{H} = \epsilon_b \es |b \rangle \langle b| + \frac{1}{2 \pi} \int_0^{\infty}\!\ud k\, \Big[\eta(k) e^{\sminus i \varphi_*(t)} |k \rangle \langle b | + \eta^*(k) e^{i \varphi_*(t)} |b \rangle \langle k | + \epsilon(k) |k \rangle \langle k | \Big]\, . \label{eq:toyHam}
        \end{equation}
        As in Section~\ref{sec:review}, the continuum states are normalized such that $\langle k | k' \rangle = 2 \pi \delta(k - k')$, while the phase is $\varphi_*(t) = \varphi_0 + \Omega_0 t + \gamma t^2/2$. A general state in the Hilbert space can be written as
        \begin{equation}
            |\psi \rangle = c_b(t) e^{\sminus i\epsilon_b t} | b \rangle + \frac{1}{2 \pi} \int_{0}^{\infty}\!\ud k \, c_k(t) e^{\sminus i \epsilon(k) t} |k \rangle\, , \label{eq:warmupState}
        \end{equation}
     where we have peeled off the standard oscillatory behavior caused by the non-zero energies of each state---this will help us isolate the effect of the interactions $\eta(k)$. The Schr\"{o}dinger equation associated to the Hamiltonian (\ref{eq:toyHam}) leads to the equations of motion
        \begin{align}
            i\hskip 1pt \frac{\ud{c}_b}{\ud t} &= \frac{1}{2 \pi} \int_{0}^{\infty}\!\ud k\, \eta^*(k) e^{i \varphi_*(t) + i (\epsilon_b - \epsilon(k)) t} c_k(t)\,, \label{eq:warmupBoundEom} \\
            i\hskip 1pt \frac{\ud {c}_k}{\ud t} &= \eta(k) e^{-i \varphi_*(t) + i(\epsilon(k) - \epsilon_b) t} c_b(t)\,. \label{eq:warmupContEom}
        \end{align}
        Our goal is to ``integrate out'' the continuum to find an approximate description of the system entirely in terms of the bound state's dynamics. We do so using the so-called Weisskopf--Wigner method; see e.g.~\cite{Weisskopf:1930bdn,scully1997quantum,Herring:2018afy}.

        \vskip 4pt
        Assuming that the system begins its life in the bound state, $c_k(t) \to 0$ as $t \to \minus \infty$, for all $k$, we can solve (\ref{eq:warmupContEom}) exactly, 
         \begin{equation}
            c_k(t) = -i \int_{\sminus \infty}^{t}\!\ud t'\, \eta(k) e^{i (\epsilon(k) - \epsilon_b) t'-i \varphi_*(t')} c_b(t')\,. \label{eq:warmupContSol}
        \end{equation}
        Substituting this into (\ref{eq:warmupBoundEom}), we find an (integro-differential) equation for the dynamics of the entire system purely in terms of the bound state amplitude,
        \begin{equation}
          i \hskip 1pt \frac{ \ud{c}_b}{\ud t} = \int_{\sminus \infty}^{t}\!\ud t'\, \Sigma_b(t, t') \hskip 1pt c_b(t')\,, \label{eq:toySelfEnergyEq}
        \end{equation} 
        where we have defined the \emph{self-energy}
        \begin{equation}
          \Sigma_b(t, t') \equiv \frac{1}{2 \pi i} \int_{0}^{\infty}\!\ud k\, |\eta(k)|^2\,  e^{i (\varphi_*(t) - \varphi_*(t'))-i (\epsilon(k) - \epsilon_b)(t - t')} \,. \label{eq:toySelfEnergyDef}
        \end{equation}
        This equation of motion is still quite complicated, but we can make significant progress via the \emph{Markov approximation}~\cite{Herring:2018afy}, wherein we integrate by parts and drop the remainder term. The bound state Schr\"{o}dinger equation then simplifies to 
        \begin{equation}
             i\hskip 1pt \frac{\ud{c}_b}{\ud t}  = \mathcal{E}_b(t) c_b(t)\,, \label{eq:toyEffSchro}
        \end{equation}
        where we have introduced the  
        \emph{induced energy}
        \begin{equation}
            \mathcal{E}_b(t) = \int_{\sminus \infty}^{t}\!\ud t' \, \Sigma_b(t, t') = \frac{1}{2 \pi i} \int_{\sminus \infty}^{t} \!\ud t' \int_0^{\infty}\!\ud k \, |\eta(k)|^2 \, e^{i(\varphi_*(t) - \varphi_*(t'))-i (\epsilon(k)- \epsilon_b)(t - t')}\,. \label{eq:effEnergy}
        \end{equation}
        As we discuss in Appendix~\ref{app:Markov}, this approximation consists of dropping terms that are higher order in $\mathcal{E}_b(t)$ and its time integrals. 
       The imaginary part of the induced energy completely determines the behavior of the bound state occupation probability, which may be approximated as
        \begin{equation}
            \frac{\ud \log |c_b(t)|^2}{\ud t} = 2 \Im \mathcal{E}_b(t) \approx - \frac{\mu |\eta(k_*(t))|^2}{k_*(t)} \, \Theta\big(k_*^2(t)\big)\,, \label{eq:toyDeoccupation}
        \end{equation}
        where $k_*(t) = \sqrt{2 \mu \left(\dot{\varphi}_*(t) + \epsilon_b\right)}$ and $\Theta(x)$ is the Heaviside function, with $\Theta\big(k^2_*(t)\big) = \Theta(\dot{\varphi}_*(t) + \epsilon_b)$. We will devote the rest of this section to understanding the time dependence of $\Im \mathcal{E}_b(t)$ and qualitatively justifying the approximation in (\ref{eq:toyDeoccupation}).

        \begin{figure}
            \centering
            \includegraphics[trim={0 2pt 0 0}]{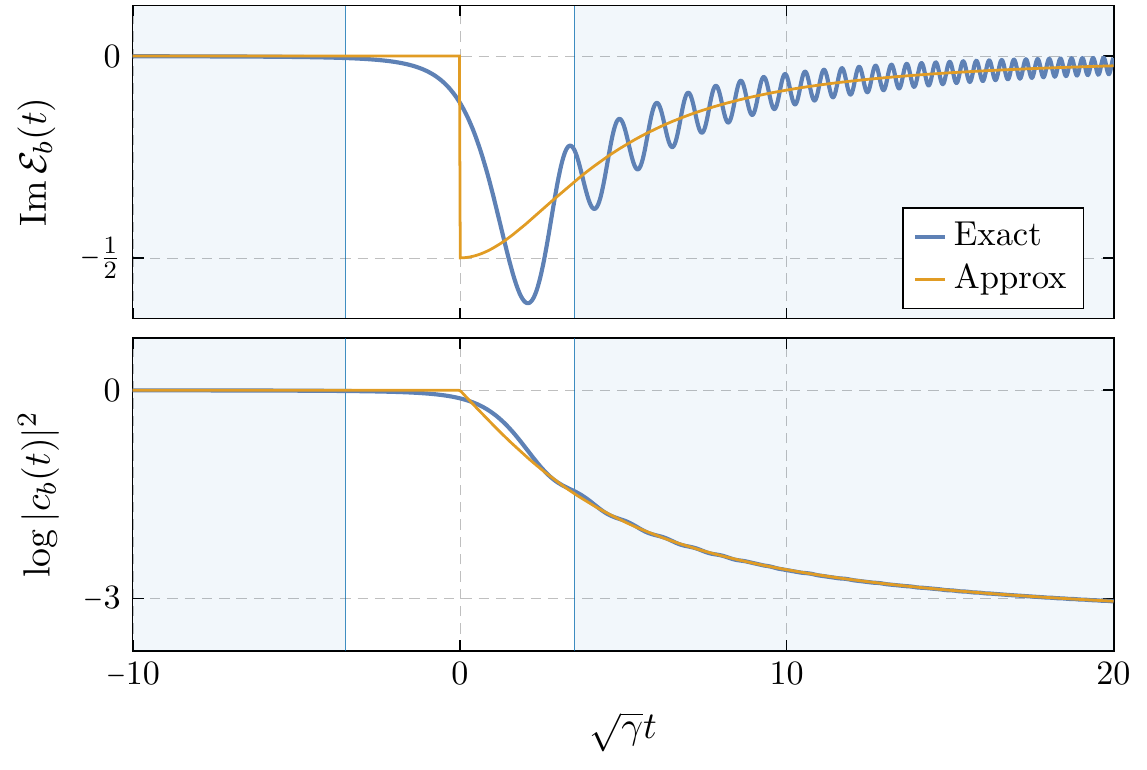}
            \caption{The imaginary part of the induced energy $\mathcal{E}_b(t)$ (\emph{top}) and the log occupation of the bound state~$\log |c_b(t)|^2$ (\emph{bottom}) as functions of dimensionless time $\sqrt{\gamma} t$, using both the exact expression (\ref{eq:effEnergy}) [{\color{Mathematica1}blue}] and our approximation (\ref{eq:toyDeoccupation}) [{\color{Mathematica2}orange}]. Here, we assume that the bound-to-continuum couplings take the form $|\eta(k)|^2 = (k/\mu)/[1 + k^4/(81 \mu^2 \gamma)]$.
            \label{fig:transientBehavior}}
        \end{figure}

\vskip 4pt
        As we might expect, the bound state only starts to significantly interact with the continuum when the frequency of the perturbation $\dot{\varphi}_*(t)$ is high enough to compensate for the bound state's binding energy, $\minus \epsilon_b$. 
        This is when the bound state starts to ``resonate" with the continuum and we can choose our time coordinate so that this resonance occurs at $t = 0$. This is not a resonance in the classic sense, but we find it useful to continue using this language. As illustrated in Figure~\ref{fig:transientBehavior}, the system (\ref{eq:toyEffSchro}) evolves on a time scale set by $\gamma^{\sminus 1/2}$ and its behavior can be divided into three distinct stages. 

        \vskip 4pt
        Far before the resonance, in the left shaded region, where $\sqrt{\gamma} t \ll -1$, the perturbation cannot provide enough energy for the bound and continuum states to interact and so the population of the bound state is, to good approximation, completely unaffected by the presence of the continuum. This changes when $|\sqrt{\gamma} t | \lesssim 1$, in the unshaded transient region, where the system goes on resonance and develops a relatively complicated time dependence. We do not need to fully understand this complicated stage, other than to note that this region serves to smoothly interpolate between the $\sqrt{\gamma} t \ll -1$ stage and the final $\sqrt{\gamma} t \gg 1$ stage.

        \vskip 4pt
        In the right shaded region, where $\sqrt{\gamma} t \gg 1$, the system approaches a type of steady state where the imaginary part of the induced energy $\mathcal{E}_b(t)$ is well-approximated by two distinct behaviors. The first is a remaining transient oscillation whose amplitude decays in time and whose properties depend only on the behavior of the coupling $|\eta(k)|^2$ as $k \to 0$. As described in Appendix~\ref{app:approx}, when $|\eta(k)|^2$ goes to zero linearly in $k$ at the edge of the continuum, these oscillations decay as $(\sqrt{\gamma} t)^{\sminus 1}$, and thus their effect on the solution $\log |c_b(t)|^2$ decays as $(\sqrt{\gamma} t)^{\sminus 2}$. As illustrated in Figure~\ref{fig:transientBehavior}, these oscillations provide a subleading correction to the dominant behavior, which is a steady and smooth deoccupation of the cloud, whose instantaneous rate depends only on the properties of the continuum state that the system is currently ``resonating'' with, i.e. the continuum state whose energy is $\frac{1}{2 \mu} k_*^2(t) = \dot{\varphi}_*(t) + \epsilon_b$. This dominant contribution (\ref{eq:toyDeoccupation}) is the only one we will consider in the text.\footnote{As we explain in Appendix~\ref{app:approx}, we can also derive the deoccupation rate (\ref{eq:toyDeoccupation}) using stationary perturbation theory, $\gamma=0$, and assuming that the obtained answer holds for $\gamma \neq 0$, if the frequency evolves slow enough.}

        \vskip 4pt 
            We will mostly be interested in the total energy that has been ionized by the perturbation, as a function of time. Assuming that the system only occupies the bound state in the far past, this ionized energy can be defined as the total energy contained within the continuum,
        \begin{equation}
        E_\lab{ion}(t) \equiv \frac{1}{2\pi} \frac{M_\lab{c}}{\mu} \int_{0}^{\infty} \!\ud k\, (\epsilon(k) - \epsilon_b)|c_k(t)|^2\,.
        \end{equation}
        As we describe in Appendix~\ref{app:ionizedEnergy}, the rate at which energy is ionized $\ud E_\lab{ion}/\ud t$, which we will call the \emph{ionization power} %JS, 
        $P_\lab{ion}$,  can be expressed in a particularly simple form by again working in the Markov approximation and ignoring subleading transient contributions, 
        \begin{equation}
          P_\lab{ion}(t) 
           \approx  \frac{M_\lab{c}}{\mu} \!\left[\dot{\varphi}_*(t) \, \frac{  \mu  \big|\eta(k_*(t))\big|^2}{k_*(t)}\right] \!\Theta\big(k^2_*(t)\big)\, |c_b(t)|^2  \, .
                \label{equ:EionApprox}
        \end{equation}
        This is clearly evocative of (\ref{eq:toyDeoccupation}) and enjoys a simple interpretation: how quickly the ionized energy grows is equal to the rate at which the bound state ``resonates'' into the state at $k_*(t)$, namely $\mu |\eta(k_*(t))|^2/k_*(t)$, weighted both by the energy difference $\epsilon(k_*(t)) - \epsilon_b = \dot{\varphi}_*(t)$ between these states and by how much is still left in the bound state at that time,  $(M_\lab{c}/\mu) |c_b(t)|^2$.

        \vskip 4pt
        Perhaps the most striking phenomenon we will encounter is the appearance of seemingly discontinuous jumps in the ionization power. We will find that these jumps occur when the perturbation begins to resonate with the continuum---that is, when the perturbation's frequency is just enough to excite the bound state into the very edge of the continuum. These discontinuities are apparent in our approximation of the time evolution (\ref{eq:toyDeoccupation}), shown in Figure~\ref{fig:transientBehavior}, and are ultimately due to the behavior of the continuum wavefunctions as $k \to 0$. 
        As we explain in Appendix~\ref{app:zeroMode}, the long-range nature of the $r^{\sminus 1}$ potential localizes this zero mode to a Bohr radius-sized region around $r = 0$ and, by a matching argument, this implies that the wavefunction's normalization scales like $\sqrt{k}$ as $k \to 0$, as do all matrix elements between the bound and continuum states. The combination $\mu|\eta(k_*(t))|^2/k_*(t)$ thus approaches a \emph{finite} limit for $k_*(t) \to 0$, when the bound state begins to resonate with the continuum, leading to an apparent discontinuity in our approximation (\ref{eq:toyDeoccupation}). Said differently, the coupling per unit energy $|\eta(\epsilon)|^2 = \ud k(\epsilon)/\ud \epsilon \hskip 1pt \big|\eta\big(k(\epsilon)\big)\big|^2$ is finite in the zero-energy limit because the zero-energy modes are still localized about the origin. Of course, this approximation does not capture the transient region shown in Figure~\ref{fig:transientBehavior}, which smooths out these apparent discontinuities on a timescale $\gamma^{\sminus 1/2}$. 
        
        \vskip 4pt
        It is instructive to compare the timescale of the transition, $\gamma^{\sminus 1/2}$, to the characteristic timescale of the inspiral, $\Omega_0/\gamma$, which measures how long it takes for the separation between the two black holes to change by a $\mathcal{O}(1)$ fraction. Using the definition of $\gamma$ in (\ref{eq:freqEom}),
        the ratio of the two timescales is\footnote{Here, we have ignored the backreaction of  ionization on the binary's dynamics, which can increase the effective chirp rate $\ddot{\varphi}_*(t) \approx \gamma$ by a factor of ${\cal O}(100)$.
        This changes the estimate (\ref{eqn:estimate-sqrtgamma}), which scales as $\gamma^{1/2}$, by an $\mathcal{O}(10)$ factor. However, for the values of $q$ and $\alpha$ we consider, this does not change the fact that these transitions are~fast.\label{ftnt:comment-backreaction}}
      \begin{equation}
   \frac{\gamma^{\sminus 1/2}}{\Omega_0/\gamma} = \sqrt{\frac{96}{5}} \frac{q^{1/2}}{(1+q)^{1/6}} \left(\frac{\alpha \Omega_0}{\mu}\right)^{5/6} \propto \sqrt{q \alpha^3}\, ,
          \label{eqn:estimate-sqrtgamma}
       \end{equation}   
       where we used that the transitions occur for $\Omega_0 \sim \mu \alpha^2$ to get the scaling in the final equality. For small $q$ and $\alpha$, the transitions therefore are very fast on the timescale of the inspiral.
% \db{Fix writing:}       \GMT{The unbackreacted value of $\gamma$ is given in (\ref{eq:freqEom}). As we will discover, however, the dynamics of the system only depends mildly on the value of $\gamma$ as long as it is small enough; we will then argue that, even with backreaction, this is the case, see (\ref{eqn:estimate-sqrtgamma}) and footnote \ref{ftnt:comment-backreaction}.}
  
\subsection{The Realistic Case} \label{sec:realisticIonization}
  
    Conceptually, extending our analysis to the realistic case of the gravitational atom requires very little extra work beyond what we have already done, the main complication being that there are simply many more states to keep track of. Our goal is again to integrate out the continuum states and encode their effects on the bound states in terms of a set of induced energies and couplings, analogous to (\ref{eq:effEnergy}). These effective interactions will be relatively complicated functions of time, but will contain a simple ``steady-state'' behavior similar to (\ref{eq:toyDeoccupation}).

\vskip 4pt
    We can write the Hamiltonian of the gravitational atom as
    \begin{equation}
      \mathcal{H} = \sum_{b} \epsilon_b(t) |b \rangle \langle b | + \sum_{a \neq b} \eta_{ab}(t) |a \rangle \langle b | + \sum_{K} \epsilon_K(t) |K \rangle \langle K| + \sum_{K, b} \big[\eta_{K b}(t) |K \rangle \langle b| + \lab{h.c.}\big]\,,
    \end{equation}
    where we use $a, b, \dots$ as a bound state multi-index,\footnote{In the previous subsection, we used the subscript $b$ to denote ``bound state'' whereas now we use it as a bound state \emph{index}, slightly abusing notation.}  $|a \rangle \equiv |n_a\es \ell_a \es m_a \rangle$ and $|b \rangle = |n_b \es \ell_b \es m_b \rangle$, while $K, L, \dots$ is a continuum state multi-index,  $|K \rangle \equiv |k; \ell \es m \rangle$. We take $\epsilon_b(t)$, $\epsilon_K(t)$, $\eta_{ab}(t)$ and $\eta_{K b}(t)$ as shorthands for the energies and couplings $\epsilon_{n_b \ell_b m_b}(t)$, $\epsilon_{\ell m}(k; t)$, $\eta_{n_a \ell_a m_a | n_b \ell_b m_b}(t)$ and $\eta_{k; \ell m | n_b \ell_b m_b}(t)$, respectively. Sums over multi-indices should be understood to include a sum over \emph{all} states of a given type. For instance, the analog of (\ref{eq:warmupState}) for a generic state is\footnote{Since the energies $\epsilon_b(t)$ and $\epsilon_K(t)$ depend on time, the appropriate ``integrating factor'' in this ansatz should be $\exp\big(\minus i \int \!\ud t' \epsilon_b(t')\big)$ instead of $\exp(\minus i \epsilon_b t)$, etc. However, the time dependence of these energies is extremely suppressed, $\dot{\epsilon}_b \sim \mathcal{O}\big(\gamma (q \alpha)^2\big)$, since it only arises from the radial dynamics of the companion. Such time-dependent terms are not critical to the resonant effects we discuss in this section, and only provide very small corrections to details like the time at which the resonance begins. We will thus ignore them.}
    \begin{equation}
      \begin{aligned}
        |\psi \rangle &= \sum_{b} c_b(t) e^{\sminus i \epsilon_b t} |b \rangle + \sum_{K} c_{K}(t) e^{\sminus i \epsilon_K t} | K \rangle \\
        &= \sum_{n, \ell, m} c_{n \ell m}(t) e^{\sminus i E_{n \ell m} t} |n \es \ell \es m \rangle + \frac{1}{2 \pi} \sum_{\ell, m} \int_0^{\infty}\!\ud k\, c_{k; \ell m}(t) e^{\sminus i \epsilon(k) t} |k; \ell \es m \rangle\,,
      \end{aligned}
    \end{equation}
    where $E_{n \ell m}$ and $\epsilon(k)$ are defined in (\ref{eq:boundEnergies}) and~(\ref{eq:contEnergies}), respectively.

    \vskip 4pt
    In this abbreviated notation, the coefficients obey the following equations of motion
    \begin{align}
       i\hskip 1pt \frac{\ud{c}_b}{\ud t}  &= \sum_{a \neq b} \eta_{b a}(t) c_{a}(t) e^{i (\epsilon_b - \epsilon_{a}) t} + \sum_{K} \eta_{b K}(t) c_{K}(t) e^{i (\epsilon_b - \epsilon_K) t} \,,\label{eq:realisticBoundEom}\\
        i\hskip 1pt \frac{\ud{c}_K}{\ud t} &= \sum_{a} \eta_{K a}(t) c_{a}(t) e^{i (\epsilon_K - \epsilon_{a}) t}\,. \label{eq:realisticContEom}
    \end{align}
    Assuming that the continuum states are completely deoccupied in the far past, $t \to \minus \infty$, we can solve (\ref{eq:realisticContEom}) exactly,
    \begin{equation}
      c_{K}(t) = -i \int_{\sminus \infty}^{t}\!\ud t' \left[\sum_{a} \eta_{K a}(t') c_{a}(t') e^{i(\epsilon_K - \epsilon_a) t'}\right] .
    \end{equation}
Substituting this into (\ref{eq:realisticBoundEom}) yields an integro-differential equation purely in terms of the bound states
    \begin{equation}
     i\hskip 1pt \frac{\ud{c}_b}{\ud t} =   \sum_{a \neq b} \eta_{b a}(t) c_a(t) e^{i(\epsilon_b - \epsilon_a) t}+ \sum_{a} \int_{\sminus \infty}^{t}\!\ud t'\,\Sigma_{ba}(t, t') c_a(t')\,, \label{eq:realisticSelfEnergyEom}
    \end{equation}
    where we have defined the self-energies
    \begin{equation}
      \Sigma_{b a}(t, t') = -i \sum_{K} \eta_{b K}(t) \eta_{K a}(t')  e^{i(\epsilon_b - \epsilon_K) t + i (\epsilon_K - \epsilon_a) t'}\,,
    \end{equation}
    which generalize (\ref{eq:toySelfEnergyEq}) to include multiple bound states. The main complication, compared to the toy model presented in Section~\ref{sec:warmup}, is that the continuum can mediate transitions between different bound states, and will thus induce off-diagonal couplings.

    \vskip 4pt
    Again working in the Markov approximation, we can rewrite (\ref{eq:realisticSelfEnergyEom}) as an effective Schr\"{o}dinger equation for the bound states
    \begin{equation}
     i\hskip 1pt \frac{\ud{c}_b}{\ud t}  = \mathcal{E}_{b}(t) c_b(t) + \sum_{a \neq b} \left[\eta_{ba}(t)e^{i(\epsilon_b - \epsilon_a)t} + \mathcal{E}_{ba}(t)\right] c_{a}(t)\,, \label{eq:realisticEffSchro}
    \end{equation}
    where we have defined both the \emph{induced couplings}
    \begin{equation}
      \mathcal{E}_{ba}(t) = -i \int_{\sminus \infty}^{t}\!\ud t'\, \sum_{K} \eta_{b K}(t) \eta_{K a}(t') e^{i( \epsilon_b - \epsilon_K) t + i (\epsilon_K - \epsilon_a) t'} \label{eq:inducedCouplings}
    \end{equation}
    and the \emph{induced energies} $\mathcal{E}_b(t) = \mathcal{E}_{bb}(t)$, the realistic analog of (\ref{eq:effEnergy}). As before, we have reduced the complicated problem of bound states interacting with a continuum to the analysis of a set of (complicated) time-dependent functions $\mathcal{E}_{ba}(t)$.

    \vskip 4pt
    These induced couplings take a much simpler form when we remember that both the bound and continuum states have definite azimuthal angular momentum, which we will denote as $m$ for the continuum state $K$ and $m_a$ or $m_b$ for the bound states $|a\rangle$ or $|b \rangle$, respectively. Since the couplings between the bound and continuum states $\eta_{K a}(t)$ reduce to a single Floquet component~(\ref{eq:couplingSelection}), we can write the induced couplings appearing in (\ref{eq:realisticEffSchro}) as\footnote{ The Floquet components $\eta_{Ka}^{\floq{m \minus m_a}}$ inherit their time dependence purely from the radial motion of the companion. Though this slow radial motion is extremely important when it forces the frequency of the perturbation to slowly increase in time and cannot be ignored there, taking the adiabatic approximation $\eta_{K a}^\floq{m \minus m_a}(t') \approx \eta_{K a}^\floq{m \minus m_b}(t)$ only requires dropping subleading terms of~$\mathcal{O}(\gamma)$, and so we will use this approximation throughout.} 
    \begin{equation}
    \begin{aligned}
      \mathcal{E}_{ba}(t) = -i \int_{\sminus \infty}^{t}\!\ud t'\, &\sum_{K} \eta^{*\floq{m \es\es \scalebox{0.9}{$-$} \es\es m_b}}_{K b}(t) \eta^{\floq{m \es\es \scalebox{0.9}{$-$} \es\es m_a}}_{K a}(t) \\
      &\times e^{i (m - m_b) \varphi_*(t) - i (m - m_a) \varphi_*(t') + i( \epsilon_b - \epsilon_K) t + i (\epsilon_K - \epsilon_a) t'}\,.
      \end{aligned}
    \end{equation}
    As we argue in Appendix~\ref{app:approx}, the off-diagonal terms oscillate as $\mathcal{E}_{ba}(t) \propto e^{i (\epsilon_b - \epsilon_a)t-i(m_b - m_a) \varphi_*(t)}$,
    just like the directly mediated transitions between the bound states
    \begin{equation}
      \eta_{ba}(t)\es e^{i( \epsilon_b - \epsilon_a)t} = \eta^\floq{m_b \es\es \scalebox{0.9}{$-$} \es\es m_a}_{ba}(t)\es e^{i (\epsilon_b - \epsilon_a)t-i(m_b - m_a) \varphi_*(t)}\,.
    \end{equation}
    The total coupling between the bound states $|a \rangle$ and $|b \rangle$, $\eta_{ba}(t) e^{i (\epsilon_b - \epsilon_a) t} + \mathcal{E}_{ba}(t)$, thus oscillates extremely rapidly unless the argument of this exponential becomes stationary, which occurs when
    \begin{equation}
      (m_b - m_a) \dot{\varphi}_*(t) = \epsilon_b - \epsilon_a\,.
    \end{equation}
    This is exactly the resonance condition (\ref{eqn:resonance-condition}) and so, even including the effects of the continuum, we can ignore transitions between bound states as long the system is not actively in resonance, cf.~\cite{Baumann:2019ztm}. That is, away from resonances the coupling between $|a \rangle$ and $|b \rangle$ oscillates rapidly enough so as to effectively average out to zero. Ignoring these resonances, we can thus dramatically simplify the effective Schr\"{o}dinger equation (\ref{eq:realisticEffSchro}) to
    \begin{equation}
     i\hskip 1pt \frac{\ud{c}_b}{\ud t} = \mathcal{E}_b(t) c_b(t)\,,
    \end{equation}
    where the induced energies,
    \begin{equation}
      \mathcal{E}_b(t) = -i \, \sum_{K}  \int_{\sminus \infty}^{t}\!\ud t'\,  \big|\eta^{\floq{m \es\es \scalebox{0.9}{$-$} \es\es m_b}}_{K b}(t)\big|^2 \,e^{i (m - m_b) ( \varphi_*(t) - \varphi_*(t'))- i(\epsilon_K - \epsilon_b) (t- t')  }\,,
    \end{equation}
    are simply the generalization of (\ref{eq:effEnergy}) to include continuum states with different angular momenta.

    \vskip 4pt
    The dynamics of this effective Schr\"{o}dinger equation are very similar to those of the toy model. Assuming that the system occupies a single bound state and ignoring the transient oscillations as we discussed in Section~\ref{sec:warmup}, we may write the analog of (\ref{eq:toyDeoccupation}) as
    \begin{equation}
      \frac{\ud \log |c_b(t)|^2}{\ud t} = 2 \, \lab{Im}\, \mathcal{E}_b(t) \approx -\sum_{\ell, g}\left[ \frac{\mu \big|\eta^\floq{g}_{K_* b}(t)\big|^2}{k_*^\floq{g}(t)} \Theta\big(k^\floq{g}_*(t)^2\big)\right] , \label{eq:realisticDeoccupation}
    \end{equation}
    with $K_* = \{k_*^{\floq{g}}(t), \ell, m = g+m_b\}$ and $k_*^\floq{g}(t) = \sqrt{2 \mu(g \dot{\varphi}_*(t) + \epsilon_b)}$, where the sum ranges from $\ell = 0,1, \dots, \infty$ and over all $g$ such that $|g + m_b| \leq \ell$. As before, the instantaneous rate of deoccupation only relies on the properties of the state that the system currently ``resonates'' with. However, in contrast to our toy model, there are two main complications. First, the perturbation oscillates at every overtone $g \in \mathbb{Z}$ of the base frequency $\dot{\varphi}_*(t)$. Second, the continuum state with energy $\frac{1}{2 \mu} k_*^2(t) = g \dot{\varphi}_* + \epsilon_b$ is infinitely degenerate. The selection rule (\ref{eq:couplingSelection}) kills the sum over overtones, but we still need to account for this infinite degeneracy, leading to the sum over total and azimuthal orbital angular momentum.

    \vskip 4pt
    The same simplifications apply to the ionization power, which we may write as
        \begin{equation}
      P_\lab{ion} \equiv  \frac{\ud E_\lab{ion}}{\ud t} \approx \sum_{\ell, g} \frac{M_\lab{c}}{\mu}\!\left[g \dot{\varphi}_*(t) \, \frac{\mu \big|\eta_{K_* b}^\floq{g}(t)\big|^2}{k_*^{\floq{g}}(t)}\right] \!\Theta\big(k_*^\floq{g}(t)^2\big)\, |c_b(t)|^2 \,, \label{eq:realisticIonizationPower}
    \end{equation}
    assuming that the system initially only occupies one bound state $|b \rangle$, where the sum is again over all states that can participate in the resonance. If the system occupies multiple bound states, we can approximate the ionization power by summing~(\ref{eq:realisticIonizationPower}) over each occupied state. 

    \begin{figure}
      \centering
      \includegraphics[trim={0 6pt 0 0}]{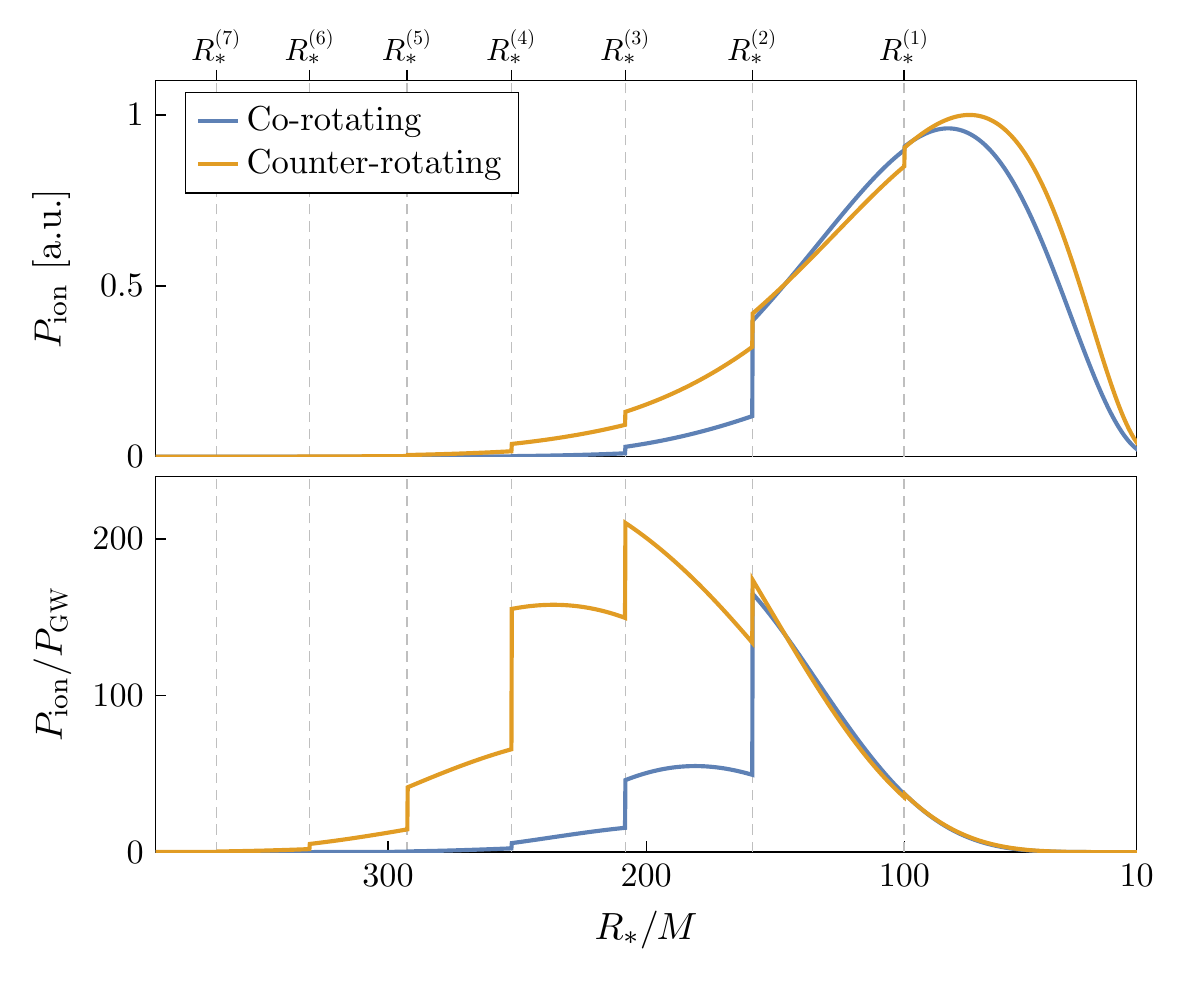}
      \caption{The ionization power (\ref{eq:realisticIonizationPower}) as a function of the binary separation $R_*$, for $\alpha = 0.2$, $q = 10^{\protect \sminus 3}$, $M_\lab{c} = 0.01 M$, and a cloud in the $|2 \es 1 \es 1 \rangle$ state. We ignore both cloud depletion and the backreaction on the orbital dynamics (see Section~\ref{sec:binary-evolution}). In the top panel, we normalize both curves by the peak ionization power of the counter-rotating orbit; so-called arbitrary units. In the bottom panel, we have normalized each curve by $P_\slab{gw}$, the energy lost due to GW emission (\ref{eq:P_GW}). We see that the energy lost due to ionization, whose overall amplitude is controlled by the mass of the cloud~$M_\lab{c} |c_b(t)|^2$, can dominate over %JS the
       GW emission. }
      \label{fig:ionization-power}
    \end{figure}
  
      \vskip 4pt  
    We plot this ionization power as a function of the binary separation $R_*$ in Figure~\ref{fig:ionization-power}. We show this ionization power normalized arbitrarily (\emph{top}) and by the energy lost due to gravitational wave emission $P_\slab{gw} \equiv \ud E_\slab{gw}/\ud t$ (\emph{bottom}), ignoring cloud depletion $|c_b(t)|^2 = 1$, for both co- and counter-rotating orbits. As we explained in the previous subsection, the discontinuous jumps that appear in both panels are due to the bound state beginning to resonate with the continuum and the fact that all couplings $|\eta_{K b}|^2$ are $\propto k$ as $k \to 0$. The fact that the perturbation now has multiple overtones means that this resonance can occur at multiple points in the orbit. Specifically, for a cloud whose initial state is $|n_b \es \ell_b \es m_b\rangle$, these discontinuities will appear at the orbital separations  
    \begin{equation}
      \frac{R_*^{(g)}}{M} = \alpha^{\sminus 2} \left[4 g^2 (1+q) n_b^4\right]^{1/3}\,,\mathrlap{\qquad g = 1, 2, \dots\,,} \label{eq:rDiscont}
    \end{equation}
    though they become progressively weaker for higher overtones $g$. 
    From the bottom panel, we also see that ionization is a \emph{large} effect compared to the energy loss due to GW emission; for $M_\lab{c}=0.01M$, $P_\lab{ion}$ can be two orders of magnitude larger than $P_\slab{gw}$. To understand this more intuitively, we note that the cloud's binding energy per unit mass, $\alpha^2/(2n^2)$, is comparable to the same quantity for the binary, $M/(2R_*)$, when $R_*\sim r_\lab{c}$. 
    If ionization reduces the cloud's mass by an amount of order the companion's mass $M_*$, this will therefore cause a large backreaction on the orbit. We confirm this intuitive expectation numerically in Section~\ref{sec:binary-evolution}.

\vskip 4pt
 It is worth noting that, for small $q$, the curves shown in Figure \ref{fig:ionization-power} exhibit a universal scaling behavior. 
  The radial wavefunctions $R_{n\ell}(r)$ and $R_{k;\ell}(r)$, given in (\ref{eq:boundWavefunctions}) and (\ref{eq:contWavefunctions}), only depend on the dimensionless variables $r/r_\lab{c}=\alpha^2r/M$ and $k r$, respectively. The wavelength $k_*$ appearing in (\ref{eq:realisticDeoccupation}) and (\ref{eq:realisticIonizationPower}) is also a function of $r/r_\lab{c}$ that scales as $\alpha^2$ and is independent of $q$, when $q \ll 1$. 
  Because the matrix elements $\big|\eta_{K_* b}^\floq{g}(t)\big|^2$ are evaluated at $k_*$, every radial variable in the overlap integrals will therefore appear in the combination $\alpha^2r/M$. The overlaps themselves thus also inherit a homogeneous $\alpha$-scaling, which can be found by power counting. For the ionization power and the deoccupation rate, this leads to
    \begin{align}
    P_\lab{ion} &=\alpha^5 q^2 \frac{M_\lab{c}}M\, \mathcal{P}(\alpha^2 R_*/M)\, , \label{eqn:scaling-Pion} \\
     \frac{\ud \log |c_b(t)|^2}{\ud t} &=\frac{\alpha^3 q^2}M\, \mathcal{R}(\alpha^2 R_*/M)\, , 
\end{align}
where $\mathcal{P}$ and $\mathcal{R}$ are universal functions for each bound state $|n_b\es \ell_b\es  m_b\rangle$ that have to be found numerically.  These relations are particularly useful when results are needed for many points in parameter space, as we now only need to compute the relatively complicated functions $\mathcal{P}$ and $\mathcal{R}$ once for a fiducial set of parameters.

\section{Accretion: Absorbing the Cloud}
\label{sec:accretion}

\begin{figure}[h!]
      \centering
      \includegraphics{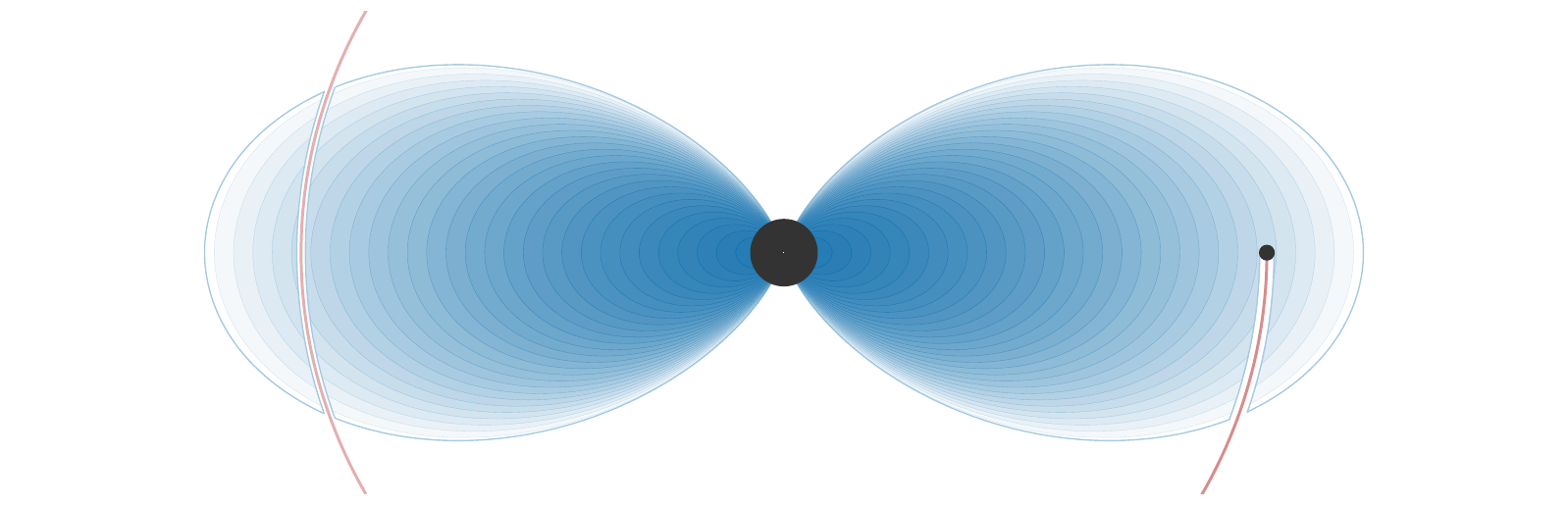}
      \caption{Cartoon illustrating the accretion of the boson cloud by the companion black hole. As explained in the text, the cloud will respond rapidly and replenish the local density behind the companion.}
      \label{fig:accretionSchematic}
    \end{figure}

So far, we have treated the perturbing object as pointlike and studied only its gravitational influence on the cloud.
In this section, we will take the finite size of the companion into account and compute its absorption of the cloud (see Figure~\ref{fig:accretionSchematic}).\footnote{
The absorption cross section of a scalar field by a black hole has been studied extensively: in the massless case for rotating black holes in~\cite{Das:1996we,Higuchi:2001si,Macedo:2013afa,Cardoso:2019dte}, in the massive case for Schwarzschild black holes in \cite{Unruh:1976fm}, and more recently, in the massive case for charged and/or rotating black holes in \cite{Benone:2014qaa,Benone:2019all}.  Our analysis will be similar to that in \cite{Benone:2019all, Unruh:1976fm}.}  
 If the secondary object is a black hole of mass $M_*$ and spin $a_*$, then this absorption will play an important role in the binary's dynamics. 
 
\subsection{Motion in a Uniform Medium}
\label{sec:uniform-medium}

We start by solving the problem in the idealized case of a black hole moving with a constant velocity in a medium with a uniform density $\rho$. If the medium were made of small particles at rest at infinity, the problem would be relatively straightforward to solve via geodesic motion in the rest frame of the black hole. In the Schwarzschild case, the energy flux takes the form~\cite{Unruh:1976fm}
\begin{equation}
\frac{\dd M_*}{\dd t}= \frac{\pi \rho \es M_*^2}{2v^3}\,\frac{\bigl(4v^2+\sqrt{8v^2+1}-1\bigr)^3}{\bigl(\sqrt{8v^2+1}-1\bigr)^2} \sim \frac{ 4 \pi\rho\, (2M_*)^2}{v}\,,\mathrlap{\quad v \to 0\,,}
\label{eqn:accretion-particles-gr}
\end{equation}
where $v$ is the asymptotic value of the relative velocity between the particles and the black hole.  The divergence at $v\to0$ signals the non-existence of a stationary configuration with $v=0$ where the density of the medium approaches a finite non-zero value at infinity.

\vskip 4pt
In the case of interest, the Compton wavelength of the medium is much larger than the gravitational radius, $r_{\lab{g},*}=M_*$, and therefore \eqref{eqn:accretion-particles-gr} does not hold. 
We expect the true answer to be smaller because the quantum pressure of the field suppresses small-scale overdensities.  
Because of the relative motion, the black hole will see the scalar field as having a  
wavenumber~$k\sim\mu v$. Besides the (reduced) Compton wavelength, $\lambda_\slab{c} = \mu^{\sminus 1}$, the other
 relevant scale in the problem is then the (reduced) de Broglie wavelength, $\lambda_\lab{dB} = k^{\sminus 1}$.  It will also be useful to define the  dimensionless ratios   $r_{\lab{g},*}/\lambda_\slab{c} = \mu M_*$ and $\lambda_\slab{c}/\lambda_\lab{dB} = k/\mu$. We are interested in the limit where both of these ratios are small,
\begin{equation}
\begin{split}
\mu M_* &\ll1\qquad\text{(``fuzzy'')\,,}\\
k/\mu&\ll 1\qquad\text{(``non-relativistic'')\,.}\\
\end{split}
\end{equation}
\label{equ:fuzzy}
We will see, in Section \ref{sec:accretion-realistic}, why these are the relevant limits in the realistic setting.

\vskip 4pt
Our goal is to compute the radial energy flux at the outer horizon $r=r_\subp$,
\beq
\frac{\dd M_*}{\dd t}=\int\! \dd\theta\dd\phi\,\sqrt{g_{\theta\theta}\es g_{\phi\phi}}\,\es T^r{}_0(r_\subp)\,,
\label{eqn:dM/dt-general}
\eeq
where the energy-momentum tensor $T_{\mu \nu}$ is that of 
the field profile $\Phi(t,\r)$.
Expanding this profile in modes with definite frequency $\omega^2 = \mu^2 + k^2$, we have (cf.~Appendix~\ref{app:heunc})
\begin{equation}
\Phi(t, \mb{r}) =\sum_{\compell,\compm}R_{k; \compell \compm}(r)S_{\compell \compm}(ka_*;\cos\theta)e^{-i\omega t+i \compm \phi}\, ,
\label{eqn:Phi-separation}
\end{equation}
where $S_{\compell \compm}(c;\cos\theta)$ are spheroidal harmonics with spheroidicity $c$, 
we can write the radial energy flux associated to this profile as
\begin{equation}
\label{eqn:Tr0-expanded}
  \begin{split}
    T^r{}_0 = \frac{2\omega(r-r_\subp)(r-r_\subm)}{r^2+a^2\cos^2\theta} \sum_{\compell,\compm} \Im(\partial_r R_{\compell \compm}^*R^{\phantom{*}}_{\compell \compm})\abs{S_{\compell \compm}}^2 + \cdots\,,
  \end{split}
\end{equation}
where the ellipses represent terms that mix different angular momenta and will vanish when integrated in (\ref{eqn:dM/dt-general}) to compute the radial energy flux. We denote the angular momentum quantum numbers measured with respect to the companion's position as $\compell$ and $\compm$, to distinguish them from those measured with respect to the parent black hole.

\vskip 4pt
The presence of the black hole deforms the field profile and determines its shape at the horizon, and thus the flux, as function of the boundary conditions at large distances. We work in the rest frame of the black hole and consider an incident monochromatic plane wave from infinity with wavevector $\mb{k}$. In Minkowski spacetime, the asymptotic field profile would be 
\begin{equation}
\Phi(t, \mb{r}) \sim\sqrt{\frac{\rho}{2\omega^2}}\,e^{i\vec{k}\cdot\mb{r}}e^{\sminus i\omega t}= \sqrt{\frac{\rho}{2\omega^2}}\,\sum_{\compell=0}^\infty (2\compell+1) i^{\compell} j_{\compell}\es\!(kr) P_{\compell}(\hat{\mb{k}} \cdot \hat{\mb{r}})\,e^{\sminus i\omega t} \,,\quad r/M_*\to\infty\,,
\label{eqn:field-uniform-asymptotic-Mink}
\end{equation}
where $\omega=\sqrt{\mu^2+k^2}$, with $k = \mu v/\sqrt{1 - v^2}$. 
 In this expression, 
 $j_{\compell}\!\es (k r)$ is the spherical Bessel function, $P_{\compell}(\hat{\mb{k}} \cdot \hat{\mb{r}})$ is the Legendre polynomial and the normalization has been chosen so that $\rho\approx T_{00}=2\omega^2\Phi^*\Phi$. The long-range nature of the gravitational field, however, deforms the field; in a spherically symmetric spacetime, we have \cite{Benone:2014qaa}
\begin{equation}
\Phi(t, \mb{r}) \sim\sqrt{\frac{\rho}{2\omega^2}}\,\sum_{\compell=0}^\infty (2\compell+1)i^{\compell} j_{\compell}\!\es\big(kr+\delta(r)\big)P_{\compell}(\hat{\mb{k}} \cdot \hat{\mb{r}}) \,e^{\sminus i\omega t}\,,\quad r/M_* \to\infty\,,
\label{eqn:field-uniform-asymptotic}
\end{equation}
where 
$\delta(r) = k M_* (1 + \omega^2/k^2) \log(2 k r) + \delta_{\compell}$, and $\delta_{\compell}$ is a constant phase shift. Although our case is not quite spherically symmetric, deviations from (\ref{eqn:field-uniform-asymptotic}) are controlled by the spheroidicity parameter, which is $ka_* \ll 1$ in the non-relativistic limit we are considering.

\begin{figure}
            \centering
            \includegraphics[scale=1, trim={0 2pt 0 0}]{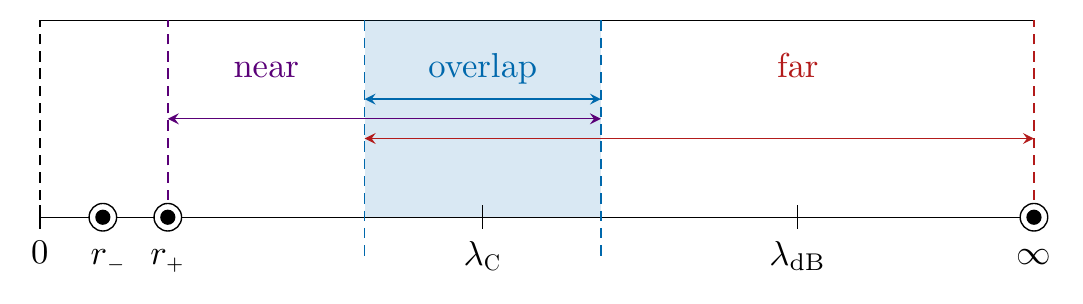}
            \caption{Schematic illustration of the near-field and far-field expansions, where $r_\pm$ are the inner and outer horizons of the black hole. The two asymptotic solutions are matched in the overlap region, $M_* \ll r\ll 1/k$.} 
            \label{fig:matching}
        \end{figure}

\vskip 4pt
To compute the energy flux at the horizon, we must understand the dependence of the near-field solution on the boundary condition (\ref{eqn:field-uniform-asymptotic}). This will be achieved by a \emph{matched asymptotic expansion}: the far-field and near-field solutions will be studied separately and matched in the overlap region, where both expansions hold. The boundary condition will then fix the overall amplitude of the solution. This procedure is schematically illustrated in Figure~\ref{fig:matching}.

\vskip 4pt
{\it Near-field solution}---With the ansatz (\ref{eqn:Phi-separation}), the Klein--Gordon equation is separable. The exact solution of the equation for $R_{k; \compell \compm}(r)$ can be expressed in terms of the confluent Heun function (see Appendix~\ref{app:heunc} and \cite{Bezerra:2013iha}). 
We expect the contributions from modes with $\compell \geq 1$  to be suppressed at radii smaller than about $\compell^2/(\mu^2M_*)$ (due to the angular momentum barrier), so that the $\compell=\compm=0$ mode dominates near the horizon. 
Expanding the confluent Heun function around $r=r_\subp$, one can show that
\begin{equation}
R_{k}(r)=C_{k}\es e^{-i\omega(\tilde r-r)-i \compm \tilde \phi}\bigl(1+\mathcal O(\mu M_*,kM_*)\bigr)\,,\quad\text{for}\quad r_\subp \le r<r\ped{max}\,,
\label{eqn:R00-near-field}
\end{equation}
where we use $R_{k}(r) = R_{k; 00}(r)$ as a shorthand, the coefficient $C_{k}= C_{k; 00}$ defines the near-horizon amplitude of the $\compell=\compm=0$ mode, $\tilde r$ and $\tilde\phi$ are the radial and angular tortoise coordinates (defined in Appendix~\ref{app:heunc}),  
and the breakdown of the expansion is at
\begin{equation}
    \frac{r_\lab{max}}{M_*} \sim \min \left\{ \frac{1}{(\mu M_*)^2}, \frac{1}{k M_*}\right\} \gg 1\,.
\end{equation}
Using the explicit expressions of the tortoise coordinates, and plugging (\ref{eqn:R00-near-field}) into 
(\ref{eqn:dM/dt-general}), we get
\begin{equation}
  \frac{\dd M_*}{\dd t}= 4 M_*\hskip 1pt  r_\subp \es\omega^2 \abs{C_{k}}^2\,.
  \label{eqn:flux-C00}
\end{equation}
We will now determine $C_{k}$ by matching (\ref{eqn:R00-near-field}) to the far-field solution.

\vskip 4pt
{\it Far-field solution}---Far from the companion, $r\gg M_*$, the equation for $R_{k}(r)$ becomes
\begin{equation}
\frac{\dd^2R_{k}}{\dd r^2}+\biggl(\frac2r+\cdots\biggr)\frac{\dd R_{k}}{\dd r}+\biggl(k^2+\frac{2M_*(\omega^2+k^2)}{r}+\cdots\biggr)R_{k}=0\,.
\end{equation}
This equation is solved by a linear combination of confluent hypergeometric functions, 
\begin{equation}
    e^{ikr}R_{k} =C_F\,{}_1F_1 \!\left(1 + i k M_* \big(1 + \omega^2/k^2\big); 2 \es ; 2ikr\right)+C_U \es U\!\left(1 + i k M_*\big(1 + \omega^2/k^2\big); 2; 2 i k r\right) . 
    \label{eqn:lin-comb-hypergeo}
\end{equation}
For $kr \ll 1$, this solution overlaps with the near-field solution (\ref{eqn:R00-near-field}). Expanding (\ref{eqn:lin-comb-hypergeo}) in this limit and matching to (\ref{eqn:R00-near-field}) then gives $C_F=C_{k}$ and $C_U\leq \mathcal O\bigl((\mu M_*)^2\bigr)$. To determine the overall amplitude of the solution, we then expand (\ref{eqn:lin-comb-hypergeo}) for $kr \gg 1$, where it reduces to a spherical Bessel function, $R_{k}(r)\propto j_0(kr+\delta(r))$, and compare it to the $\compell=0$ mode of the boundary condition (\ref{eqn:field-uniform-asymptotic}).  This gives
\begin{equation}
C_F = C_{k}= \frac{\sqrt{2 \pi \rho}}{\omega}  \left|\es\es \Gamma\!\left(1+i k M_* \big(1 + {\omega^2}/{k^2}\big) \vphantom{\tfrac{\omega^2}{k^2}}\right) e^{\frac{1}{2} \pi k M_* (1 + {\omega^2}/{k^2})}\right| .
\end{equation}
Plugging this back into \eqref{eqn:flux-C00}, we get
\begin{equation}
\frac{\dd M_*}{\dd t}=\mathcal{A}_* \es\es \rho\es \left|\es\es \Gamma\!\left(1+i k M_* \big(1 + {\omega^2}/{k^2}\big) \vphantom{\tfrac{\omega^2}{k^2}}\right)\right|^2 e^{\pi k M_* (1 + {\omega^2}/{k^2})}\, ,
\label{eqn:flux-analytical-gamma}
\end{equation}
where $\mathcal{A}_* \equiv 8\pi M_* r_{\subp,*}$ is the area of the outer horizon of the Kerr black hole. This is our final answer for the mass accretion rate.

\begin{figure}[t!]
\centering
\includegraphics{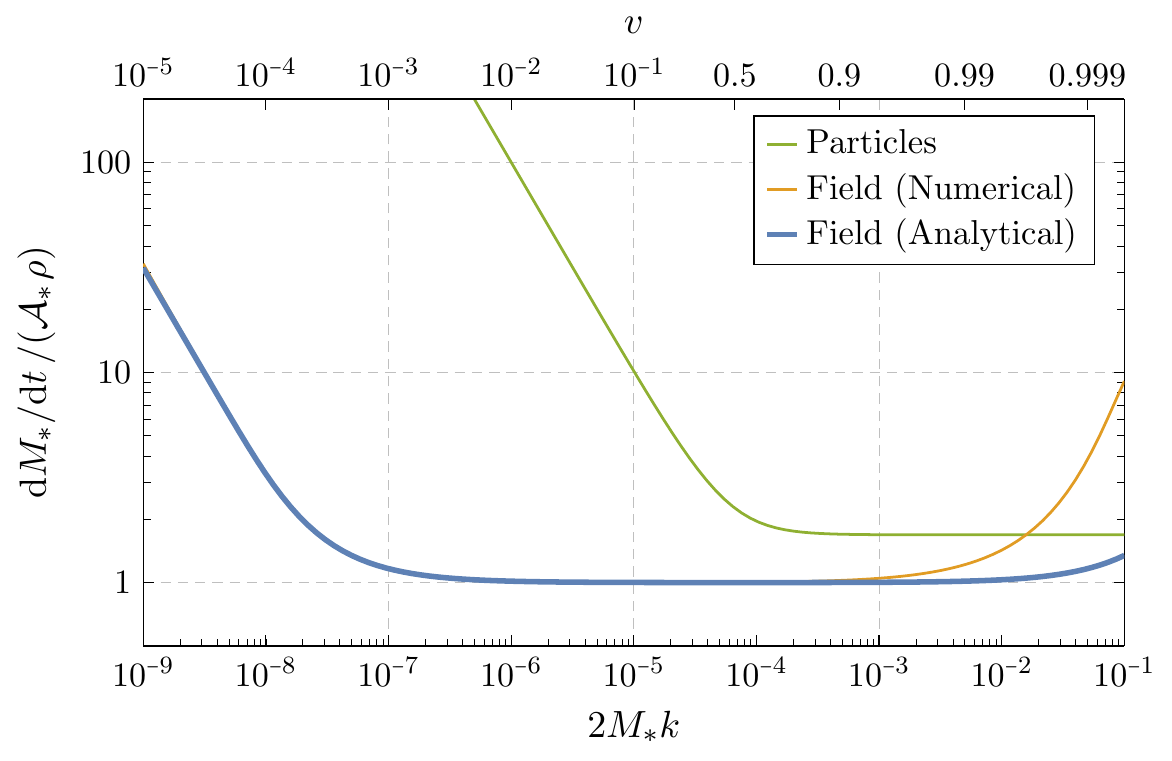}
\caption{Mass accretion rate of a Schwarzschild black hole computed analytically---from (\ref{eqn:flux-analytical-gamma})---and numerically for a scalar field with $2\mu M_* =10^{\protect \sminus 4}$. Shown for comparison is also the accretion rate for particles given by (\ref{eqn:accretion-particles-gr}). }
\label{fig:accretion}
\end{figure}

\vskip 4pt
The result is shown in Figure~\ref{fig:accretion} for $2\mu M_* =10^{\sminus 4}$. As anticipated, the flux is smaller than for particles, but still divergent for $v\to0$. For non-relativistic momenta, $k < \mu$, we can identify two different regimes 
\begin{equation}
\frac{\dd M_*}{\dd t}=\mathcal{A}_*\es\rho  \begin{cases}
\quad\,\, 1 & \text{for}\quad k \gg2\pi\mu^2M_* \,,\\[6pt]
\displaystyle \frac{2\pi\mu^2M_*}{k}\ & \text{for}\quad k\ll2\pi\mu^2M_*\,.
\end{cases}
\label{eqn:accretion-cases}
\end{equation}
It is worth noting that, at the cross-over point $k=2\pi\mu^2M_*$, the de Broglie wavelength of the scalar field equals the Bohr radius of the gravitational atom, $2\pi/k=r\ped{c}$.\footnote{To give an interpretation of this result, recall that a particle with impact parameter $b$ and velocity $v$ is scattered by an angle $\sim M/(v^2b)$ by the Coulomb interaction. Taking $b\sim\lambda_\lab{dB}$, we get an order-one deflection angle for $\lambda_\lab{dB} \sim r\ped{c}$. Scattering of waves with more (less) energy will be less (more) effective.} 
For $k\ll2\pi\mu^2M_*$, the energy flux diverges as $1/v$, just like in the particle case, but with a smaller normalization. For $k \gg2\pi\mu^2M_*$, instead, the energy flux is  independent of $v$ and takes the very natural form~$\mathcal{A}_*\es \rho c$, if we restore a factor of $c$. 
This indeed matches the result for the low-energy cross section for a massless field \cite{Das:1996we,Higuchi:2001si,Macedo:2013afa,Cardoso:2019dte}. The regime holds until relativistic corrections kick in at $k\sim\mu$, and our derivation breaks down.

\vskip 4pt 
{\it Numerical solution}---Figure~\ref{fig:accretion} also shows the result of a numerical approach to the problem. In the Schwarzschild case, we numerically integrated the confluent Heun function for different values of $k$ and $\compell$, with the main goal of confirming that  the $\compell= \compm=0$ mode indeed dominates in the fuzzy limit. This allowed us to determine the near-horizon amplitudes $C_{k; \compell \compm}$ of modes with $\compell \geq 1$ as a function of the asymptotic density $\rho$ by comparing the asymptotic limit of the confluent Heun function with the partial wave expansion of the boundary conditions (\ref{eqn:field-uniform-asymptotic}). 
The results are in remarkable agreement with the analytical estimate for all $\mu M_*\ll1$ and $k\ll\mu$, and explicitly show the suppression of $C_{k; \compell \compm}$ for $\compell\ge1$.

\subsection{Application to the Realistic Case}
\label{sec:accretion-realistic}

So far, we have studied an idealized model of a black hole moving through a uniform scalar field mass density. However, we would like to apply these results to the case we are actually interested in: a companion black hole of mass $M_* = q M$ moving through a non-uniform cloud that is bound to its parent black hole. This more realistic scenario has a few major complications over its idealized counterpart and in this section we confront them.

\vskip 4pt
First and foremost, the scalar field mass density can have nontrivial azimuthal structure and so the companion can experience different densities along a single orbit. For instance, if the cloud is composed of a \emph{real} scalar field occupying the $|2 \es 1 \es 1 \rangle$ state, its mass density (\ref{eq:massDensity}) behaves as $\rho(\mb{r}) \propto \cos^2 \phi$. In contrast, if it is a  \emph{complex} scalar field occupying the same state (or any other pure eigenstate), its mass density does not vary along the orbit, $\rho(\mb{r}) = \rho(r, \theta)$. When the mass density has nontrivial $\phi$-dependence, we will assume that we can replace it with its azimuthal average, $\rho(r, \theta) = \frac{1}{2 \pi} \! \es \int_{0}^{2 \pi} \!\ud \phi\, \rho(r, \theta, \phi)$. In this case, both real and complex scalar fields are treated equally and give identical predictions. We do not expect this to be a bad approximation, as it is roughly akin to only tracking quantities that have been averaged over an orbit, like those we work with in Section~\ref{sec:binary-evolution}.

\vskip 4pt
Even assuming that we can azimuthally average the scalar field density, it is still non-uniform in the radial direction and the relative asymptotic velocity between the companion and scalar field is ill-defined.  We will assume that accretion occurs dynamically in a region that is much smaller than the size of the cloud, so that we can define this velocity ``locally.'' We will later justify this assumption. This dynamical region is mesoscopic, in the sense that the dynamics is only sensitive to the local properties of the cloud (like its density and velocity), but the region is still much larger than the  size of the companion object.   In place of the asymptotic fluid density, we can then use the local density $\rho(\vec R_*)$ of the cloud at the position of the companion. Similarly, we define the local velocity to be the ratio of the probability current to the probability density, 
\begin{samepage}
\begin{equation}
\vec v_\lab{c}(\vec R_*)=\frac{i}{2\mu|\psi|^2}\bigl(\psi \nabla\psi^*-\psi^* \nabla\psi\bigr)=\frac{m}{\mu R_*^{\sperp}}\,\vec{\hat\phi}\,,
\label{eqn:v_c}
\end{equation}
where $m$ is the azimuthal angular momentum of the cloud and $R_*^\sperp$ is the length of the projection \end{samepage} of $\mb{R}_*$ on the equatorial plane, so that the difference between (\ref{eqn:v_c}) and the orbital velocity of the companion, $\mb{v}_* \sim \pm \sqrt{M/R_*} \es \hat{\bm{\phi}}$, is the relative fluid-black hole velocity. For equatorial circular orbits, with $R_*^\sperp=R_*$, this relative velocity is
\begin{equation}
  v  = \biggl|\sqrt{\frac{M}{R_*}}\mp\frac{m}{\mu R_*}\biggr| = \frac{\alpha}{\sqrt{R_*/r_\lab{c}}}\left|1 \mp \frac{m}{\sqrt{R_*/r_\lab{c}}}\right| ,
  \label{eqn:relative-v-c-BH}
\end{equation}
where the $-$ ($+$) sign  
refers to co-rotating (counter-rotating) orbits and $r_\lab{c} = (\mu \alpha)^{\sminus 1}$ is the typical radius of the cloud. We stress that the quantities $\rho(\mb{R}_*)$ and $\mb{v}_\lab{c}(\mb{R}_*)$ are computed without taking the backreaction of the companion into account. For small $q$, this is a good approximation.

\vskip 4pt
Under these assumptions, and for the systems we study, the mass accretion flux is approximately independent of velocity,
\begin{equation}
    \frac{\ud M_*}{\ud t} \approx \mathcal{A}_* \, \rho(\mb{R}_*)\,,\label{eqn:accretion-law}
\end{equation}
where $\mathcal{A}_* \approx 4 \pi(2 q M)^2$ is the area of the companion's horizon. From the discussion of the previous section (see the ``plateau'' in Figure~\ref{fig:accretion}), this approximation is valid as long as the relative fluid velocity is neither too slow nor too fast,
\begin{equation}
    2\pi q\alpha\ll v\ll1\,. \label{eqn:limits-on-v}
\end{equation}
From (\ref{eqn:relative-v-c-BH}), we see that this condition can be violated when either the orbital separation is very small, $R_* \sim \alpha^2 r_\lab{c}$, in which case the fluid is moving too quickly, $v \sim 1$, or when the orbital separation is very large, $R_* \sim r_\lab{c}/q^2$, in which case the fluid is moving too slowly, $v \ll 2 \pi \alpha q^2$. Both of these cases occur during a typical inspiral. However, for small $q$ and $\alpha$, the cloud is extremely dilute whenever (\ref{eqn:limits-on-v}) is violated, because the companion is either too close\footnote{We have assumed that the cloud has nontrivial angular momentum, which pushes the density of the cloud away from the parent black hole. This is a fair assumption, as these are the types of states prepared by superradiance. Moreover, we do not expect accretion to be significant for $\ell=0$ states anyway, since the time spent by the companion in the region $R_*\lesssim\alpha^2 r_\lab{c}$ is very short.} or too far away from the parent black hole to see an appreciable density, and so accretion is negligible whenever (\ref{eqn:accretion-law}) does not apply.\footnote{This reasoning can fail when the relative velocity (\ref{eqn:relative-v-c-BH}) vanishes and the companion orbits the parent black hole at the same local speed as the cloud, which occurs for co-rotating orbits at $R_* = m^2 r_\lab{c}$. 
In an orbital band of width $\Delta R_*\sim\pi qm^3r_\lab{c}$ around this special orbit, the constraint $2\pi q\alpha\ll v$ is violated and (\ref{eqn:accretion-law}) cannot be applied. Rather, the low-velocity limit of (\ref{eqn:accretion-cases}) must be used instead and accretion is enhanced.}

\vskip 4pt
Finally, let us now check that the accretion process actually happens in a mesoscopic region where we can assume that the companion sees a uniform medium. 
The mass absorption formula (\ref{eqn:accretion-law}) can be written as
\begin{equation}
\frac{\dd M_*}{\dd t}=\big(\pi b\ped{max}^2\big)\es v\rho\,, 
\label{eqn:accretion-formula-M*}
\end{equation}
where $b\ped{max} \equiv 4qM/\sqrt{v}$ is the radius of the absorption cross section, or the maximum impact parameter for absorption in a particle analogy. To apply the idealized derivation, we need to satisfy two conditions: (1) the density and velocity of the cloud are approximately constant over a region of size $b\ped{max}$ and (2) the region of size $b\ped{max}$ is gravitationally dominated by the companion, i.e.~it is smaller than the radius of the Hill sphere $r\ped{Hill}=R_*(q/3)^{1/3}$.
These two conditions then require that
\begin{align}
&(1) \ \  
b\ped{max}\ll r_\lab{c} \hspace{9pt} \implies\ \frac{R_*}M\ll \big(4q\alpha^2\big)^{\sminus 4}\, ,
\label{eqn:condition-a} \\
 &(2) \ \ b\ped{max}\ll r\ped{Hill} \implies\ \frac{R_*}M\gg \big(8 q/{\sqrt{3}}\big)^{8/9}\, .
\label{eqn:condition-b}
\end{align}
Both of these conditions are  easily satisfied for the typical values of $\alpha$, $q$ and $R_*$ that we are interested in.

\vskip 4pt
There are two ways the companion can fail to see such a uniform medium. The first is simply if the azimuthally-averaged density $\rho(\mb{R}_*)$ vanishes, or changes dramatically, at a particular orbital separation. This can occur when the cloud occupies a state $|n \es \ell \es m \rangle$, with $\ell \neq n -1$, for which the radial wavefunction has zeros away from the origin. In this case, we can think of the density that the companion sees as simply being the averaged density within a Hill sphere about the companion. Similarly, as illustrated in Figure~\ref{fig:accretionSchematic}, the companion itself changes the local density---it vacuums up the scalar field as it passes through the cloud and leaves an empty ``tube'' of diameter $\mathcal{O}(M_*)$. However, the cloud will respond and replenish this local density on a relatively short timescale. This perturbation excites modes with typical wavelength of $\mathcal{O}(M_*)$, whose frequencies $\omega^2 = \mu^2 + k^2$ scale as $\mathcal{O}\big(\mu/(\alpha q)\big)$. These modes respond extremely quickly, and we expect that this empty ``tube'' is rapidly filled in before companion can complete an orbit and encounter this locally depleted region again. So, the companion should see a relatively uniform medium throughout the inspiral, and we will thus use the approximation~(\ref{eqn:accretion-law}) throughout Section~\ref{sec:binary-evolution} to capture the effect accretion has on the binary's dynamics.

\newpage
\section{Backreaction on the Orbit}
\label{sec:binary-evolution}

We will now study the effect that both ionization and accretion have on a binary inspiral. We are mostly interested in intermediate or extreme mass ratio inspirals, where the light companion moves inside the cloud of the much heavier parent black hole. 
 In Section~\ref{sec:evolution-equations}, we describe the system and its evolution equations, 
  while in Section~\ref{sec:solution-evolution} we show numerical solutions to these equations for a few representative examples.

\subsection{Evolution Equations}
\label{sec:evolution-equations}

Chronologically, the first resonant transitions in the inspiral are those with the lowest frequency. These typically happen before the separation becomes comparable to $\rBC$. During those resonances, the state of the cloud can be transformed to decaying states. For example, for an initial $\ket{2\es 1\es 1}$ state, the first resonances mediated by the quadrupolar perturbation ($\ell_*=2$) connect it to the $\ket{2\es 1\hskip 2pt \minus1}$ and $\ket{3\es 1\hskip 2pt \minus1}$ states in the co- and counter-rotating cases, respectively. It is nontrivial to understand whether or not the cloud survives after these transitions, though it has recently been shown that it can \cite{Takahashi:2021eso} in some cases.

\vskip 4pt
Our main goal in this section is to understand the physics of the subsequent inspiral, away from resonances, and under the hypothesis that the cloud is still present when ionization and accretion kick in. Our results should thus not be read as a fully realistic solution of the dynamics of the system, as that would require  including the resonances (and their impact on the evolution of the cloud).
Rather, we present an example of the impact of ionization and accretion only, and their interplay. We will restrict to quasi-circular, equatorial orbits, and study separately orbits that are co- and counter-rotating with respect to the cloud. The gravitational field of the cloud will also be neglected, as it gives a correction of order $M_\lab{c}/M$ to the orbital quantities, which, as we will see, is subdominant with respect to the impact of ionization and accretion. We will numerically solve the time evolution of three quantities: the companion's mass $M_*$, the cloud's mass $M_\lab{c}$, and the separation~$R_*$.

\vskip 4pt
The evolution of $M_*$ and $M_\lab{c}$ is determined by mass conservation. As we discussed in Section~\ref{sec:accretion}, the mass of the companion increases by accretion, while
the mass of the cloud decreases by the corresponding amount. In addition, the cloud loses mass through ionization. We therefore have
\begin{align}
\frac{\dd M_*}{\dd t} &= 4\pi (2M_*)^2 \, \rho(\vec R_*)\,, \label{equ:q-evolve} \\[4pt]
\frac{\dd M_\lab{c}}{\dd t} &= - \frac{\dd M_*}{\dd t}-M_\lab{c}\sum_{\ell, g}\left[ \frac{\mu \big|\eta^\floq{g}_{K_* b}(t)\big|^2}{k_*^\floq{g}(t)} \Theta\big(k^\floq{g}_*(t)^2\big)\right]\,, \label{equ:qc-evolve}
\end{align}
where $\rho(\vec R_*) =M_\lab{c}\big|R_{n\ell}(R_*) Y_{\ell m}(\theta_*,\phi_*)\big|^2$ 
is the local density of the cloud at the position of the companion.\footnote{Here, we have ignored the possibility that accretion is enhanced, as it is for co-rotating orbits at $R_*=m^2r_\lab{c}$. This enhancement occurs in a region that is too narrow to be resolved for the values of $q$ we consider. We also assumed that either the scalar field is complex, or that we have azimuthally averaged the mass density of the real scalar field. Both provide the same result. } 
The accretion formula (\ref{equ:q-evolve}) holds for a non-rotating black hole, while for a rotating black hole it has to be rescaled to account for the reduced area of the horizon.
The last term of (\ref{equ:qc-evolve}) is the ionization rate, defined in (\ref{eq:realisticDeoccupation}). %\db{John, please go over this:} \GMT{Note that we are assuming that accretion is not responsible for additional transitions in the state of the cloud, which are only mediated by gravitational scattering. While this is certainly an approximation, we do not expect this effect to be significant, due to the small fraction of the cloud undergoing accretion, compared with the one undergoing scattering.}
  
  \vskip 4pt
  To determine the backreaction on the inspiral, we use the conservation of angular momentum. The system carries angular momentum in the form of the orbital angular momentum of the binary and the spin of the cloud, which are given by
  \begin{equation}
    L \equiv \frac{M_* \Omega R_*^2}{1+q} \quad {\rm and} \quad S_\lab{c} \equiv \frac{m M_\lab{c}}\mu\, ,
  \end{equation}
 where $\Omega^2 R_*^3=(1+q)M$ for quasi-circular Keplerian orbits.
   Gravitational waves carry angular momentum to infinity at a rate $\dd L_\slab{gw}/\dd t = P_\slab{gw}/\Omega$, where $P_\slab{gw}$ is given in (\ref{eq:P_GW}). In the vacuum solution, these gravitational waves are the reason for the shrinking orbit.  Ionization leads to an additional loss of angular momentum through the emission of scalar waves. These waves carry angular momentum to infinity at a rate given by an expression analogous to (\ref{eq:realisticIonizationPower}),
\begin{equation}
      \frac{\ud L_\lab{out}}{\ud t} = \sum_{\ell, g}\left[(m+g) \, \frac{\mu \big|\eta_{K_* b}^\floq{g}(t)\big|^2}{k_*^{\floq{g}}(t)}\right] \!\Theta\big(k_*^\floq{g}(t)^2\big)\es\es |c_b(t)|^2 \,.\label{eq:realisticIonizationAngularMomentumFlux}
    \end{equation}
The conservation of the total angular momentum then implies
\begin{equation}
\frac{\dd L}{\dd t} + \frac{\dd S_\lab{c}}{\dd t} =- \!\es\left(\frac{P_\slab{gw}}{\Omega} + \frac{\dd L\ped{out}}{\dd t}\right)\, .
\label{eqn:conservation-L}
\end{equation}
  Using (\ref{equ:qc-evolve}) for the evolution of $M_\lab{c}$ in $\dd S_\lab{c}/\dd t$, we can express the difference between its last term and $\dd L_\lab{out}/\dd t$ in terms of the ionization power, $P_\lab{ion}$, defined in (\ref{eq:realisticIonizationPower}). This leads to an equation for the evolution of the binary's separation,\footnote{Note that this expression neglects the transient oscillations associated with the discontinuities. As we showed in Section~\ref{sec:ionization}, these oscillations decay over a very narrow region of $R_*$ in the small backreaction limit. This region remains small even in the cases studied here, as even though the backreaction is strong and the instantaneous chirp rate $\ddot{\varphi}_*(t)$, c.f.~Appendix~\ref{app:nonlinearChirp}, is enhanced roughly by a factor of $P_\lab{ion}/P_\slab{gw} \sim \mathcal{O}(100)$, this narrow region scales as $\gamma^{1/2}$ and, especially for the parameters we are interested in, this region is still small enough to ignore the effect that the transient oscillations and varying chirp rate $\ddot{\varphi}_*(t)$
   can have on (\ref{eqn:evolution-R}).}
  \begin{equation}
\begin{aligned}
    \frac{qM^2}{2R_*^2}\frac{\dd R_*}{\dd t}={}&{-P_\slab{gw}}-P_\lab{ion}-
    \left[\frac{2+q}{2(1+q)^{3/2}} \sqrt{M R_*}\mp\frac{mM}{\alpha}\right]\! M|\Omega |\frac{\dd q}{\dd t}\,,
    \end{aligned}
    \label{eqn:evolution-R}
\end{equation}
where the minus (plus) sign refers to co-rotating (counter-rotating) orbits.  We see that the inspiral dynamics is determined by three different ``forces.'' The first two have the obvious interpretation of the drag induced by the energy lost in gravitational waves and scalar waves, respectively. The third term, instead, is the accretion of momentum that comes along with the accretion of mass. The sign of this force depends on whether the cloud is locally rotating faster or slower than the companion. Not surprisingly, the two behaviors are separated by $R_*=m^2r_\lab{c}$ (in the small-$q$ limit), corresponding to the special co-rotating orbit identified in Section~\ref{sec:accretion-realistic} where the relative velocity vanishes.

\vskip 4pt
As a final note, we observe that the backreaction of the gravitational interaction between an object and the medium it is moving through is known in the literature as ``dynamical friction." For uniform density media, the interpretation of the effect is simple: the wake of the overdensity behind the moving object exerts a gravitational pull on it, creating a drag force. The effect has been computed for a light field in~\cite{Hui:2016ltb}, and there have been some recent attempts to apply it to the case of the gravitational atom in~\cite{Zhang:2019eid,Traykova:2021dua}. The length of the wake and the intensity of the drag force depend on the history of the system, with divergent results found for stationary configurations in asymptotically uniform media. The gravitational atom, however, is special in two ways: first, it is localized in space, providing a natural regulation for the divergence mentioned previously; second, its spectrum is composed of bound and unbound states, but only the latter can carry (angular) momentum to infinity \cite{Annulli:2020ilw}. Despite these complications, the physical origin of the drag force is the same. It is therefore a question of semantics whether one calls the drag induced by the backreaction of ionization ``dynamical friction." In any case, because bound and unbound states together form a complete set, the description of the evolution of their occupations, and the associated backreaction, either in the form of resonances or drag, provides a full description of the interaction between the cloud and the moving object.

\begin{figure}
            \centering
            \includegraphics{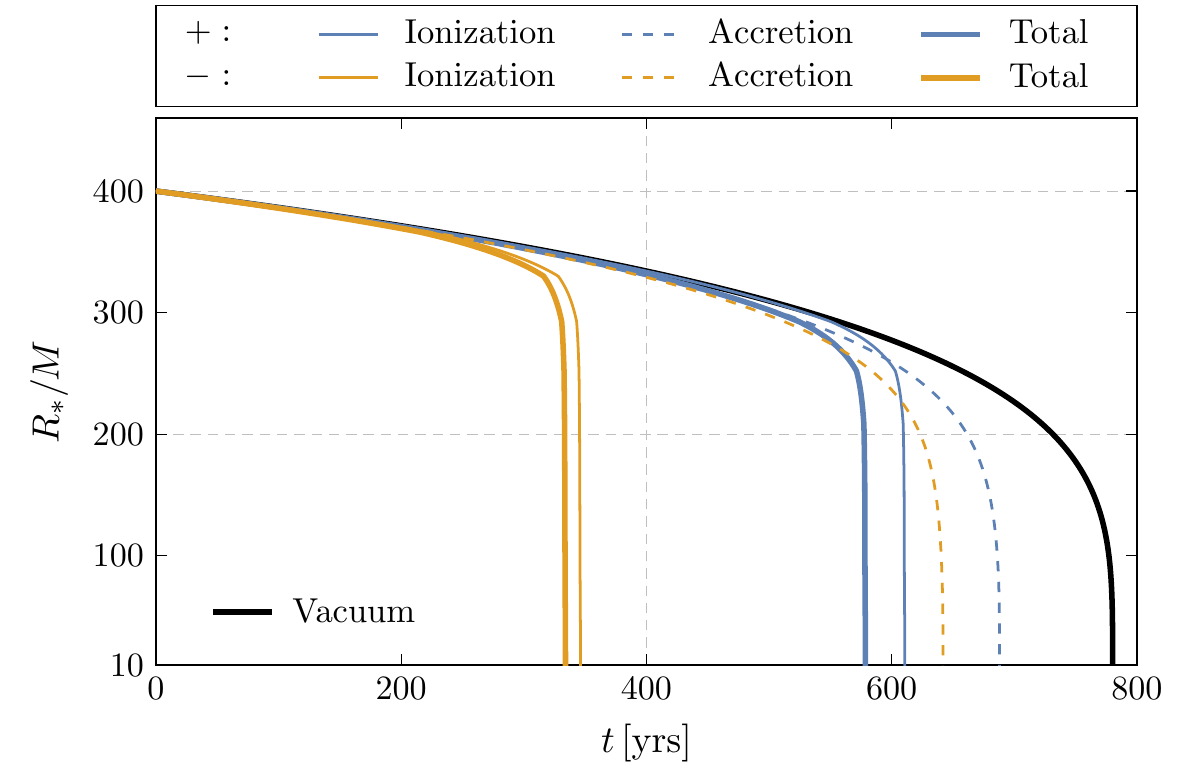}
            \caption{Evolution of the separation $R_*$, for $M=10^4M_\odot$ and $\alpha=0.2$, with initial values of $R_*=400M$, $q=10^{\protect \sminus 3}$ and $M_\lab{c}/M=0.01$ in a $|2 \es 1 \es 1\rangle$ state. Shown are the results for both co-rotating ($+$) and counter-rotating ($-$) orbits. The vacuum system, where no cloud is present, is shown for comparison. We see that accretion and ionization significantly reduce the merger time. }
    \label{fig:separation}
\end{figure}

\subsection{Numerical Results}
\label{sec:solution-evolution}

The system of equations (\ref{equ:q-evolve}), (\ref{equ:qc-evolve}) and (\ref{eqn:evolution-R}) determines the evolution of $M_*$, $M_\lab{c}$ and $R_*$. In this section, 
we solve these equations numerically for some benchmark parameters.

\vskip 4pt
It is first useful to comment on our choice of fiducial parameters and their astrophysical plausibility. 
 To make a strong observational case, we choose parameters for which ionization and accretion occur mostly in-band for a future space-based detector like LISA. At the same time, we must require that $q \ll 1$
 in order for our perturbative treatment to be applicable (see Appendix~\ref{app:Markov}). We thus consider 
 intermediate mass ratio inspirals, with $M=10^4M_\odot$ and $q=10^{\sminus 3}$, as we want the companion to be a reasonably-sized black hole. In order for the discontinuities in the ionization power $P_\lab{ion}$ to appear in the LISA band, we take $\alpha = 0.2$. This allows very fast superradiant growth of the cloud, but also makes it decay relatively rapidly to gravitational waves when the scalar field is real.
 The exact depletion rate depends on the initial mass of the cloud, but  
 for these parameters $M_\lab{c}/M$ is expected to fall to $0.01$ after $10^5$ years and to $0.001$ after $10^6$ years, with an extremely strong dependence on $\alpha$. It is thus not unreasonable to take $M_\lab{c}/M=0.01$ as a reference point for its initial value when ionization and accretion kick in; however, we will also show that even for $M_\lab{c}/M=0.001$ the impact of the cloud is still very large.

\vskip 4pt
It is possible to adjust the values of $M$, $q$ and $\alpha$. For example, we could reduce the value 
of $\alpha$ to make the cloud longer-lived. If we want the ionization features of the signal to stay in the LISA band, then we would have to simultaneously reduce the value of $M$ (which would increase $q$ if we keep $M_*$ fixed). However, in this work we only want to illustrate that ionization has a \emph{large} and \emph{sharp} effect on the inspiral, and we therefore do not attempt to find the region of parameter space with the most observational relevance. In the same vein, we fix the initial state of the cloud to $\ket{2\es 1\es 1}$ for simplicity. As previously mentioned, there is an uncertainty in the initial bound state due to the previous history of the system, both from past resonant transitions and the superradiant growth of other modes like the $|3 \es 2 \es 2 \rangle$ state which becomes relevant for larger values of $\alpha$. More concretely, for counter-rotating orbits, the state $\ket{2\es 1\es 1}$ cannot undergo any hyperfine transitions and the first Bohr transition (to the state $|3 \es 1 \es \sminus 1 \rangle$) occurs around $R_*/M\sim200$, when ionization is already a large effect (see Fig.~\ref{fig:ionization-power}). For co-rotating orbits, the hyperfine transition to the state $|2 \es 1 \es \sminus 1 \rangle$ can be significant and would have to be included in the analysis. We do not expect that choosing a different initial state would qualitatively affect our conclusions, but leave a more detailed analysis for future work.

\vskip 4pt
Let us now describe the numerical results. To understand the magnitude of the different effects, we show in Figure~\ref{fig:separation} the evolution of the parameters separately under the effects of ionization and accretion 
and then both combined, starting from a separation of $R_*=400M$. In all cases, we observe a very significant shortening of the time to merger, with the orbits suddenly \emph{sinking} as soon as the ionization energy losses overcome those in gravitational radiation. The dynamical evolution of the system is thus \emph{driven}, and not just perturbed, by the interaction of the binary with the cloud. 
The binary merges faster for counter-rotating orbits, since the ionization power is larger at large $R_*$ and the accretion force is opposite to the motion, c.f. Figure~\ref{fig:ionization-power} and~(\ref{eqn:evolution-R}).

\begin{figure}
            \centering
            \includegraphics{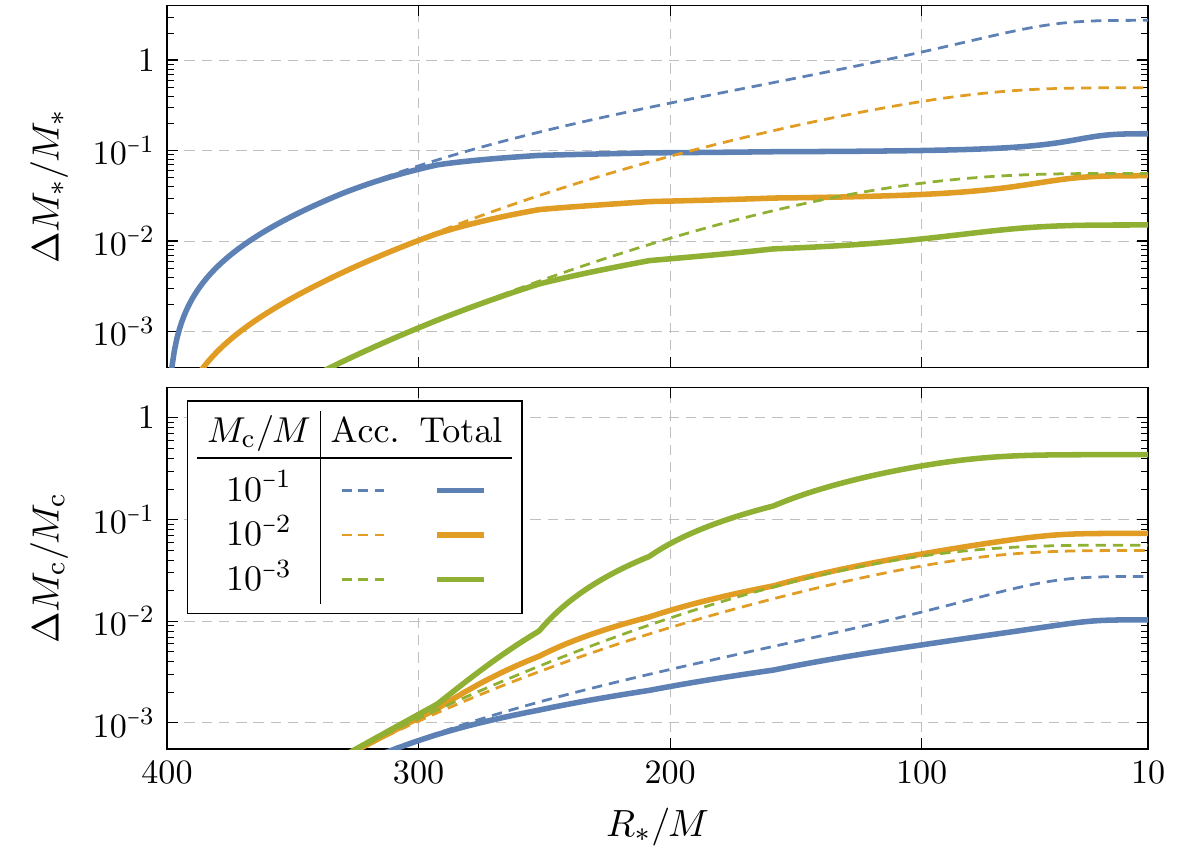}
            \caption{Fractional changes of the mass of the companion $M_*$ and the mass of the cloud $M_\lab{c}$, for $M=10^4M_\odot$ and $\alpha=0.2$, with initial values of $R_*=400M$, $M_* =10^{\protect \sminus 3} M$. Shown are the results for three different initial values of $M_\lab{c}$.  All curves refer to co-rotating orbits and a $|2\es1\es1\rangle$ bound state.}
    \label{fig:q_and_M_c}
\end{figure}

\vskip 4pt
In the top panel of Figure~\ref{fig:q_and_M_c}, we show the fractional change of the mass of the companion~$M_*$ as function of the separation $R_*$, for co-rotating orbits and three different initial values of $M_\lab{c}$. We observe that accretion takes place throughout the entire inspiral, without a clear hierarchy between the timescales of accretion and merger. Not surprisingly, we see that the accreted mass is very sensitive to $M_\lab{c}/M$, with the total $\Delta M_*/M_*$ being roughly proportional to it, at least in the early stages of the inspiral. In fact, in the case with only accretion, the final value of $\Delta M_*/M_*$ can be predicted from a simple order-of-magnitude estimate: multiplying the average accretion flux, (\ref{eqn:accretion-law}), by the time-to-merger in vacuum, we get $\Delta M_*/M_*\sim(M_\lab{c}/M)(r_\lab{c}/M)$, which is in good agreement with the numerical results shown in Figure~\ref{fig:q_and_M_c}. What is maybe more surprising is that the inclusion of ionization strongly limits the accretion of mass. This phenomenon can be explained by noting that ionization does not have a big effect on the accretion rate (\ref{equ:q-evolve}), which only depends on $\rho$ and $q$, but significantly reduces the time spent inside the cloud, and therefore the total accreted mass.

\vskip 4pt
In the bottom panel of Figure~\ref{fig:q_and_M_c}, we show the fractional change of the mass of the cloud~$M_\lab{c}$. We see that the cloud is partially depleted during the inspiral, due to both ionization and accretion. The hierarchy between the two effects depends on the initial value of $M_\lab{c}$. For more massive clouds, the primary mechanism of mass loss is accretion, which is limited by the inclusion of ionization due to the reduced time spent inside the cloud. Instead, for lighter clouds, ionization is the primary mechanism of mass loss. We see that the total mass loss does not seem to depend sensitively on the initial value of $M_\lab{c}$, so that the fractional mass loss is larger for smaller clouds. In our example, with $M_\lab{c}/M=0.1$, only about 1\% of the initial mass is lost at the end of the inspiral; instead, more than 50\% would be depleted for an initial $M_\lab{c}/M = 10^{\sminus 3}$.

\begin{figure}
            \centering
            \includegraphics{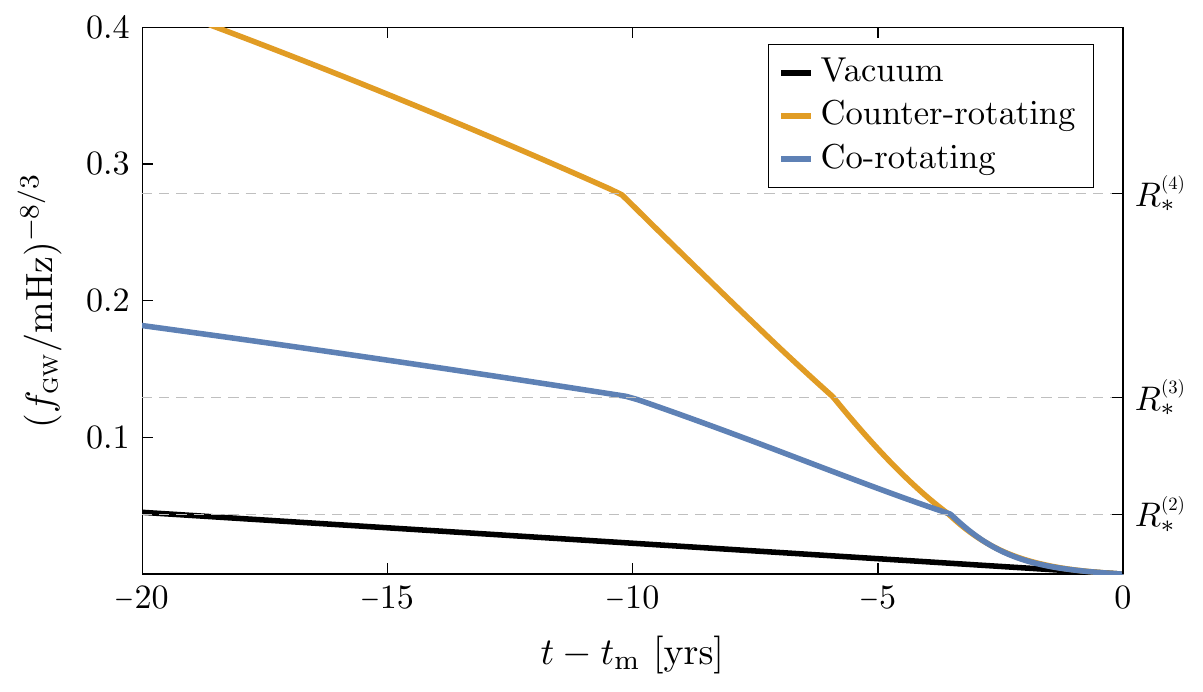}
            \caption{Evolution of the GW frequency as a function of the remaining time to merger, $t-t_\lab{m}$, for $M=10^4M_\odot$ and $\alpha=0.2$, with initial values of $R_*=400M$, $q=10^{\protect \sminus 3}$ and $M_\lab{c}/M = 10^{\protect \sminus 3}$
             in a $|2\es1\es 1\rangle$ state. The central region of the range shown on the $y$ axis corresponds to a few millihertz, falling inside the LISA sensitivity band. The ``kinks'' at separations $R_*^{(g)}$ correspond to the discontinuities in the ionization power, see Figure~\ref{fig:ionization-power}.}
    \label{fig:evolution-frequency}
\end{figure}

\vskip 4pt
It is natural to wonder how degenerate the observables are with the expected signal from a binary in vacuum with different parameters. Although we postpone a systematic study of this issue to future work, it is useful to compare the evolution of the GW frequency $f_\slab{gw}$ as a function of the time to merger. This is done in Figure~\ref{fig:evolution-frequency} for the very conservative case of initial $M_\lab{c}/M = 10^{\sminus 3}$, demonstrating that even a tiny cloud can have a strong impact on the inspiral. In the plot, the scale of the frequency axis has been chosen such that the non-relativistic vacuum evolution, $f_\slab{gw}\propto(t_\lab{m}-t)^{\sminus 3/8}$, where $t_\lab{m}$ is the merger time, becomes a straight line. It is apparent that the shape of $f_\slab{gw}(t)$ deviates significantly from a straight line: a decisive role is played by the ``kinks'' appearing at the frequencies where the ionization power $P_\lab{ion}$ is discontinuous, cf.~Figure~\ref{fig:ionization-power}. From (\ref{eq:rDiscont}), kinks appear at the frequencies
\begin{equation}
  \begin{aligned}
    f_\slab{gw}^{(g)} &= \frac{6.45\,\text{mHz}}{g}\left(\frac{10^4M_\odot}{ M}\right)\!\left(\frac{\vphantom{10^{4} M_\odot}\alpha}{0.2 \vphantom{M}}\right)^{\!3}\!\left(\frac{2}{n_b}\right)^{\!2} \\
    &=\frac{33.5 \,\text{mHz}}{g} \left(\frac{M}{10^4M_\odot}\right)^{\!2}\!\left(\frac{\vphantom{M}\mu}{10^{\sminus 14}\,{\rm eV}}\right)^{\!3}\!\left(\frac{2}{n_b}\right)^{\!2}\! ,
  \end{aligned}
  \label{eqn:f-discontinuities}
\end{equation}
where the overtone number $g$ ranges over positive integers and $n_b$ is the principal number of the cloud's initial state. These kinks thus  constitute a sharp observational signature of ionization caught in the act. If only a region between two kinks is observed, then the evolution is likely to be more degenerate with a signal from a vacuum system, whose parameters would however differ from the true parameters of the binary.

\subsection{Open Problems}

We now address a number of unresolved questions regarding the phenomenology of the system, discussing the limitations of our analysis and some future prospects.
    
\vskip 4pt
{\it Gravitational field of the cloud}---By using the simple Keplerian relation $\Omega^2 R_*^3=(1+q)M$, we have neglected the backreaction due to the gravitational field of the cloud. This backreaction would manifest itself as $\mathcal{O}(M_\lab{c}/M)$ corrections to the orbital dynamics. At the Newtonian level, the effect of the cloud is two-fold: the enclosed mass ``seen'' by the companion varies with $R_*$, due to the diffuse nature of the cloud, and the nontrivial angular structure of the cloud generates higher mass multipoles. The first effect is only relevant when the companion orbits inside the cloud, $R_*\sim r_\lab{c}$, while the second can also provide corrections at large distances.

\vskip 4pt 
{\it Angular structure of the cloud}---Similarly, we have ignored the angular structure of the cloud in our treatment of accretion, where we azimuthally averaged the mass density and assumed that the accretion process was accurately captured by averaging over each orbit. This implicitly assumes that the orbit remains quasi-circular even after we include accretion effects. However, we  expect that this assumption can break down at certain points during the inspiral, like when the relative velocity between the cloud and companion vanishes for co-rotating orbits and the companion has enough time to develop nontrivial eccentricity.

\vskip 4pt
{\it Inclination and eccentricity}---For simplicity, we have only studied equatorial quasi-circular orbits. The phenomenology of inclined orbits is potentially much richer, as the transfer of angular momentum between the cloud and orbit can cause the orbital plane to precess. The companion would also explore regions of the cloud with different densities, resulting in an uneven distribution of the ``forces" appearing in (\ref{eqn:evolution-R}) over the course of an orbit, potentially causing the orbit to become more eccentric. 
Taking into account 
eccentricity 
is necessary for a more complete analysis even in the simple case of equatorial orbits, especially in situations where the forces in (\ref{eqn:evolution-R}) have a nontrivial $R_*$ dependence. For example, for co-rotating orbits the accretion force does not always act as a drag, changing sign with $R_*$. 

\vskip 4pt
    {\it Resonances}---We have not studied the 
    interplay of the resonances between bound states with the ionization and accretion processes.
The effect of resonances on the dynamics is twofold. First, they introduce periods of either accelerated (``sinking'' orbits), or decelerated (``floating'' orbits) inspiral: these would appear as distinctive features in the evolution of the separation and frequency.
     Second, the resonances can change the state of the cloud. 
      Both of these effects can interact nontrivially with ionization and accretion, as the total mass accreted or ionized depends on the time spent at a given orbital separation, and on the state of the cloud. For instance, the effects of both ionization and accretion will be enhanced during a floating orbit, while a rapidly sinking orbit can break many the various approximations we have relied on in our analysis.
Furthermore, when the cloud transitions to an excited state it becomes easier to ionize, so this dependence on the evolution of the state has to included in a self-consistent analysis of the ionization.
  It would be interesting to study the state dependence of the ionization signal in more detail.

\vskip 4pt
 {\it Equal mass ratios}---We have only studied the case of a large mass ratio $q \ll 1$, where  
 the gravitational influence of the companion could be treated perturbatively. 
 The parameter~$q$ is one of the main order parameters in our perturbative analysis and many of our approximations 
 do not hold when $q \sim 1$. It would be interesting to develop a formalism that is able to treat the case of equal mass ratios,\footnote{For a recent attempt to describe this regime see~\cite{Takahashi:2021yhy}.} 
where ionization
can be efficient enough to 
completely evaporate the cloud before the merger.

\vskip 4pt
    {\it Transient oscillations}---Our analysis relied on replacing the dynamics of the ionization process with its ``steady state'' behavior (\ref{eq:realisticDeoccupation}). However, as we described in Section~\ref{sec:warmup}, there is interesting transient behavior that occurs when a bound state just begins to resonate with a continuum band. How are these transient oscillations modified when we include the cloud's backreaction on the orbit? Do these oscillations also affect the orbital dynamics, and can we observe them in the resulting gravitational wave signal? These are interesting questions for the future that require a different formalism to answer.

\vskip 4pt
  {\it Relativistic corrections}---Our treatment was non-relativistic, both in the derivation of the mass accretion and ionization, and in the orbital evolution. Hence, our results do not apply in the final phase of the inspiral---closer to the merger---when the velocity approaches the speed of light and the post-Newtonian expansion breaks down. This period of the inspiral is notoriously difficult to model even for vacuum systems, especially for large mass ratios. 
  However, close to the merger, we expect the effects of the cloud to fade in comparison to the increasingly strong nonlinearities of the vacuum evolution (see Figures~\ref{fig:ionization-power} and \ref{fig:evolution-frequency}). The region where resonances, ionization and accretion are most relevant is thus  within the applicability of the non-relativistic approximation.

%\newpage
\section{Conclusions}
\label{sec:conclusions}

Gravitational waves provide an interesting new window into the weak-coupling frontier of particle physics, giving us access to physics that is invisible to traditional collider experiments~\cite{Baumann:2019ztm}. 
Such weakly-coupled sectors arise in the string landscape as ultralight axions~\cite{Arvanitaki:2009fg, Mehta:2021pwf, Mehta:2020kwu,Demirtas:2018akl} and are also interesting dark matter candidates~\cite{Hui:2016ltb}. 
In this paper, we have studied the dynamical effects of clouds of ultralight scalars around black holes when they are part of binary systems.  We have calculated two novel types of cloud-binary interactions:  
the ionization of the cloud due to the gravitational perturbation from the binary companion and the accretion of mass onto the secondary object, in the case it is a black hole.

\vskip 4pt
When unbound states are excited by the gravitational perturbation due to the companion, the cloud loses mass to outgoing scalar waves. This ionization induces a backreaction on the orbit of the binary, which loses energy and angular momentum to the scalar field. These losses are notable for two reasons: (1) they can significantly exceed the energy loss due to GW emission, thus dominating the dynamics of the system, and (2) they contain sharp features (see Figure~\ref{fig:ionization-power}), which carry detailed information about the microscopic structure of the cloud.

\vskip 4pt
During the inspiral, the companion will move inside the scalar cloud. If the companion is a black hole, then its event horizon will absorb parts of the cloud. Due to the high densities reachable by superradiantly-generated clouds, the mass of the secondary object can significantly increase during the inspiral, which impacts the dynamics of the system. The momentum accreted by the object is also non-negligible.

\vskip 4pt
Both ionization and accretion affect the orbital dynamics of the binary.
We studied this backreaction numerically, finding that the deviation from the expectations for a vacuum system can be rather dramatic. 
The inspiral happens much faster than in the absence of the cloud, and both the mass of the companion and of the cloud evolve significantly in time. Even with conservative choices of parameters, the frequency evolution is quantitatively and qualitatively modified, especially due to the discontinuities in the ionization power producing ``kinks'' in the frequency evolution of the gravitational waves (see Figure~\ref{fig:evolution-frequency}). These features are a new and distinctive signature of gravitational atoms in black hole binaries.

\vskip 4pt
Our analysis made a number of simplifying assumptions. First, we restricted ourselves to extreme mass ratio inspirals on quasi-circular orbits in the equatorial plane. We expect that qualitatively new behavior appears for equal mass ratios, and that both inclination and eccentricity can lead to a rich phenomenology in the presence of the cloud. These are both interesting directions for future work. Similarly, we did not explore the interplay between bound state resonances and both ionization and accretion, nor did we account for the interesting transient phenomena that occur when the ionization process begins. A more complete analysis should take both of these into account.

\vskip 4pt
A combined treatment of the resonances studied in~\cite{Baumann:2018vus,Baumann:2019ztm}, together with the ionization and accretion discussed in this work, is required to achieve a complete understanding of the phenomenology of gravitational atoms in binaries. This in turn will serve as a starting point to devise suitable strategies to discover and characterize these systems with upcoming gravitational wave detectors.  Current data analysis techniques mostly rely on matched filtering, where waveform templates are compared to observations. Waveforms based on vacuum systems may thus produce a very low signal-to-noise ratio when applied to our case, because of the drastically different evolution of the observables. 
In a template-based approach, dedicated searches are thus needed to not miss inspirals involving gravitational atoms and to distinguish them from other kinds of environmental effects, like dark matter overdensities~\cite{Eda:2013gg,Eda:2014kra, Kavanagh:2020cfn,Coogan:2021uqv}. We postpone a systematic study of these phenomenological issues to future work.

\paragraph{Acknowledgements}
DB and JS are grateful to Horng Sheng Chia and Rafael Porto for previous collaborations on this topic.
We thank Thomas Spieksma for verifying some of the numerical results in this paper.
DB receives funding from a VIDI grant of the Netherlands Organisation for Scientific Research~(NWO) and is part of the Delta-ITP consortium.  
DB is also supported by a Yushan Professorship at National Taiwan University funded by the Ministry of Science and Technology (Taiwan).  JS is supported by NASA grant \texttt{80NSSC20K0506}.

\newpage
\appendix
\addtocontents{toc}{\protect\vskip24pt}
     \section{Integrating out the Continuum}
      \label{app:approx} 

As explained in the main text, the dynamics of the gravitational atom in a binary, including both bound and continuum states,  can be captured by integrating out the continuum and incorporating its effects in terms of a set of induced couplings and energies for the bound states alone. This process yields an effective Schr\"{o}dinger equation for the bound states that describes the behavior of the entire system. In this appendix, we justify the approximations we used to derive these continuum-induced couplings. First, we explain how our approximation for the fractional deoccupation rate (\ref{eq:toyDeoccupation}) in the toy model arises from the large time asymptotics of the induced energy. This derivation relies on ignoring the transitions between continuum states, so we then justify this assumption. Next, we discuss the complications that arise in the more realistic case, which includes many more bound and continuum states. We then describe an alternative, albeit uncontrolled, derivation of (\ref{eq:toyDeoccupation}) using stationary perturbation theory. Finally, we conclude with a discussion of the effects a nonlinearly ramping frequency $\dot{\varphi}_*(t)$ has on our approximations.

\subsection{Saddle Point Approximation}
        
 We are interested in the asymptotic behavior of the induced energy 
          \begin{equation}
            \mathcal{E}_b(t) = \int_{\sminus \infty}^{t}\!\ud t' \, \Sigma_b(t, t') = \frac{1}{2 \pi i} \int_{\sminus \infty}^{t} \!\ud t' \int_0^{\infty}\!\ud k \, |\eta(k)|^2 \, e^{-i (\epsilon(k)- \epsilon_b)(t - t') + i(\varphi_*(t) - \varphi_*(t'))}\,, \label{eq:inducedEnergyApp}
        \end{equation}
where $\epsilon(k) = k^2/(2 \mu)$.
        Without loss of generality, we can absorb the bound state energy into our reference frequency, $\varphi_*(t) = - \epsilon_b t + \gamma t^2/2$, and assume that $\gamma > 0$. The bound state then begins to ``resonate'' with the continuum for $t \gtrsim 0$, and we would like to determine the asymptotic behavior of this function before and after this time,~$|\sqrt{\gamma} t| \gg 1$, as a way of approximating its behavior away from the complicated transient region around $t = 0$. 

        \vskip 4pt 
        There are two representations of this function that will be useful. We can either first perform the integral over $t'$ to find
        \begin{equation}
                \mathcal{E}_b(t) = \frac{1}{\sqrt{8 \pi \gamma}} \int_0^{\infty}\!\ud \epsilon\, |\eta(\epsilon)|^2 \es \exp\!\left[\frac{i(\epsilon-\gamma t )^2}{2 \gamma} - \frac{3 \pi i }{4}\right] \erfc\!\left[\frac{e^{\frac{i \pi}{4}} (\epsilon-\gamma t)}{\sqrt{2 \gamma}}\right], \label{eq:selfEnergyErfc}
        \end{equation}
        or we can define $z = \sqrt{\gamma}(t - t')$ and write 
        \begin{equation}
          \mathcal{E}_b(t) = \frac{1}{2 \pi i \sqrt{\gamma}} \int_{0}^{\infty}\!\ud z\, e^{i \tau z}\, \mathcal{K}(z)\,, \label{eq:selfEnergyLaplace}
        \end{equation}
        where we introduced the dimensionless time $\tau \equiv \sqrt{\gamma} t$ and the kernel
        \begin{equation}
          \mathcal{K}(z) \equiv e^{\sminus \frac{1}{2} i z^2} \int_0^{\infty} \!\ud \epsilon \, e^{\sminus i \epsilon z/\sqrt{\gamma}} \,|\eta(\epsilon)|^2\,. \label{eq:laplaceKernel}
        \end{equation}
        The former has the benefit of making the ``resonance'' behavior much clearer, while the latter is useful for understanding the large time $|\tau| \gg 1$ asymptotics since it has the form of a standard Laplace-like integral. In both representations, we have transformed the integral over momenta $k$ into an integral over the energy $\epsilon$ and defined $|\eta(\epsilon)|^2 = \ud k(\epsilon)/\ud \epsilon\, \big|\eta(k(\epsilon))\big|^2 = \mu |\eta(k)|^2/k$. In the cases of interest, $|\eta(\epsilon)|^2$ approaches a constant as $\epsilon \to 0$ and decays algebraically as $\epsilon \to \infty$, so that the ``total coupling'' of the bound state to the continuum $\int_{0}^{\infty}\!\ud\epsilon \, |\eta(\epsilon)|^2$ is finite.

        \begin{figure}
          \centering 
          \includegraphics[trim={0 6pt 0 0}]{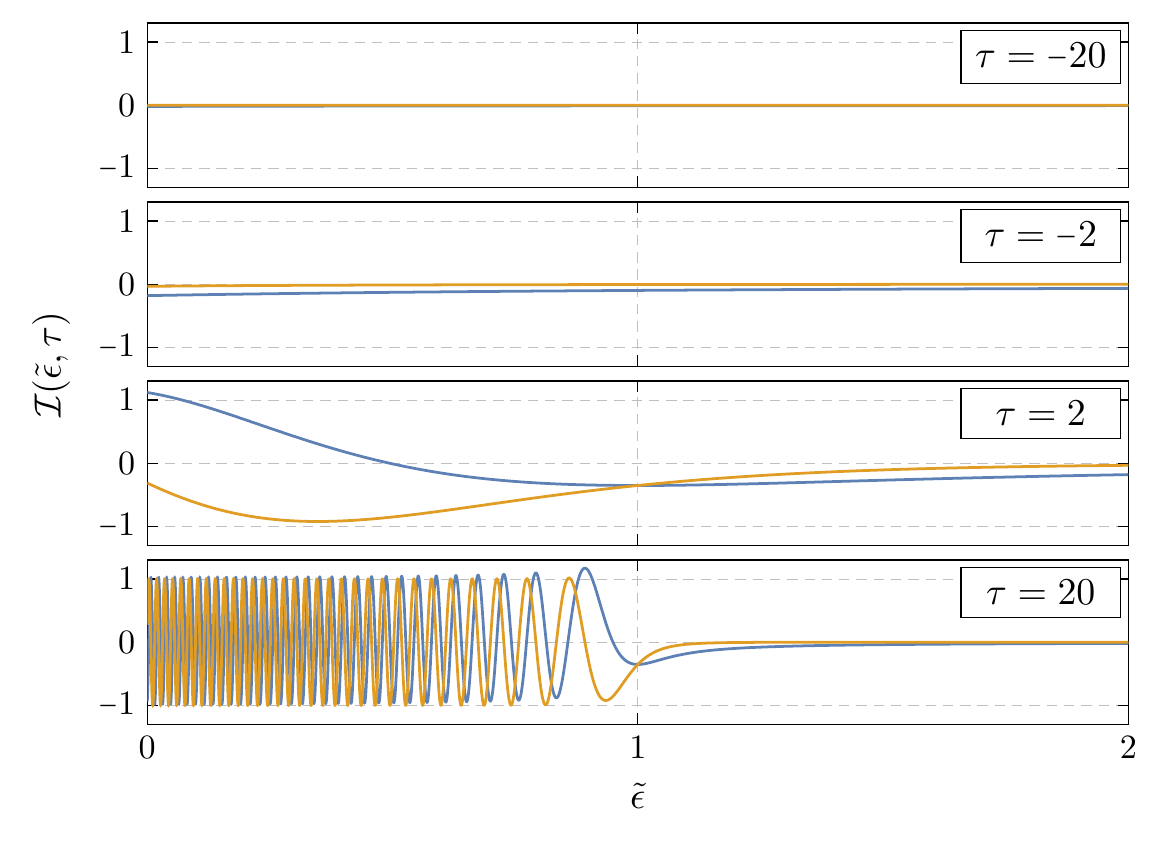}
          \caption{The real {\color{Mathematica1} [blue]} and imaginary {\color{Mathematica2} [orange]} parts of the modulating function $\mathcal{I}(\tilde{\epsilon}, \tau)$, for several values of the dimensionless time $\tau$. For large negative values of $\tau$, the integrand of (\ref{eq:inducedEnergyMapped}) is highly suppressed for $\tilde{\epsilon} \in [0, \infty)$. For large positive times $\tau \gg 1$, the integrand oscillates rapidly when $\tilde{\epsilon} \in [0, 1]$, slowing down when $\tilde{\epsilon} \sim 1$, and is then again highly suppressed for $\tilde{\epsilon} \gg 1$. \label{fig:intPlot}} 
        \end{figure}

        \vskip 4pt
        To get a sense for the behavior of this function, it is useful to first rescale the integral in (\ref{eq:selfEnergyErfc}) by taking $\epsilon \to \sqrt{\gamma} |\tau| \tilde{\epsilon}$\,,
        \begin{equation} 
            \mathcal{E}_b(\tau) = \frac{|\tau|}{\sqrt{2 \pi}} \int_{0}^{\infty}\!\ud \tilde{\epsilon} \, \big|\eta(\sqrt{\gamma} |\tau| \tilde{\epsilon})\big|^2 \,\es \mathcal{I}(\tilde{\epsilon}, \tau)\,, \label{eq:inducedEnergyMapped}
        \end{equation}
        where we defined the kernel
        \begin{equation}
            \mathcal{I}(\tilde{\epsilon}, \tau) \equiv \frac{1}{2}\es e^{\frac{i \tau^2}{2} (\tilde{\epsilon}- \sgn \tau )^2 - \frac{3 \pi i}{4}} \erfc\!\left[\tfrac{|\tau|}{\sqrt{2}}{e^{\frac{i \pi}{4}}(\tilde{\epsilon}-\sgn \tau)}\right] .
        \end{equation}
        We plot this kernel for several values of $\tau$ in Figure~\ref{fig:intPlot}. We see that, for $\tau \to \minus \infty$, the integrand of (\ref{eq:inducedEnergyMapped}) is strongly suppressed throughout the entire integration region, and so both the real and imaginary parts of the induced energy will be small. In the opposite limit, $\tau \to + \infty$, the integrand oscillates rapidly in the interval $\tilde{\epsilon} \in (0, 1)$, so we expect only the end point~$\tilde{\epsilon}=0$ and the region around $\tilde{\epsilon} = 1$
to contribute to the integral. For $\tilde{\epsilon} \in (1, \infty)$, the integrand no longer oscillates, but instead decays algebraically. The integrand---and especially the real part in {\color{Mathematica1} [blue]}---has a very heavy tail which the saddle point approximation is not able to fully capture. Instead, we will need to use the Laplace-like form (\ref{eq:selfEnergyLaplace}) to compute these additional contributions.

        \vskip 4pt
        Keeping in mind that the saddle point approximation does not capture the full behavior of the induced energy as $\tau \to \infty$, we will apply it anyway. As stated before, there are two relevant contributions---from the endpoint at $\tilde{\epsilon} = 0$ and from the ``saddle point'' at $\tilde{\epsilon} = 1$. From Figure~\ref{fig:intPlot}, we expect that the contribution at $\tilde{\epsilon} = 0$ produces an oscillatory \emph{ringing} that is left over from when the bound state first hits the edge of the continuum, and how quickly these oscillations decay depends on how the bound state couples to the lowest energy continuum modes, i.e.~how $|\eta(\epsilon)|^2$ scales as $\epsilon \to 0$. In contrast, the saddle point at $\tilde{\epsilon} = 1$ gives a non-oscillatory decay which only depends on the coupling between the bound state and the particular continuum state it is ``resonating with,''~$|\eta(\epsilon = \gamma t)|^2$. Assuming that $|\eta(\epsilon)|^2$ approaches a constant $|\eta|^2$ as $\epsilon \to 0$, we find that
        \begin{equation}
            \mathcal{E}_b(t) \sim - \frac{i \mu \big|\eta(k_*(t))\big|^2}{2 k_*(t)} -\frac{|\eta|^2 \lab{e}^{\frac{1}{2} i \gamma t^2 - \frac{i\pi}{4}}}{2\sqrt{2 \pi \gamma} t} \left[1 + \erf\!\left(\frac{\lab{e}^{\frac{i \pi}{4}} \!\sqrt{\gamma} t}{\sqrt{2}} \right)\right]  ,  \quad \sqrt{\gamma} t \to +\infty\,,
        \end{equation}
        where we have switched back to parameterizing the system in terms of the momentum and introduced $k_*(t) = \sqrt{2 \mu \gamma t}$, the momentum of the state at the saddle point. 

        \vskip 4pt
        To find the dominant behavior of $\Re\mathcal{E}_b(t)$ as $\tau \to \pm \infty$, we can use (\ref{eq:selfEnergyLaplace}) and repeatedly integrate by parts in $z$ to generate an expansion in powers of $\tau^{\sminus 1}$. However, the aforementioned heavy tail can hinder this iterative process. Each integration by parts generates higher derivatives of the kernel evaluated at $z = 0$, 
        but these derivatives are not necessarily finite. From (\ref{eq:laplaceKernel}), we see that $\partial^k_z \mathcal{K}(z)|_{z = 0}$ contains a term proportional to $\int_{0}^{\infty}\!\ud \epsilon \, \epsilon^k |\eta(\epsilon)|^2$, and since $|\eta(\epsilon)|^2$ decays only algebraically, sufficiently high derivatives will diverge. This signals that $\mathcal{K}(z)$ has terms of the form $z^k \log^n z$, which produce asymptotic behavior of the form $\log^n \tau/\tau^{k+1}$, i.e. logarithmic behavior that is not captured in the standard saddle point approximation. 

        \vskip 4pt
        For our purposes, we will only concentrate on the leading order $|\tau| \to \infty$ behavior. This is governed by the total coupling $\mathcal{K}(0) = \int_0^{\infty}\!\ud \epsilon\, |\eta(\epsilon)|^2 = \int_{0}^{\infty}\!\ud k\, |\eta(k)|^2$, and direct integration yields
        \begin{equation}
          \mathcal{E}_b(t) \sim \frac{1}{2 \pi \gamma t} \left[\int_{0}^{\infty}\!\ud k \, |\eta(k)|^2\right] + \cdots\,.
        \end{equation}
        As $\tau \to \minus \infty$, this is the dominant contribution and gives an accurate approximation---as the effective energy gap between the bound and continuum states shrinks, the coupling to the continuum induces a correction to the bound state's energy. There is, however, no appreciable deoccupation of the bound state until after the transition at $\tau = 0$. As $\tau \to +\infty$, the integral picks up an additional saddle point and the induced energy is well approximated by
        \begin{equation}
          \mathcal{E}_b(t) \sim - \frac{i \mu \big|\eta(k_*(t))\big|^2}{2 k_*(t)} -\frac{|\eta|^2 \lab{e}^{\frac{1}{2} i \gamma t^2 - \frac{i \pi}{4}}}{2\sqrt{2 \pi \gamma} t} \left[1 + \erf\!\left(\frac{\lab{e}^{\frac{i \pi}{4}} \!\sqrt{\gamma} t}{\sqrt{2}} \right)\right]  + \frac{1}{2 \pi \gamma t} \left[\int_{0}^{\infty}\!\ud k \, |\eta(k)|^2\right] + \cdots\, .\label{eq:inducedApproximation}
        \end{equation} 
        Since we are mainly concerned with the imaginary part of this expression, we use the first term in (\ref{eq:inducedApproximation})  throughout the main text.

    \newpage
    \subsection{Unbound-Unbound Transitions}
  \label{sec:Unbound}
    
      It will be helpful to address our assumption that we can ignore the transitions between the continuum states in our analysis of the ionization process. We will do so in the toy model studied above and in Section~\ref{sec:warmup}. Numerical experiments show that the bound state's dynamics are relatively unaffected if we include these transitions and is still well-described by the first term in (\ref{eq:inducedApproximation}).  We can understand better why they may be ignored, and justify our assumption, by including  these couplings in the toy Hamiltonian (\ref{eq:toyHam}) and arguing that they should, at least at weak coupling, provide a subleading correction to the effective Schr\"{o}dinger equation (\ref{eq:toyEffSchro}).

      \vskip 4pt
      A nontrivial coupling between continuum states $\eta(k, k') = \langle k | \mathcal{H} | k'\rangle$, for $k \neq k'$, changes the solution (\ref{eq:warmupContSol}) for the continuum amplitudes to
         \begin{equation}
         \begin{aligned}
            c_{k}(t) = &-i \int_{\sminus \infty}^{t}\!\ud t'\, \eta(k)\, e^{- i \varphi_*(t') + i(\epsilon(k) - \epsilon_b) t'} c_b(t') \\
            &+ \frac{1}{2 \pi i}\int_{\sminus \infty}^{t}\! \ud t' \int_0^\infty\!\ud k'\, \eta(k, k') \,e^{i (\epsilon(k) - \epsilon(k'))t'} c_{k'}(t')\,. 
            \end{aligned}\label{eq:warmupContSolMod}
        \end{equation}
        Importantly, both the bound-to-unbound couplings $\eta(k)$ and unbound-to-unbound couplings $\eta(k, k'; t)$ are $\mathcal{O}(q \alpha)$ and we work exclusively in the $q \alpha \ll 1$ regime. By plugging this solution back into itself, we can generate a solution purely in terms of the bound state amplitude, with the first correction to the $\eta(k, k') \to  0$ limit of (\ref{eq:warmupContSolMod}) being
        \begin{equation}
          c_k(t) \supset - \frac{1}{2 \pi }\int_{\sminus \infty}^{t}\! \ud t_1 \int_{\sminus \infty}^{t_1}\!\ud t_2 \int_0^\infty\!\ud k'\, \eta(k, k')\eta(k') \,e^{i (\epsilon(k) - \epsilon(k'))t_1 - i \varphi_*(t_2) + i(\epsilon(k') - \epsilon_b) t_2}c_b(t_2)\,,
        \end{equation}
        which is $\mathcal{O}(q^2 \alpha^2)$, while other corrections are higher order. 

        \vskip 4pt
        In the bound state Schr\"{o}dinger equation (\ref{eq:toyEffSchro}), this correction contributes a term involving the chain of matrix elements $\langle b | \mathcal{H} | k \rangle \langle k | \mathcal{H} | k' \rangle \langle k' | \mathcal{H} |b\rangle$, while the leading-order solution only involves the chain of elements~$\langle b | \mathcal{H} | k \rangle \langle k | \mathcal{H} |b \rangle$. Clearly, the leading-order contribution only accounts for the system transitioning into the continuum and then back to the bound state, while higher-order corrections involve the system going into the continuum and then bouncing around between different continuum states before returning to the bound state. Each of these transitions is thus penalized by an additional factor of $q \alpha$ and so we expect that they provide a subleading effect, especially at weak coupling $q \alpha \ll 1$.

        \vskip 4pt
        We might worry that, over long times, a substantial enough continuum population can be built up so that the second term in (\ref{eq:warmupContSolMod}) can overcome its $\mathcal{O}\big(q^2 \alpha^2\big)$-suppression and compete with the first. However, this sort of coherent effect is extremely unlikely in light of the oscillatory factors in (\ref{eq:warmupContSolMod}), which serve to randomize the ``direction'' of this perturbation and suppress its effects on long time scales. These arguments can be trivially extended to the more realistic case discussed in the next section, so we will ignore continuum-to-continuum transitions throughout our analysis and focus only on how the bound states interact with the continuum.

    \newpage
    \subsection{Extension to the Realistic Case}

      The main complication in going to the more realistic case is that there are many more bound and continuum states, and the continuum now mediates transitions between different bound~states. These effects appear in the form of off-diagonal induced couplings:
      \begin{equation}
        \mathcal{E}_{ba}(t) = -i \sum_{K} \eta^{*\floq{\Delta m_b}}_{K b}(t) \eta^{\floq{\Delta m_a}}_{K a}(t) \int_{\sminus \infty}^{t}\!\ud t'\,  e^{i \Delta m_b \varphi_*(t) - i \Delta m_a \varphi_*(t') + i( \epsilon_b - \epsilon_K) t + i (\epsilon_K - \epsilon_a) t'}\, , \label{eq:appInducedCouplings}
      \end{equation}
       where we have introduced the shorthand $\Delta m_a \equiv m - m_a$ and $\Delta m_b \equiv m - m_b$. We would like to understand the general behavior of these off-diagonal terms and argue that they can be ignored whenever the resonance condition between the states $|a \rangle$ and $|b \rangle$ is not satisfied. On resonance, they provide a small correction compared to the direct coupling between these states and so they can be neglected.

      \vskip 4pt
      Assuming that the frequency $\dot{\varphi}_*(t)$ is linear, we can again define the variable $z \equiv t - t'$ and write (\ref{eq:appInducedCouplings}) as
      \begin{equation}
        \begin{aligned}
        \mathcal{E}_{ba}(t) = &\, e^{i (\epsilon_b - \epsilon_a) t - i (m_b - m_a) \varphi_*(t)} \\
          & \times \left[-i \sum_{K}  \int_{0}^{\infty}\!\ud z\,  e^{-\frac{1}{2} i \Delta m_a \gamma z^2 + i (\Delta m_a \dot{\varphi}_*(t) - (\epsilon_K - \epsilon_a)) z } \eta^{*\floq{\Delta m_b}}_{K b}(t) \eta^{\floq{\Delta m_a}}_{K a}(t)\right] . \label{eq:appInducedCouplingsLaplace}
        \end{aligned}
      \end{equation}
      The term in braces is of a similar form to the induced energy (\ref{eq:selfEnergyLaplace}), whose behavior we have already analyzed in (\ref{eq:inducedApproximation}). It contains both oscillating and smoothly decaying  terms. Ignoring these oscillating terms for now, we see that the induced couplings oscillate rapidly with phase $\exp\!\big[i (\epsilon_b - \epsilon_a) t - i (m_b - m_a) \varphi_*(t)\big]$. As we argue in Section~\ref{sec:realisticIonization}, the direct couplings between $|a\rangle$ and $|b \rangle$ also oscillate with this phase, and if these oscillations are too rapid the contribution to the bound state solution will quickly average out. Of course, this oscillation slows down when the resonance condition $(m_b - m_a) \dot{\varphi}_*(t) = (\epsilon_b - \epsilon_a)$ is satisfied, but again these induced couplings, which are $\mathcal{O}(q^2 \alpha^2)$, must compete with the $\mathcal{O}(q \alpha)$ direct couplings $\eta_{ba}$, and so even then they have a small effect on the behavior of the resonance for $q \alpha \ll 1$.

      \vskip 4pt
      We might worry about the oscillations that arise in (\ref{eq:inducedApproximation}) as transients when the state $|a \rangle$  begins to resonate with the continuum might spoil this story, and that these induced couplings might become relevant. Fortunately, this is not the case. These transient oscillations ``start'' when the companion can excite $|a \rangle$ into the continuum, $\Delta m_a \dot{\varphi}_*(t) = -\epsilon_a$, and if they are present they modify the overall exponential in (\ref{eq:appInducedCouplingsLaplace}) to
      \begin{equation}
        \exp\!\left[-i (m_b - m_a) \varphi_*(t) + i (\epsilon_b - \epsilon_a) t +i (\Delta m_a\dot{\varphi}_*(t) + \epsilon_a)^2/(2 \Delta m_a \gamma)\right] .
      \end{equation}
      This term can  contribute appreciably when the argument of the exponential slows down, that is when $\Delta m_b \dot{\varphi}_*(t) = -\epsilon_b$. 
      The two conditions $\Delta m_i \dot{\varphi}_*(t) = - \epsilon_i$, for $i=a,b$,
      can only simultaneously satisfied when $(m_b - m_a) \dot{\varphi}_*(t) = \epsilon_b - \epsilon_a$, i.e.~exactly on resonance. So, the transient oscillatory terms in (\ref{eq:inducedApproximation}) may ``smear out'' the resonance slightly, but again since they are $\mathcal{O}(q^2 \alpha^2)$ and must compete with the $\mathcal{O}(q \alpha)$ direct couplings $\eta_{ba}(t)$, we do not expect that they provide a qualitative change in behavior in the dynamics, and away from resonance we can ignore the induced couplings entirely.

      \vskip 4pt
      With this out of the way, we can focus entirely on the diagonal terms, $\mathcal{E}_{b}(t) \equiv \mathcal{E}_{bb}(t)$, which are much simpler:
      \begin{align}
         \mathcal{E}_{b}(t) &= -i  \int_{\sminus \infty}^{t}\!\ud t'\, \sum_{K}  \big|\eta^{\floq{\Delta m_b}}_{K b}(t)\big|^2 e^{i \Delta m_b (\varphi_*(t) -  \varphi_*(t')) - i( \epsilon_K - \epsilon_b) (t-t')} \nonumber \\
         &= \frac{1}{2 \pi i} \sum_{\ell, m} \int_{\sminus \infty}^{t}\!\ud t'\int_{0}^{\infty}\!\ud k \, \big|\eta^{\floq{\Delta m_b}}_{K b}(t)\big|^2 e^{i \Delta m_b (\varphi_*(t) -  \varphi_*(t')) - i( \epsilon(k) - \epsilon_b) (t-t')}\,.
      \end{align}
      This is nothing more than a sum over integrals of the form we have already analyzed, and we can use the same techniques as before to attack this. In particular, the integral over $t'$ yields
      \begin{align}
        \mathcal{E}_{b}(t)&=   \frac{1}{2 \pi}\sum_{\ell, m}\sqrt{\frac{\pi }{2 \Delta m_b \gamma}}  \int_{0}^{\infty}\!\ud k\,    \big|\eta_{K b}^{\floq{\Delta m_b}}(t)\big|^2  \exp\!\left(\frac{i(\Delta m_b \dot{\varphi}_*(t) - (\epsilon(k) - \epsilon_b))^2}{2 \Delta m_b \gamma} - \frac{3 \pi i}{4} \right) \nonumber \\
         &\qquad\qquad \times  \left[ \lab{sgn}\, \Delta m_b \gamma + \erf\!\left(\frac{e^{\frac{i \pi}{4}} (\Delta m_b \dot{\varphi}_*(t) - (\epsilon(k)- \epsilon_b))}{\sqrt{2 \Delta m_b \gamma}}\right)\right] .
      \end{align}
      As discussed previously, we can think of the imaginary part as getting a saddle point contribution at $k_*^\floq{g}(t) = \sqrt{2 \mu(g \dot{\varphi}_*(t) + \epsilon_b)}$, which again only contributes if $k_*^\floq{g}(t)^2 > 0$. For this to ever happen (since $\epsilon_b < 0$), we must have that $\Delta m_b \gamma = (m - m_b) \gamma > 0$. Thus, ignoring the oscillatory terms and other transients, we have
      \begin{equation}
        \mathcal{E}_b(t) \approx -\sum_{\ell, g}\left[ \frac{i\mu \big|\eta^\floq{g}_{K_* b}(t)\big|^2}{2k_*^\floq{g}(t)} \Theta\big(k^\floq{g}_*(t)^2\big)\right] , \label{eq:realisticDeoccupationApp}
      \end{equation}
      with $K_* = \{k_*^{\floq{g}}(t), \ell, m = g+m_b\}$ and $k_*^\floq{g}(t) = \sqrt{2 \mu(g \dot{\varphi}_*(t) + \epsilon_b)}$, where the sum ranges from $\ell = 0,1, \dots, \infty$ and over all $g$ such that $|g + m_b| \leq \ell$. This is the extension of the first term in (\ref{eq:inducedApproximation}) to include other sectors of continuum states, with different angular momenta, connected to the bound state by perturbations that oscillate at different frequencies.

    \subsection{Stationary Perturbation Theory}
\label{ssec:SPT}

        We can get a better sense for the origin of  the first term in (\ref{eq:inducedApproximation}) 
        by deriving it via stationary perturbation theory. We start with the toy Hamiltonian (\ref{eq:toyHam}), with $\varphi_*(t) = \Omega_0 t$, so that
        \begin{equation}
           \mathcal{H} = \epsilon_b \es |b \rangle \langle b| + \frac{1}{2 \pi} \int_0^{\infty}\!\ud k\, \Big[\eta(k) e^{\sminus i \Omega_0t} |k \rangle \langle b | + \eta^*(k) e^{i \Omega_0t} |b \rangle \langle k | + \epsilon(k) |k \rangle \langle k | \Big]\, . \label{eq:toyHam-const-freq}
        \end{equation}
The transition rate from the bound to the unbound states is then computed with  Fermi's Golden Rule, which states that the transition probability per unit time per unit phase space volume is
        \begin{equation}
          \ud \Gamma=2\pi\es |\eta(k)|^2\,\delta\bigl(\epsilon(k)-\epsilon_b-\Omega_0\bigr)\,\frac{\ud k}{2\pi}\,.
        \end{equation}
        Using $\epsilon(k)=k^2/(2\mu)$, the fractional change in the bound state population is 
        \begin{samepage}
        \begin{equation}
        \frac{\ud \log |c_b(t)|^2}{\ud t} =  -\!\int \! \ud \Gamma= - \frac{\mu |\eta(k_*)|^2}{k_*} \Theta(k_*^2) \,,
        \label{eqn:1pt-toy}
        \end{equation}
        where $k_*=\sqrt{2\mu\,(\Omega_0+\epsilon_b)}$ and the $\Theta$ function ensures that this is only non-zero when $k_*$ is real.\end{samepage} This is the same as (\ref{eq:toyDeoccupation}), with $\gamma=0$, and is equivalent to the quantum mechanical derivation of the cross section in the photoelectric effect. 

        \vskip 4pt
        We see that the first term in (\ref{eq:inducedApproximation}) has a simple interpretation---it represents the ``steady state'' deoccupation of the bound state into the continuum that is captured by assuming the perturbation's frequency does not change in time. We can extend this  to the case of interest by adiabatically increasing the frequency $\dot{\varphi}_*(t) = \Omega_0 + \gamma t$ in~(\ref{eqn:1pt-toy}). It is not clear from Fermi's Golden Rule how slowly this frequency change needs to be in order for (\ref{eqn:1pt-toy}) to be valid, but we see from (\ref{eq:inducedApproximation}) that this stationary picture accurately captures the most important aspect of the true dynamics we use throughout the main text. 

        \vskip 4pt
        In the realistic case, the companion connects the states $|b \rangle$ and $|K \rangle$, each with definite azimuthal angular momentum $m_b$ and $m$, respectively, with a perturbation that oscillates with definite frequency, $\eta_{K b} \propto \exp[-i (m - m_b) \varphi_*(t)]$. It is trivial to extend the above discussion to the case where there are many such decay channels for the bound state, in which case we sum (\ref{eqn:1pt-toy}) over all of them. Once we adiabatically restore the frequency's time dependence, we find that this stationary perturbation theory approach recovers (\ref{eq:realisticDeoccupationApp}).

    \subsection{Nonlinear Chirp Frequency} \label{app:nonlinearChirp}

      Throughout this work, we have assumed that we can linearize the frequency and write the phase as  $\varphi_*(t) = -\epsilon_b t + \gamma t^2/2$. It will be useful to justify this approximation.

      \vskip 4pt
      Let us return to (\ref{eq:inducedEnergyApp}) and try to understand the behavior of the $t'$ integral,
      \begin{equation}
        \int_{\sminus \infty}^{t}\!\ud t'\, e^{i( \epsilon - \epsilon_b) t' - i \varphi_*(t')}\,,
      \end{equation}
      for a phase $\varphi_*(t)$ with general time dependence. This integral has essentially two contributions. One comes from the end point, which we can isolate through integration by parts,
      \begin{equation}
        \int_{\sminus \infty}^{t}\!\ud t'\, e^{i( \epsilon - \epsilon_b) t' - i \varphi_*(t')} \supset \frac{ie^{i( \epsilon - \epsilon_b) t - i \varphi_*(t)}}{\dot{\varphi}_*(t) - (\epsilon - \epsilon_b)}  + \cdots\,,
      \end{equation}
      while another can arise if $\dot{\varphi}(t_*) = \epsilon - \epsilon_b$
      for some $t' = t_*$ in the integration interval.  When such a time exists, the integral receives an additional contribution
      \begin{equation}
        \int_{\sminus \infty}^{t}\!\ud t'\, e^{i( \epsilon - \epsilon_b) t' - i \varphi_*(t')} \supset \sqrt{\frac{2 \pi}{\ddot{\varphi}_*(t_*)}} e^{i(\epsilon - \epsilon_b) t_* - i \varphi_*(t_*) - \frac{i \pi}{4}}\,,
      \end{equation}
      which we should divide in half when $t = t_*$. We obtain a rough approximation for the $t'$ integral,
      \begin{equation}
         \int_{\sminus \infty}^{t}\!\ud t'\, e^{i( \epsilon - \epsilon_b) t' - i \varphi_*(t')} \approx
         \begin{dcases} 
         \frac{ie^{i( \epsilon - \epsilon_b) t - i \varphi_*(t)}}{\dot{\varphi}_*(t) - (\epsilon - \epsilon_b)}\,, & t < t_* \\[4pt] \sqrt{\frac{\pi}{2 \ddot{\varphi}_*(t_*)}} e^{i(\epsilon - \epsilon_b) t_* - i \varphi_*(t_*) - \frac{i \pi}{4}}\,, & t = t_* \\[4pt]
          \frac{ie^{i( \epsilon - \epsilon_b) t - i \varphi_*(t)}}{\dot{\varphi}_*(t) - (\epsilon - \epsilon_b)} + \sqrt{\frac{2\pi}{\ddot{\varphi}_*(t_*)}} e^{i(\epsilon - \epsilon_b) t_* - i \varphi_*(t_*) - \frac{i \pi}{4}}\,, & t > t_* 
         \end{dcases} \,, \label{eq:approxGenTime}
      \end{equation}
      by adding these different contributions.

      \vskip 4pt
      If we use $\varphi_*(t) = -\epsilon_b t + \gamma t^2/2$ and consider the exact answer, we find that
      \begin{equation}
        \sqrt{\frac{\pi}{2 \gamma}} e^{\frac{i \epsilon^2}{2 \gamma} - \frac{i \pi}{4}} \lab{erfc}\!\left[\frac{e^{\frac{i \pi}{4}} (\epsilon - \gamma t)}{\sqrt{2 \gamma}}\right]\approx
         \begin{dcases} 
         \frac{ie^{-\frac{1}{2} i \gamma t^2 + i \epsilon t}}{\gamma t}\,, & t \ll t_* \\[4pt] \sqrt{\frac{\pi}{2 \gamma}} e^{\frac{i\epsilon^2}{2 \gamma}- \frac{i \pi}{4}}\,, & t = t_* \\[4pt] \frac{ie^{-\frac{1}{2} i \gamma t^2 + i \epsilon t}}{\gamma t} + \sqrt{\frac{2\pi}{ \gamma}} e^{\frac{i\epsilon^2}{2 \gamma}- \frac{i \pi}{4}}\,, & t \gg t_* 
         \end{dcases} \,,
      \end{equation}
      where $t_* = \epsilon(k)/\gamma$. We see that (\ref{eq:approxGenTime}) accurately captures the large $|t|$ asymptotics of the integral, and that the complicated error function is merely present to interpolate between these three regimes. Furthermore, the relevant chirp rate for the induced energy (\ref{eq:inducedEnergyApp}) is just the instantaneous chirp rate $\ddot{\varphi}_*(t)$ which we can, to excellent approximation, replace with the chirp rate defined in (\ref{eq:freqEom}) associated to the frequency $\Omega_0 = -\epsilon_b$ of the energy gap between the bound state and the continuum.

      \vskip 4pt
      We see then that the linearization of $\dot \varphi_*(t)$ is not such a dramatic approximation. The integrand in (\ref{eq:selfEnergyErfc}) will still have a similar form as to the one considered there, and we would still be able to do a saddle point computation isolating the large $|\tau|$ asymptotics and get effectively the same results we have found in the main text, up to corrections in the (small) nonlinearities we have ignored.

\newpage
\section{Markov Approximation} \label{app:Markov} 
    
    In the main text, we studied how the cloud is ionized by first constructing an effective Schr\"{o}dinger equation (\ref{eq:realisticEffSchro}) for the bound states, fully integrating out the dynamics of the continuum states and incorporating their effects in the induced couplings (\ref{eq:inducedCouplings}). This was valid in the so-called ``Markov approximation," which we justify in this appendix.

    \vskip 4pt
    Let us review how the Markov approximation comes about for a single bound state interacting with the continuum. We argued in Section~\ref{sec:realisticIonization} that we can ignore the continuum-induced interactions between the bound states off-resonance, and so this truncation to a single bound state still accurately captures the true dynamics of the system, especially when the orbital frequency is too high for any resonance to occur. By solving (\ref{eq:realisticContEom}) for the continuum state amplitudes and plugging the result into  (\ref{eq:realisticBoundEom}), we arrive at a single equation for the bound state amplitude
    \begin{equation}
      i \frac{\ud c_b}{\ud t} = \int_{\sminus \infty}^{t}\!\ud t'\, \Sigma_{b}(t, t') c_b(t')\,, \label{eq:appIntegDiff}
    \end{equation}
    in terms of the self-energy
    \begin{equation}
      \Sigma_{b}(t, t') \equiv -i \sum_{K} \eta_{b K}(t) \eta_{K b}(t') e^{-i(\epsilon_{K} - \epsilon_b)(t - t')}\,.
    \end{equation}
   Assuming that the couplings between the continuum states vanish and ignoring the transitions into other bound states, this equation of motion is exact. We then implement the Markov approximation by first integrating by parts,
    \begin{equation}
      i \frac{\ud c_b}{\ud t} = \mathcal{E}_{b}(t) c_b(t) - \int_{\sminus \infty}^{t}\!\ud t_1\, \mathcal{E}_b(t, t_1) \, \frac{\ud c_b(t_1)}{\ud t_1}\,,\label{eq:eomMarkov}
    \end{equation}
    and dropping the second term, which we will argue can be neglected. Here, we have defined
    \begin{equation}
      \mathcal{E}_b(t, t') = \int_{\sminus \infty}^{t'}\!\ud t_1\, \Sigma_{b}(t, t_1)\,,
    \end{equation}
    and the induced energy $\mathcal{E}_b(t) \equiv \mathcal{E}_b(t, t)$.

    \begin{figure}
      \centering
      \includegraphics[trim={0 6pt 0 0}]{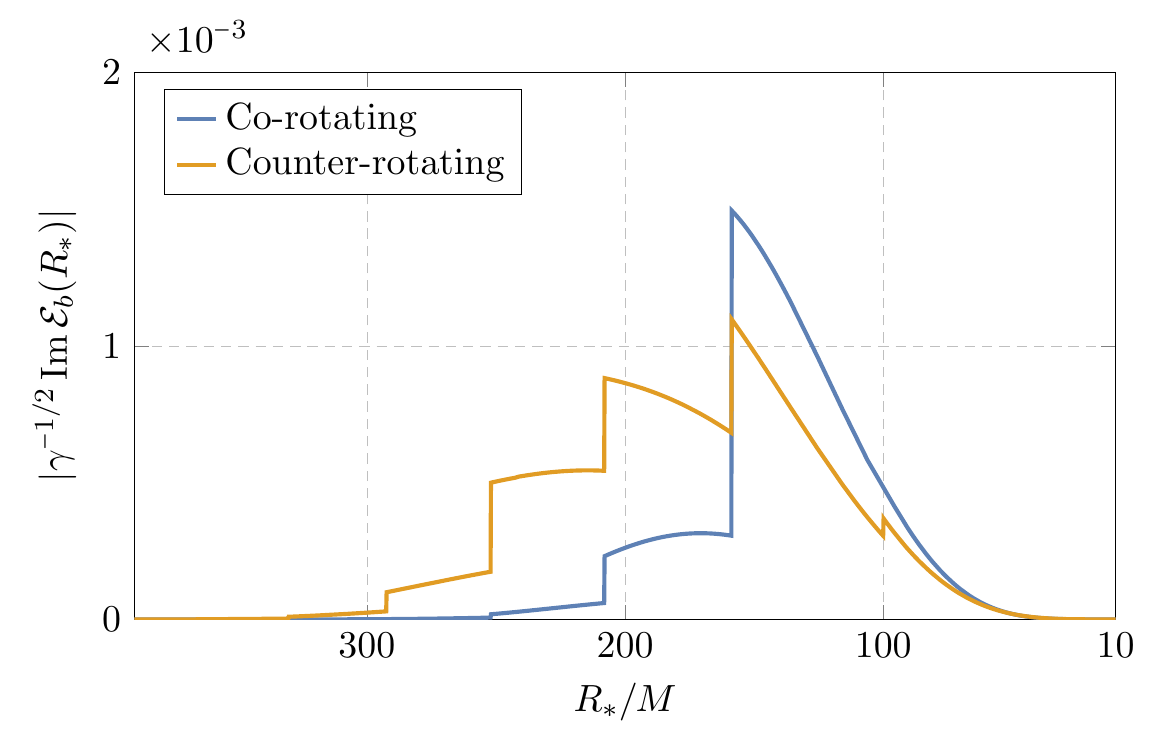}
      \caption{The dimensionless ratio $\big|\gamma^{\protect \sminus 1/2} \Im \mathcal{E}_b(R_*)\big|$ as a function of the orbital separation $R_*$, using our approximation (\ref{eq:realisticDeoccupation}) as an estimate, for an inspiral with $q = 10^{\protect \sminus 3}$ and $\alpha = 0.2$, where $\gamma$ is the instantaneous chirp rate $\gamma = \ddot{\varphi}_*(t)$, defined in (\ref{eq:freqEom}) with $\Omega_0^2 R_*^3 = (1+q)M$.  \label{fig:approxPlot}}
    \end{figure}

    \vskip 4pt
    Our goal now is to estimate the effect of the second term in (\ref{eq:eomMarkov}). To do this, we first strip off the first-order behavior by defining $\tilde{c}_b(t) = e^{i \varphi_b(t)} c_b(t)\,,$
    where $\varphi_b(t) = \int_{\sminus \infty}^{t}\!\ud t_1 \, \mathcal{E}_b(t)$ is the time-dependent phase induced at first order by the continuum. Plugging this into (\ref{eq:eomMarkov}) yields
    \begin{equation}  
      i \frac{\ud \tilde{c}_b(t)}{\ud t} = i \int_{\sminus \infty}^{t}\!\ud t_1 \, e^{i \varphi_b(t) - i \varphi_b(t_1)} \left[\mathcal{E}_b(t, t_1) \mathcal{E}_b(t_1, t_1) \tilde{c}_b(t_1) + i \mathcal{E}_b(t, t_1) \dot{\tilde{c}}_b(t_1) \right] . \label{eq:eom2Markov}
    \end{equation}
    %\JS{This equation has a similar structure to (\ref{eq:appIntegDiff}), so it will again be useful to integrate by parts.}
%
 %   \newpage
    Defining the second-order induced energy
    \begin{equation}
      \mathcal{E}_b^\floq{2}(t, t') = i \int_{\sminus \infty}^{t'}\!\ud t_1\, e^{i \varphi_b(t) - i \varphi_b(t_1)} \mathcal{E}_b(t, t_1) \es \mathcal{E}_b(t_1, t_1)\,,
    \end{equation}
    with $\mathcal{E}^\floq{2}_b(t) \equiv \mathcal{E}^\floq{2}_b(t, t)$, integrating the first term in (\ref{eq:eom2Markov}) by parts, and dropping terms that 
    \newpage
    \noindent contain factors of $\ud \tilde{c}_b/\ud t$, (\ref{eq:eom2Markov}) reduces to
    \begin{equation}
        i\frac{\ud \tilde{c}_b}{\ud t} = \mathcal{E}_b^{\floq{2}}(t)\es \tilde{c}_b(t)\,,
    \end{equation}
   % \JS{which is exactly the same form as (\ref{eq:toyEffSchro}).}
    As long as we can argue that this contribution is small compared to the first-order motion, this step of dropping terms containing $\ud \tilde{c}_b/\ud t$ is consistent. In principle, we could also iterate this process to find ever more accurate approximations to the true dynamics.

    \vskip 4pt
    It will be helpful to write the second-order induced energy as
    \begin{equation}
      \mathcal{E}_b^\floq{2}(t) = i \int_{\sminus \infty}^{t}\!\ud t_1\, e^{-\Im\left[\varphi_b(t) - \varphi_b(t_1)\right] + i \Re\left[\varphi_b(t) - \varphi_b(t_1)\right]} \, \mathcal{E}_b(t, t_1) \es\mathcal{E}_b(t_1, t_1)\,.
    \end{equation}
    Of particular importance is the oscillating phase factor, which depends on the real part of the induced phase difference $\Re\left[\varphi_b(t) - \varphi_b(t_1)\right]$. Contributions to this integral will cancel unless $t_1$ is close to $t$. Since the relevant time scale of the transition is 
    \begin{equation}
        \gamma^{\sminus 1/2} =  \sqrt{\frac{5}{96}} \frac{\alpha }{\mu}\frac{q^{\sminus \frac{1}{2}}}{(1 + q)^{\frac{3}{4}}}\! \left(\!\es\frac{ \mu R_*}{\alpha}\!\es\right)^{\!\frac{11}{4}}\,,
    \end{equation}
    we can think of $\mathcal{E}_b^\floq{2}(t)$ as being on the same order as $\gamma^{\sminus 1/2} \mathcal{E}_b(t, t)^2$. These second-order corrections are thus small as long as $\big|\gamma^{\sminus 1/2} \es \mathcal{E}_b(t)^2\big| \ll |\es \mathcal{E}_b(t)|$. Since there is typically not a hierarchy between the real and imaginary parts of $\mathcal{E}_b(t)$, we can instead write this condition as $\big|\gamma^{\sminus 1/2} \Im \mathcal{E}_b(t)\big| \ll 1$. We plot this quantity in Figure~\ref{fig:approxPlot} for the parameter values we consider in the main text and we see that it is comfortably small, so the Markov approximation is justified.

\newpage
\section{Ionization Power} \label{app:ionizedEnergy}
  
    In this appendix, we justify our approximation of the ionization power $P_\lab{ion} \equiv \ud E_\lab{ion}/\ud t$ in the toy model of Section~\ref{sec:warmup}. The extension to the realistic case is conceptually trivial.

    \vskip 4pt
    The total ionized energy is defined as
    \begin{equation}
        E_\lab{ion}(t) = \frac{1}{2 \pi}\frac{M_\lab{c}}{\mu}\int_{0}^{\infty}\!\ud k\, (\epsilon(k) - \epsilon_b) |c_k(t)|^2\,,
    \end{equation}
    where $M_\lab{c}/\mu$ represents the total occupation number of the cloud. We will set this to one and restore it at the end of the calculation.
    By taking a single time derivative we can express the ionization power as,
    \begin{equation}
  P_\lab{ion}  = \frac{1}{2 \pi}\int_{0}^{\infty}\!\ud k\, (\epsilon(k) - \epsilon_b) \left[\dot{c}_k^*(t) c_k(t) + c_k^*(t) \dot{c}_k(t)\right] .
    \end{equation} 
    and inserting both the Schr\"{o}dinger equation (\ref{eq:warmupContEom}) and the solution (\ref{eq:warmupContSol}), we can find an equation of motion for the ionized energy purely in terms of the bound state
    \begin{equation}
        P_\lab{ion}= \frac{1}{2 \pi}\int_{0}^{\infty}\!\ud k \int_{\sminus \infty}^{t}\!\ud t'  \left[   (\epsilon(k) - \epsilon_b) |\eta(k)|^2 e^{i(\varphi_*(t) - \varphi_*(t')) - i (\epsilon(k) - \epsilon_b)(t- t')} c^*_b(t) c_b(t')+ \lab{c.c.}\right] .
    \end{equation}
    This has a very similar flavor to the effective bound state equation of motion (\ref{eq:toySelfEnergyEq}), and we can implement the Markov approximation by integrating by parts and dropping the remainder,
    \begin{equation}
           P_\lab{ion} = 2\Re\!\left[\frac{1}{2 \pi}\int_{0}^{\infty}\!\ud k \int_{\sminus \infty}^{t}\!\ud t'\, (\epsilon(k) - \epsilon_b)\,  |\eta(k)|^2  e^{i(\varphi_*(t) - \varphi_*(t')) - i (\epsilon(k) - \epsilon_b)(t- t')}\right] \!|c_b(t)|^2 \,\,. \label{eq:appEIonApprox}
    \end{equation}
    This equation of motion is very similar to (\ref{eq:toyDeoccupation}), though now the term analogous to the induced energy $\mathcal{E}_b(t)$ is weighted with the energy difference $\epsilon(k) - \epsilon_b$. 

    \vskip 4pt
    This expression for the ionization power can be analyzed with the same techniques as used in Appendix~\ref{app:approx}---ignoring the transient region around $\dot{\varphi}_*(t) + \epsilon_b = 0$ and the subleading oscillatory terms, we can approximate (\ref{eq:appEIonApprox}) with its steady-state growth
    \begin{equation}
         P_\lab{ion} \approx \frac{M_\lab{c}}{\mu}\left[\frac{\mu \dot{\varphi}_*(t) |\eta(k_*(t))|^2}{k_*(t)}\right] |c_b(t)|^2 \,\Theta(k_*(t))\,,
    \end{equation}
    where we have replaced $\epsilon(k_*(t)) - \epsilon_b = \dot{\varphi}_*(t)$.

    \vskip 4pt
    One of the main benefits of the derivation of the deoccupation rate using stationary perturbation theory, presented in Section~\ref{ssec:SPT}, is that it makes inferring rates like the ionization power trivial. For instance, the amount of energy it takes to ionize the bound state $|b\rangle$ into the continuum state $|k; \ell m \rangle$ is $\epsilon(k) - \epsilon_b$. However, this only can happen if the perturbation's frequency matches this difference, $\epsilon(k) - \epsilon_b = g \dot{\varphi}_*$ with $g$ an integer. The rate at which energy is ionized is then determined by the rate at which the bound state is ionized into the continuum state (\ref{eqn:1pt-toy}), weighted by this energy difference and the total occupation of the bound state $|c_b(t)|^2$, and then summed over all different decay channels. This is the content of the final expression (\ref{eq:realisticIonizationPower}).

\newpage
      \section{Zero Mode} \label{app:zeroMode}

\begin{figure}[b!]
          \centering 
          \includegraphics[trim={0 6pt 0 0}]{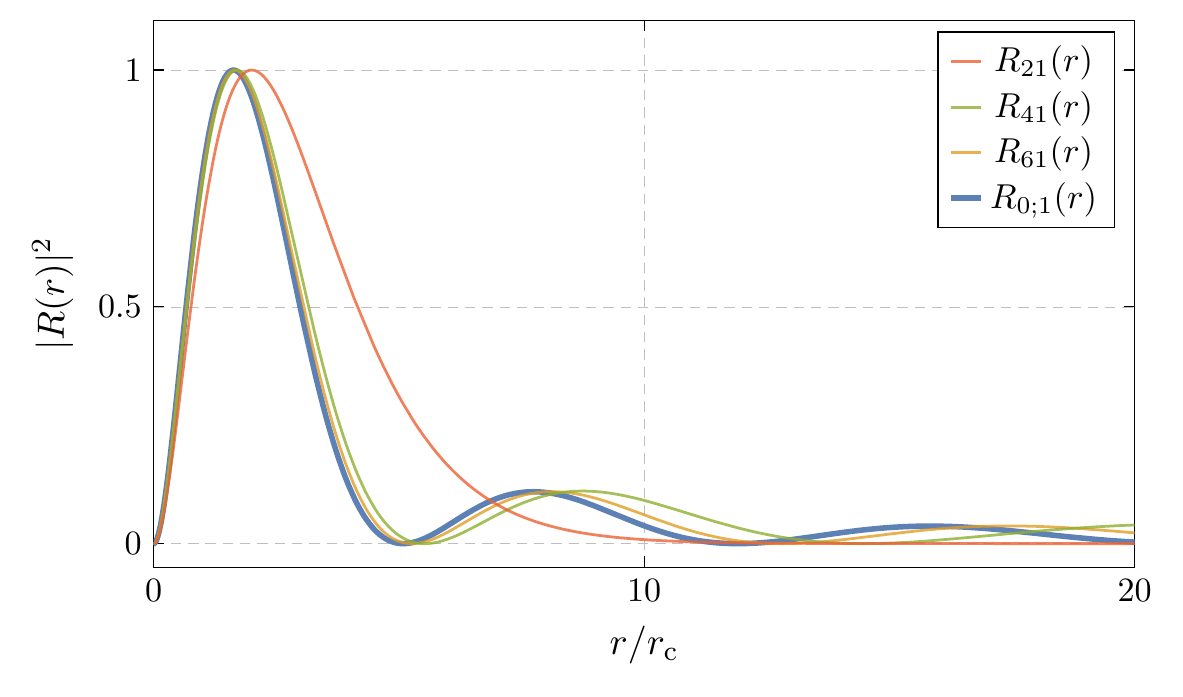}
          \caption{The radial zero mode density $\lim_{k \to 0} | k^{\protect \sminus 1/2} R_{k; 1}(r)|^2$ compared to several bound state densities, all with orbital angular momentum $\ell = 1$. Here, $r_\lab{c} = (\mu \alpha)^{\protect \sminus 1}$ is the typical radius of the cloud, and we have normalized each density so that it has unit maximum. Ignoring the overall normalization, the zero mode wavefunction can also be thought of as the limit of the bound state wavefunctions as $n \to \infty$. \label{fig:zeroMode}}  
        \end{figure}

        As we explained in the main text, the dramatic ``discontinuous'' behavior of the ionization power~$P_\lab{ion}$ is due to the fact that the coupling function $|\eta(k)|^2$ goes to zero linearly in $k$ as $k \to 0$. We mentioned there that this is because the long-range Coulombic potential keeps the zero mode relatively well-localized about the origin, as illustrated in Figure~\ref{fig:zeroMode}, such that the couplings in energy $|\eta(\epsilon)|^2 \equiv \ud k(\epsilon)/\ud \epsilon\, |\eta(k(\epsilon))|^2$ are finite as $\epsilon \to 0$. In this appendix, we discuss the zero mode of the hydrogen atom, its normalization, and the role the long-ranged $1/r$ potential plays in its radial behavior.

        \vskip 4pt
        In order to determine the overall normalization of the zero mode, we begin by writing the normalized continuum radial wavefunctions (\ref{eq:contWavefunctions}) as
        \begin{equation}  
          R_{k; \ell}(r) = \frac{2 k \es i^{\ell} e^{\frac{\pi \mu \alpha}{2 k}} \big|\Gamma\big(\ell + 1 + \tfrac{i \mu \alpha}{k}\big)\big|}{(\minus 2 i k r)^{\frac{1}{2}}\es\Gamma\big(\ell+1 + \frac{i \mu \alpha}{k}\big)} \,  e^{\sminus i k r}\! \int_{0}^{\infty}\!\ud \zeta\, e^{\sminus \zeta + \frac{i \mu \alpha}{k} \log \zeta} \zeta^{\sminus \frac{1}{2}} J_{2 \ell+1}\!\left(2 \sqrt{\minus 2 i k r \zeta}\right) ,
        \end{equation}
        where we have used a standard integral representation of the confluent hypergeometric function in terms of the Bessel function of the first kind $J_{\nu}(z)$. 
        As $k \to 0$, the integral is localized around its saddle point $\zeta = i \mu \alpha/k$ and asymptotes to
        \begin{equation}
          R_{k; \ell}(r) \sim \sqrt{\frac{4 \pi k}{r}} J_{2 \ell+1}\big(2 \sqrt{2 \mu \alpha r} \hskip 1pt\big)\,,\mathrlap{\qquad k \to 0\,.} \label{eq:zeroAsympt}
        \end{equation}
      It is then clear that any matrix element between a continuum state and a bound state will also scale as $\sqrt{k}$ for $k \to 0$, so that $|\eta(k)|^2/k$ approaches a finite, non-zero limit as $k \to 0$.

        \vskip 4pt
        We can understand this scaling in a less opaque way by considering the Schr\"{o}dinger equation with a potential that asymptotes to a generic power law, $V(r) \sim 1/r^{\Delta}$ as $r \to \infty$, with $\Delta > 0$. Defining $\rho = 1/r$, the radial Schr\"{o}dinger equation for a state with energy $\epsilon(k) = {k^2}/{2 \mu}$ can then be written as
        \begin{equation}
          \left(-\frac{\ud^2}{\ud \rho^2} + \frac{\ell(\ell+1)}{\rho^2} - \frac{2 \alpha \mu^{2}}{\rho^{4}} \frac{\rho^\Delta}{\mu^\Delta} - \frac{k^2}{\rho^4}\right)R_{k;\ell}(\rho) = 0 \,, \label{eq:schroInv}
        \end{equation}
where we have introduced additional factors of $\mu$ to keep $\alpha$ dimensionless.
        We will only be concerned with the behavior of the solutions as $\rho \to 0$ or, analogously, as $r \to \infty$, so we have replaced the potential with its dominant long-distance behavior. If $\Delta > 2$, then the potential term is subleading to the centrifugal $\ell(\ell+1)/r^2$ term and the asymptotics of $R_{k; \ell}(\rho)$ are identical to that of a free particle.

        \vskip 4pt 
        For long-ranged potentials, $0 < \Delta < 2$, we can determine the overall normalization of the continuum wavefunctions as $k \to 0$ via a matching procedure. The basic idea is that the potential singularity $2 \alpha \mu^{2 -\Delta}/\rho^{4 - \Delta}$ in (\ref{eq:schroInv}) dominates over the energy singularity $k^2/\rho^4$ in the region $\rho \gtrsim \mu \big[ (k/\mu)^2/\alpha\big]^{1/\Delta}$. When $\rho$ is smaller than this, the energy singularity dominates, so we can construct asymptotic approximations to $R_{k;\ell}(\rho)$ that are valid in these two different regions. When $k$ is very small, the region $\rho \gtrsim \mu \big[ (k/\mu)^2/\alpha\big]^{1/\Delta}$ comprises most of space, and so this is the relevant solution in the $k \to 0$ limit. However, the overall normalization of the continuum wavefunctions is set for $\rho \lesssim \mu \big[ (k/\mu)^2/\alpha\big]^{1/\Delta}$, and so we must deduce the overall normalization in the $k \to 0$ limit by matching. Our goal then is to first determine the asymptotic behavior of $R_{k;\ell}(\rho)$ around each of these singularities and then match them.

        \vskip 4pt
        Depending on the value of $\Delta$, the asymptotic behavior of $R_{k;\ell}(\rho)$ in the region near the energy singularity can be relatively complicated,
        \begin{equation}
          R_{k; \ell}(\rho) \sim A \es \rho \es \sin \!\left(\frac{k}{\rho} + \sum_{n = 1}^{n \Delta \leq 1} \frac{(\minus \frac{1}{2}\alpha)^n (2n)!}{(2n-1)(n!)^2} \frac{(k/\mu)^{1- 2n}}{n \Delta -1} \frac{\rho^{n \Delta -1}}{\mu^{n \Delta-1}} + \delta\right) , \quad  \rho \lesssim \mu \left[ \frac{(k/\mu)^2}{\alpha}\right]^{1/\Delta},\label{eq:zeroModeLess}
        \end{equation}
        where the sum is over all $n$ such that $n \Delta \leq 1$, and a $n \Delta = 1$ term should be understood to give a logarithmic correction. Here, $A$ and $\delta$ are the overall normalization and phase, respectively. For example, the asymptotic behavior of wavefunctions for the Coulombic potential, with $\Delta = 1$, is  
        \begin{equation}
          R_{k; \ell}(\rho) \sim A \es \rho \es \sin \!\left(\frac{k}{\rho} + \frac{\mu \alpha}{k} \log \frac{k}{\rho} + \delta\right) ,
        \end{equation}
        and demanding these wavefunctions are appropriately normalized, $\langle k; \ell\es m| k'; \ell \es m \rangle = 2\pi\delta(k - k')$, sets the overall amplitude in this region to $A = 2$. In contrast, the asymptotic behavior of $R_{k;\ell}(\rho)$ in the region where the potential singularity dominates is relatively simple,
        \begin{equation}
          R_{k; \ell}(r) \sim A' \es \rho^{1 - {\Delta/4}} \sin \!\left(\frac{2 \sqrt{2 \alpha (\rho/\mu)^{\Delta - 2}}}{2 - \Delta} + \delta'\right) , \ \quad \rho \gtrsim \mu \left[ \frac{(k/\mu)^2}{\alpha}\right]^{1/\Delta}\,, \label{eq:zeroModeMore}
        \end{equation}
        where again $A'$ and $\delta'$ are an undetermined amplitude and phase.

        \vskip 4pt

        In the limit $k \to 0$, the region of (\ref{eq:zeroModeLess})'s validity, $\rho \lesssim \mu \big[ (k/\mu)^2/\alpha\big]^{1/\Delta}$, shrinks to a point, and the continuum wavefunctions are well approximated by (\ref{eq:zeroModeMore}) as $\rho \to 0$. However, we do not yet know its amplitude $A'$ or, specifically, the $k$-scaling of its amplitude. We can determine this scaling by matching the amplitudes of (\ref{eq:zeroModeLess}) and (\ref{eq:zeroModeMore}) in the  region where both expansions apply,~$\rho \sim \mu \big[ (k/\mu)^2/\alpha\big]^{1/\Delta}$. We find that the continuum wavefunctions then behave as
        \begin{equation}
          R_{k; \ell}(r) \propto \frac{\sqrt{k}}{r^{\frac{1}{4}(4-\Delta)}} \sin\!\left(\frac{2 \sqrt{2\alpha (\mu r)^{2 - \Delta}}}{2 - \Delta} + \tilde{\delta}\right), \mathrlap{\qquad \begin{aligned} k &\to 0 \\[-6pt] r &\to \infty \end{aligned}\ ,}
        \end{equation}
        for arbitrary $0 < \Delta < 2$, with $\tilde{\delta}$ an undetermined phase. As long as the potential is sufficiently long-ranged, $\Delta < 2$, the continuum wavefunctions therefore asymptote to a fixed radial function multiplied by an overall factor of $\sqrt{k}$ as $k \to 0$. This implies that, for $\Delta < 2$, the potential is sufficiently long-ranged enough to localize the zero mode.
        We can compare this general result with the asymptotic expansion of (\ref{eq:zeroAsympt}), in which case $\Delta = 1$ and 
        \begin{equation}
          R_{k; \ell}(r) \sim \frac{2 \sqrt{k}}{(2 \mu \alpha)^{1/4} r^{3/4}} \sin \!\left(2 \sqrt{2 \mu \alpha r} - \pi \ell - \frac{\pi }{4}\right), \mathrlap{\qquad \begin{aligned} k &\to 0 \\[-6pt] r &\to \infty \end{aligned}\,\,,}
        \end{equation}
        in agreement with our predicted scaling.

        \vskip 4pt
        This scaling can be contrasted with that of a free particle. In this case, the effective potential due to angular momentum $\ell(\ell+1)/\rho^2$ dominates the $\rho \to 0$ limit, and, for $k \to 0$, the radial wavefunction behaves as
        \begin{equation}
          R_{k; \ell}(\rho) \sim  C_1 \rho^{\ell + 1} + C_2 \rho^{\sminus \ell}\,, \mathrlap{\qquad \begin{aligned} k &\to 0 \\[-6pt] \rho&\to 0 \end{aligned}\,\,.}       
        \end{equation}
        The appropriate $k \neq 0$ continuum wavefunctions are, instead, just the spherical Bessel functions,
        \begin{equation}
          R_{k; \ell}(\rho) = 2 k j_{\ell} (k/\rho)\,,
        \end{equation}
        which obey the asymptotic scaling
        \begin{equation}
          R_{k; \ell}(\rho) \sim \frac{2^\ell \ell! \, k}{\big(\ell + \frac{1}{2}\big) (2 \ell)!} \left(\frac{k}{\rho}\right)^{\ell}\,,\mathrlap{\qquad k \to 0\,.}
        \end{equation}
        Unlike for potentials with $0 < \Delta < 2$, these continuum wavefunctions do not have a normalization that scales as $\sqrt{k}$ as $k \to 0$, and indeed are \emph{not} localized near the origin. We see that $\Delta = 2$ represents a qualitative dividing line in the behavior of the continuum modes in the $k \to 0$ limit. The matrix elements between a bound state and the zero mode of a potential with $\Delta \geq 2$ obeys $|\eta(k)|^2/k \to 0$, while this approaches a finite limit for potentials with $0 < \Delta < 2$.

\newpage
\section{More on Scalars around Kerr}
\label{app:heunc}

The aim of this appendix is to present self-contained overview of the exact solutions for the definite frequency modes of a massive scalar field around a Kerr black hole.

\subsection{Definite Frequency Solutions}

  The Kerr geometry has two relevant isometries: time translations and azimuthal rotations. This suggests that we choose an ansatz for the scalar field profile, with a definite frequency, $\omega$, and azimuthal angular momentum, $m \in \mathbb{Z}$: %under these two isometries, so we will search for scalar field profiles with the form
  \begin{equation}
    \Phi(t, \mb{r}) = e^{-i \omega t + i m \phi} R(r) S(\theta)\, .\label{eqn:Phi-separation-no-sum}
  \end{equation}
  It is a special property of the Kerr background that this ansatz separates the Klein--Gordon equation (\ref{eq:kgEom}) into the angular spheroidal equation
  \begin{equation}
      \left(-\frac{1}{\sin \theta} \frac{\ud}{\ud \theta}\!\left(\sin \theta \frac{\ud}{\ud \theta}\right) - k^2 a^2 \cos^2 \theta + \frac{m^2}{\sin^2 \theta}\right) S(\theta) = \lambda S(\theta) \,,
  \end{equation}
  and the radial equation
  \begin{equation}
    \begin{aligned}
      0 =\ & \frac{1}{\Delta R} \frac{\ud}{\ud r}\!\left(\!\Delta \frac{\ud R}{\ud r}\right) + k^2  + \frac{P_\subp^2}{(r - r_\subp)^2} + \frac{P_\subm^2}{(r - r_\subm)^2}  \\
     & - \frac{A_\subp}{(r_\subp - r_\subm) (r - r_\subp)} + \frac{A_\subm}{(r_\subp - r_\subm)(r - r_\subm)}\,,
   \end{aligned}\label{eqn:radial}
\end{equation}
where we have introduced $k^2 = \omega^2 - \mu^2$, the eigenvalue of the spheroidal equation $\lambda$, and the parameter combinations
\begin{equation}
  \begin{aligned}
    P_{\subpm} &= \frac{m a - 2 M \omega r_\subpm}{r_\subp - r_\subm} \,, \\
    A_\subpm &= P_\subp^2 + P_\subm^2 + \gamma_{\subpm}^2 + \lambda \,,
  \end{aligned} \label{eq:radialParameters}
\end{equation}
with $\gamma^2_\subpm = \mu^2 r_\subpm^2  - \omega^2 (4 M^2 + 2 M r_\subpm + r_\subpm^2)$.

\vskip 4pt
Requiring the solution to be regular at $\theta = 0$ and $\pi$, forces the spheroidal eigenvalue $\lambda = \lambda_{\ell m}(c)$ to take a set of discrete values, depending on the spheroidicity parameter $c = k a$ and labeled by $\ell =0, 1, \dots$ and $|m| \leq \ell$. The corresponding angular functions $S(\theta) = S_{\ell m}(c; \cos \theta)$ are the ``spheroidal harmonics," which reduce to the ordinary spherical harmonics for $c=0$.

  \vskip 4pt
The radial equation (\ref{eqn:radial}) has three singularities: one at the outer horizon $r = r_\subp$ controlled by the parameter $P_\subp^2$, one at the inner horizon $r = r_\subm$ controlled by $P_\subm^2$, and an irregular singularity at $r = \infty$ controlled by $k^2$, which can be understood as the confluence of two regular singularities. This uniquely identifies the radial equation as a form of the ``confluent Heun equation," and we expect the radial solutions $R(r)$ to be proportional to the confluent Heun function, which we will define now.\footnote{
  It is useful to compare this to the Schr\"{o}dinger equation of the hydrogen atom, which has both a regular singularity at $r = 0$ and an irregular singularity at $r = \infty$ that can also be understood as the confluence of two regular singularities. Any linear differential equation with three regular singular points can be mapped to the hypergeometric equation with singularities at $z = 0$, $1$ and $\infty$. The solution to this equation that is regular about $z = 0$ is the familiar hypergeometric function ${}_2 F_1(a, b; c;z)$. Upon the confluence of the singularities at $z = 1$ and $z = \infty$, this turns into the confluent hypergeometric equation, and the regular solution ${}_2 F_{1}(a, b; c; z)$ turns into the confluent hypergeometric function ${}_1 F_{1}(a; c; z)$. An analogous story applies to the radial equation in the Kerr background, except it has an additional regular singularity at the inner horizon $r= r_\subm$. Any linear differential equation with four regular singular points can be mapped to the Heun equation, and upon a confluence of two singularities this reduced to the confluent Heun equation.}

\vskip 4pt
  Our goal is to find solutions on $r \in [r_\subp, \infty)$ that are purely ingoing at the outer horizon $r = r_\subp$, since no physical mode can escape from the black hole. Near the outer horizon, the singularity forces solutions to behave as $R(r) \sim (r - r_\subp)^{\pm i P_{\vphantom{|}\scalebox{0.6}{$+$}}}$, where the plus sign in the exponent corresponds to purely ingoing modes. Similarly, the singularity at $r = \infty$ forces the modes to behave as $R(r) \sim e^{\pm i k r}$. It will be convenient to define $z \equiv -(r - r_\subp)/(r_\subp - r_\subm)$ and peel these asymptotic behaviors from the solution,
  \begin{equation}
    R(r) = e^{- i k(r - r_\subp)} z^{i P_{\vphantom{|}\scalebox{0.6}{$+$}}} (z - 1)^{\sminus i P_{\vphantom{|}\scalebox{0.55}[0.6]{$-$}}} H(z)\,.
  \end{equation} 
  The function $H(z)$ then satisfies the confluent Heun equation \cite{ronveaux1995heun,Fiziev:2009kh}:
  \begin{equation}
    \frac{\ud^2 H}{\ud z^2} + \left( \alpha + \frac{1 + \beta}{z} + \frac{1 + \gamma}{z - 1}\right) \frac{\ud H}{\ud z} + \left(\frac{\mu}{z} + \frac{\nu}{z - 1}\right) H = 0 \,, \label{eqn:heunc-usual}
  \end{equation}
  where
  \beq
  \begin{aligned}
  \mu &= \frac{1}{2}(\alpha - \beta - \gamma + \alpha \beta - \beta \gamma) - \eta \,,\\
  \nu &= \frac{1}{2}( \alpha + \beta + \gamma + \alpha \gamma + \beta \gamma) + \delta + \eta\,,
  \end{aligned}
 \eeq
 with $ \alpha = 2 i k (r_\subp - r_\subm)\,,$ $\beta = 2 i P_\subp\,,$ $ \gamma = -2 i P_\subm\,,$ $\delta = A_\subp - A_\subm\,,$ and  $\eta = \minus A_\subp\,.$
Equation~(\ref{eqn:heunc-usual}) has a solution that is regular at the origin, $H(0) = 1$, called the confluent Heun function, $H(z) = \HeunC(\alpha, \beta, \gamma, \delta, \eta; z)$, and one which behaves as $z^{\sminus 2 i P_{\vphantom{|}\scalebox{0.55}[0.6]{$-$}}}$ as $z \to 0$. Since we impose purely ingoing boundary conditions, we discard the latter and find that
  \begin{equation}
    \begin{aligned}
      \Phi(t, \mb{r}) &= R_{k; \ell m}(r) S_{\ell m}(k a; \cos \theta) e^{-i \omega t+i m \phi} \\
      &= \mathcal{C} \es e^{-i \omega t - i k(r -r_\subp) + i m \phi} z^{i P_{\vphantom{|}\scalebox{0.6}{$+$}}} (z - 1)^{\sminus i P_{\vphantom{|}\scalebox{0.55}[0.6]{$-$}}} \HeunC(\alpha, \beta, \gamma, \delta, \eta; z) S_{\ell m}(k a; \cos \theta)\,, \label{eq:phiExactSol}
    \end{aligned}
  \end{equation}
  where $\mathcal{C}$ is a normalization constant.

  \vskip 4pt
Using the tortoise coordinates, 
  \begin{equation}
    \begin{aligned}
      \tilde{r} &=  \frac{2 M}{r_\subp - r_\subm} \left[ r_\subp \log \!\left(\frac{r - r_\subp}{r_\subp - r_\subm}\right) - r_\subm \log\!\left(\frac{r - r_\subm}{r_\subp - r_\subm}\right)\right] + r \,, \\
      \tilde{\phi} &= \frac{a}{r_\subp - r_\subm}\left[ \log\!\left(\frac{r - r_\subp}{r_\subp - r_\subm}\right) - \log\!\left(\frac{r - r_\subp}{r_\subp - r_\subm}\right) \right] ,
    \end{aligned}
  \end{equation}
the solution can be written as
  \begin{equation}
    \Phi(t, \bm{r}) = \mathcal{C}\es e^{-i k(r -r_\subp) - i \omega(t + \tilde{r} - r) + i m (\phi + \tilde{\phi})} \HeunC(\alpha, \beta, \gamma, \delta, \eta; z) S_{\ell m} (k a; \cos \theta)\,.
  \end{equation}
  Since the combination $\tilde{r} - r$ increases as we move away from the outer horizon, this mode indeed represents a purely ingoing wave.

  \vskip 4pt
  There are two classes of solutions that we use throughout the main text. The first are the quasi-bound states, which are purely ingoing at the outer horizon and exponentially decaying as $r \to \infty$. These two boundary conditions can only be satisfied for a discrete set of frequencies $\omega_{n \ell m} = E_{n \ell m} + i \es \Gamma_{n \ell m}$, cf. (\ref{eq:boundEnergies}), and so these mode only come in a discrete set. The second are the unbound continuum states, which are purely ingoing at the outer horizon, but oscillate as $r \to \infty$. Since we impose only one boundary condition, these unbound modes comprise a continuous set with frequencies $\omega^2 = \mu^2 + k^2$.

\subsection{Non-Relativistic Limit}

The first four parameters of the confluent Heun function %(\ref{eq:HeunParameters}) 
are either first order ($\alpha$, $\beta$, $\gamma$) or second order ($\delta$) in the dimensionless combinations $\mu M$ and $k M$. The fifth parameter, on the other hand, is generally $\eta = \mathcal{O}(1)$, because
\begin{equation}
    \lambda_{\ell m}(c) = \ell(\ell + 1) - \frac{1}{2}\left[1 - \frac{(2 m - 1)(2m+1)}{(2 \ell - 1) (2 \ell + 3)}\right] c^2 + \mathcal{O}\big(c^4\big)\,.
\end{equation}
The only exception is when $\ell=0$, where $\eta$ is second order in both $\mu M$ and $kM$. Modes with non-zero angular momentum see a centrifugal barrier which forces the field away from the black hole, suppressing its amplitude at radii below $\sim \ell^2/\big(\mu^2 M\big)$. This is not the case for the $\ell = 0$ mode, whose amplitude is not suppressed near the horizon. 

\vskip 4pt
In the main text, we need the profile of the $\ell = 0$ mode in the non-relativistic  ($k M \ll 1$) and fuzzy~($\mu M \ll 1$) limits. In this case, the confluent Heun function can be expanded to second order in $\mu M$ and $k M$, but at fixed $z$, as\footnote{Here, $\dilog(1-z)=\Li_2(z)=\sum_{n=1}^\infty z^n/n^2$.}${}^,$\footnote{The procedure consists in finding a recurrence relation among the coefficients of the power series $\HeunC(\alpha, \beta, \gamma, \delta, \eta; z) =\sum_{n=0}^\infty a_nz^n$, of the form $P_na_n=Q_na_{n-1}+R_na_{n-2}$, see e.g.~\cite{ronveaux1995heun,Hui:2019aqm}. After solving it to second order in $\alpha$, $\beta$, $\gamma$ and first order in $\delta$, $\eta$, the series can be resummed to give (\ref{eqn:heunc-expansion}).}
\begin{equation}
  \begin{split}
      \HeunC(\alpha,\beta&,\gamma,\delta,\eta;z)= 1-\frac{1}{2} \alpha z +\frac{1}{6} \alpha^2 z^2- \frac{1}{24} \big(\alpha^2 +12 \delta\big)z +\frac{1}{4}\big(\alpha\beta+\alpha\gamma\big)\,z\log(1-z) \\
      & - \frac{1}{2}(\beta + \gamma) \log(1 - z)  +\frac{1}{4}\big(\gamma^2-\beta^2\big)\dilog(1-z) + \frac{1}{4}\big(\beta\gamma+ \gamma^2\big)\log^2(1-z)\\
      & - \frac{1}{24} \big(\alpha^2 - 6 \beta^2 - 6 \gamma^2 + 24 \eta + 12 \delta\big) \log(1 - z) +\cdots  \, .\label{eqn:heunc-expansion}
    \end{split}
\end{equation}
In the first line, we have grouped terms that are dominant as $z \to - \infty$, while the next two lines contain terms that are subdominant and can be ignored. Given that $\alpha\sim \mathcal{O}(kM)$ and $\delta\sim\beta^2\sim\gamma^2\sim \mathcal{O}(\mu^2M^2)$, we see that the confluent Heun function is approximately constant
\begin{equation}
\HeunC(\alpha, \beta, \gamma, \delta, \eta; z)  \sim 1+\mathcal O\bigl(\mu M,kM\bigr)\,, \qquad r_\subp \le r<r\ped{max}\,,
\end{equation}
until the linear or quadratic terms in the first line of (\ref{eqn:heunc-expansion}) become $\mathcal{O}(1)$. This occurs at the radius
\begin{equation}
\frac{r\ped{max}}{M} \sim \min\biggl\{\frac1{(\mu M)^2},\frac1{kM}\biggr\}\gg 1\,.
\end{equation}
We use this approximation to derive the accretion rate in Section~\ref{sec:accretion}.

\newpage

\clearpage

\phantomsection

\addtocontents{toc}{\protect\vskip24pt}
\addcontentsline{toc}{section}{References}

\makeatletter

\interlinepenalty=10000

{\linespread{1.09}
\bibliographystyle{utphys}
\bibliography{main}
}
\makeatother

\end{document}